\documentclass[acmtocl]{acmtrans2m}

\usepackage{times}
\usepackage{amsmath}
\usepackage{amsfonts}
\usepackage{latexsym}
\usepackage{xspace}
\usepackage{url}
\usepackage{color}
\usepackage{epsfig}
\usepackage[dvips]{rotating}
\usepackage{paralist}
\usepackage{verbatim} 
\usepackage{proof}

\newcommand\noabmixmu{\eta}
\newcommand\abmixmu{\mu}

\newcommand{\ignore}[1]{}
\newcommand{\sref}[1]{\S{}\ref{#1}}

\definecolor {snow}                {rgb}{1.00,0.98,0.98}
\definecolor {ghostwhite}          {rgb}{0.97,0.97,1.00}
\definecolor {whitesmoke}          {rgb}{0.96,0.96,0.96}
\definecolor {gainsboro}           {rgb}{0.86,0.86,0.86}
\definecolor {floralwhite}         {rgb}{1.00,0.98,0.94}
\definecolor {oldlace}             {rgb}{0.99,0.96,0.90}
\definecolor {linen}               {rgb}{0.98,0.94,0.90}
\definecolor {antiquewhite}        {rgb}{0.98,0.92,0.84}
\definecolor {papayawhip}          {rgb}{1.00,0.94,0.84}
\definecolor {blanchedalmond}      {rgb}{1.00,0.92,0.80}
\definecolor {bisque}              {rgb}{1.00,0.89,0.77}
\definecolor {peachpuff}           {rgb}{1.00,0.85,0.73}
\definecolor {navajowhite}         {rgb}{1.00,0.87,0.68}
\definecolor {moccasin}            {rgb}{1.00,0.89,0.71}
\definecolor {cornsilk}            {rgb}{1.00,0.97,0.86}
\definecolor {ivory}               {rgb}{1.00,1.00,0.94}
\definecolor {lemonchiffon}        {rgb}{1.00,0.98,0.80}
\definecolor {seashell}            {rgb}{1.00,0.96,0.93}
\definecolor {honeydew}            {rgb}{0.94,1.00,0.94}
\definecolor {mintcream}           {rgb}{0.96,1.00,0.98}
\definecolor {azure}               {rgb}{0.94,1.00,1.00}
\definecolor {aliceblue}           {rgb}{0.94,0.97,1.00}
\definecolor {lavender}            {rgb}{0.90,0.90,0.98}
\definecolor {lavenderblush}       {rgb}{1.00,0.94,0.96}
\definecolor {mistyrose}           {rgb}{1.00,0.89,0.88}
\definecolor {white}               {rgb}{1.00,1.00,1.00}
\definecolor {black}               {rgb}{0.00,0.00,0.00}
\definecolor {darkslategray}       {rgb}{0.18,0.31,0.31}
\definecolor {dimgray}             {rgb}{0.41,0.41,0.41}
\definecolor {slategray}           {rgb}{0.44,0.50,0.56}
\definecolor {lightslategray}      {rgb}{0.47,0.53,0.60}
\definecolor {gray}                {rgb}{0.75,0.75,0.75}
\definecolor {lightgrey}           {rgb}{0.83,0.83,0.83}
\definecolor {midnightblue}        {rgb}{0.10,0.10,0.44}
\definecolor {navy}                {rgb}{0.00,0.00,0.50}
\definecolor {cornflowerblue}      {rgb}{0.39,0.58,0.93}
\definecolor {darkslateblue}       {rgb}{0.28,0.24,0.55}
\definecolor {slateblue}           {rgb}{0.42,0.35,0.80}
\definecolor {mediumslateblue}     {rgb}{0.48,0.41,0.93}
\definecolor {lightslateblue}      {rgb}{0.52,0.44,1.00}
\definecolor {mediumblue}          {rgb}{0.00,0.00,0.80}
\definecolor {royalblue}           {rgb}{0.25,0.41,0.88}
\definecolor {blue}                {rgb}{0.00,0.00,1.00}
\definecolor {dodgerblue}          {rgb}{0.12,0.56,1.00}
\definecolor {deepskyblue}         {rgb}{0.00,0.75,1.00}
\definecolor {skyblue}             {rgb}{0.53,0.81,0.92}
\definecolor {lightskyblue}        {rgb}{0.53,0.81,0.98}
\definecolor {steelblue}           {rgb}{0.27,0.51,0.71}
\definecolor {lightsteelblue}      {rgb}{0.69,0.77,0.87}
\definecolor {lightblue}           {rgb}{0.68,0.85,0.90}
\definecolor {powderblue}          {rgb}{0.69,0.88,0.90}
\definecolor {paleturquoise}       {rgb}{0.69,0.93,0.93}
\definecolor {darkturquoise}       {rgb}{0.00,0.81,0.82}
\definecolor {mediumturquoise}     {rgb}{0.28,0.82,0.80}
\definecolor {turquoise}           {rgb}{0.25,0.88,0.82}
\definecolor {cyan}                {rgb}{0.00,1.00,1.00}
\definecolor {lightcyan}           {rgb}{0.88,1.00,1.00}
\definecolor {cadetblue}           {rgb}{0.37,0.62,0.63}
\definecolor {mediumaquamarine}    {rgb}{0.40,0.80,0.67}
\definecolor {aquamarine}          {rgb}{0.50,1.00,0.83}
\definecolor {darkgreen}           {rgb}{0.00,0.39,0.00}
\definecolor {darkolivegreen}      {rgb}{0.33,0.42,0.18}
\definecolor {darkseagreen}        {rgb}{0.56,0.74,0.56}
\definecolor {seagreen}            {rgb}{0.18,0.55,0.34}
\definecolor {mediumseagreen}      {rgb}{0.24,0.70,0.44}
\definecolor {lightseagreen}       {rgb}{0.13,0.70,0.67}
\definecolor {palegreen}           {rgb}{0.60,0.98,0.60}
\definecolor {springgreen}         {rgb}{0.00,1.00,0.50}
\definecolor {lawngreen}           {rgb}{0.49,0.99,0.00}
\definecolor {green}               {rgb}{0.00,1.00,0.00}
\definecolor {chartreuse}          {rgb}{0.50,1.00,0.00}
\definecolor {mediumspringgreen}   {rgb}{0.00,0.98,0.60}
\definecolor {greenyellow}         {rgb}{0.68,1.00,0.18}
\definecolor {limegreen}           {rgb}{0.20,0.80,0.20}
\definecolor {yellowgreen}         {rgb}{0.60,0.80,0.20}
\definecolor {forestgreen}         {rgb}{0.13,0.55,0.13}
\definecolor {olivedrab}           {rgb}{0.42,0.56,0.14}
\definecolor {darkkhaki}           {rgb}{0.74,0.72,0.42}
\definecolor {khaki}               {rgb}{0.94,0.90,0.55}
\definecolor {palegoldenrod}       {rgb}{0.93,0.91,0.67}
\definecolor {lightgoldenrodyellow} {rgb}{0.98,0.98,0.82}
\definecolor {lightyellow}         {rgb}{1.00,1.00,0.88}
\definecolor {yellow}              {rgb}{1.00,1.00,0.00}
\definecolor {gold}                {rgb}{1.00,0.84,0.00}
\definecolor {lightgoldenrod}      {rgb}{0.93,0.87,0.51}
\definecolor {goldenrod}           {rgb}{0.85,0.65,0.13}
\definecolor {darkgoldenrod}       {rgb}{0.72,0.53,0.04}
\definecolor {rosybrown}           {rgb}{0.74,0.56,0.56}
\definecolor {indianred}           {rgb}{0.80,0.36,0.36}
\definecolor {saddlebrown}         {rgb}{0.55,0.27,0.07}
\definecolor {sienna}              {rgb}{0.63,0.32,0.18}
\definecolor {peru}                {rgb}{0.80,0.52,0.25}
\definecolor {burlywood}           {rgb}{0.87,0.72,0.53}
\definecolor {beige}               {rgb}{0.96,0.96,0.86}
\definecolor {wheat}               {rgb}{0.96,0.87,0.70}
\definecolor {sandybrown}          {rgb}{0.96,0.64,0.38}
\definecolor {tan}                 {rgb}{0.82,0.71,0.55}
\definecolor {chocolate}           {rgb}{0.82,0.41,0.12}
\definecolor {firebrick}           {rgb}{0.70,0.13,0.13}
\definecolor {brown}               {rgb}{0.65,0.16,0.16}
\definecolor {darksalmon}          {rgb}{0.91,0.59,0.48}
\definecolor {salmon}              {rgb}{0.98,0.50,0.45}
\definecolor {lightsalmon}         {rgb}{1.00,0.63,0.48}
\definecolor {orange}              {rgb}{1.00,0.65,0.00}
\definecolor {darkorange}          {rgb}{1.00,0.55,0.00}
\definecolor {coral}               {rgb}{1.00,0.50,0.31}
\definecolor {lightcoral}          {rgb}{0.94,0.50,0.50}
\definecolor {tomato}              {rgb}{1.00,0.39,0.28}
\definecolor {orangered}           {rgb}{1.00,0.27,0.00}
\definecolor {red}                 {rgb}{1.00,0.00,0.00}
\definecolor {hotpink}             {rgb}{1.00,0.41,0.71}
\definecolor {deeppink}            {rgb}{1.00,0.08,0.58}
\definecolor {pink}                {rgb}{1.00,0.75,0.80}
\definecolor {lightpink}           {rgb}{1.00,0.71,0.76}
\definecolor {palevioletred}       {rgb}{0.86,0.44,0.58}
\definecolor {maroon}              {rgb}{0.69,0.19,0.38}
\definecolor {mediumvioletred}     {rgb}{0.78,0.08,0.52}
\definecolor {violetred}           {rgb}{0.82,0.13,0.56}
\definecolor {magenta}             {rgb}{1.00,0.00,1.00}
\definecolor {violet}              {rgb}{0.93,0.51,0.93}
\definecolor {plum}                {rgb}{0.87,0.63,0.87}
\definecolor {orchid}              {rgb}{0.85,0.44,0.84}
\definecolor {mediumorchid}        {rgb}{0.73,0.33,0.83}
\definecolor {darkorchid}          {rgb}{0.60,0.20,0.80}
\definecolor {darkviolet}          {rgb}{0.58,0.00,0.83}
\definecolor {blueviolet}          {rgb}{0.54,0.17,0.89}
\definecolor {purple}              {rgb}{0.63,0.13,0.94}
\definecolor {mediumpurple}        {rgb}{0.58,0.44,0.86}
\definecolor {thistle}             {rgb}{0.85,0.75,0.85}
\definecolor {snow2}               {rgb}{0.93,0.91,0.91}
\definecolor {snow3}               {rgb}{0.80,0.79,0.79}
\definecolor {snow4}               {rgb}{0.55,0.54,0.54}
\definecolor {seashell2}           {rgb}{0.93,0.90,0.87}
\definecolor {seashell3}           {rgb}{0.80,0.77,0.75}
\definecolor {seashell4}           {rgb}{0.55,0.53,0.51}
\definecolor {antiquewhite1}       {rgb}{1.00,0.94,0.86}
\definecolor {antiquewhite2}       {rgb}{0.93,0.87,0.80}
\definecolor {antiquewhite3}       {rgb}{0.80,0.75,0.69}
\definecolor {antiquewhite4}       {rgb}{0.55,0.51,0.47}
\definecolor {bisque2}             {rgb}{0.93,0.84,0.72}
\definecolor {bisque3}             {rgb}{0.80,0.72,0.62}
\definecolor {bisque4}             {rgb}{0.55,0.49,0.42}
\definecolor {peachpuff2}          {rgb}{0.93,0.80,0.68}
\definecolor {peachpuff3}          {rgb}{0.80,0.69,0.58}
\definecolor {peachpuff4}          {rgb}{0.55,0.47,0.40}
\definecolor {navajowhite2}        {rgb}{0.93,0.81,0.63}
\definecolor {navajowhite3}        {rgb}{0.80,0.70,0.55}
\definecolor {navajowhite4}        {rgb}{0.55,0.47,0.37}
\definecolor {lemonchiffon2}       {rgb}{0.93,0.91,0.75}
\definecolor {lemonchiffon3}       {rgb}{0.80,0.79,0.65}
\definecolor {lemonchiffon4}       {rgb}{0.55,0.54,0.44}
\definecolor {cornsilk2}           {rgb}{0.93,0.91,0.80}
\definecolor {cornsilk3}           {rgb}{0.80,0.78,0.69}
\definecolor {cornsilk4}           {rgb}{0.55,0.53,0.47}
\definecolor {ivory2}              {rgb}{0.93,0.93,0.88}
\definecolor {ivory3}              {rgb}{0.80,0.80,0.76}
\definecolor {ivory4}              {rgb}{0.55,0.55,0.51}
\definecolor {honeydew2}           {rgb}{0.88,0.93,0.88}
\definecolor {honeydew3}           {rgb}{0.76,0.80,0.76}
\definecolor {honeydew4}           {rgb}{0.51,0.55,0.51}
\definecolor {lavenderblush2}      {rgb}{0.93,0.88,0.90}
\definecolor {lavenderblush3}      {rgb}{0.80,0.76,0.77}
\definecolor {lavenderblush4}      {rgb}{0.55,0.51,0.53}
\definecolor {mistyrose2}          {rgb}{0.93,0.84,0.82}
\definecolor {mistyrose3}          {rgb}{0.80,0.72,0.71}
\definecolor {mistyrose4}          {rgb}{0.55,0.49,0.48}
\definecolor {azure2}              {rgb}{0.88,0.93,0.93}
\definecolor {azure3}              {rgb}{0.76,0.80,0.80}
\definecolor {azure4}              {rgb}{0.51,0.55,0.55}
\definecolor {slateblue1}          {rgb}{0.51,0.44,1.00}
\definecolor {slateblue2}          {rgb}{0.48,0.40,0.93}
\definecolor {slateblue3}          {rgb}{0.41,0.35,0.80}
\definecolor {slateblue4}          {rgb}{0.28,0.24,0.55}
\definecolor {royalblue1}          {rgb}{0.28,0.46,1.00}
\definecolor {royalblue2}          {rgb}{0.26,0.43,0.93}
\definecolor {royalblue3}          {rgb}{0.23,0.37,0.80}
\definecolor {royalblue4}          {rgb}{0.15,0.25,0.55}
\definecolor {blue2}               {rgb}{0.00,0.00,0.93}
\definecolor {blue4}               {rgb}{0.00,0.00,0.55}
\definecolor {dodgerblue2}         {rgb}{0.11,0.53,0.93}
\definecolor {dodgerblue3}         {rgb}{0.09,0.45,0.80}
\definecolor {dodgerblue4}         {rgb}{0.06,0.31,0.55}
\definecolor {steelblue1}          {rgb}{0.39,0.72,1.00}
\definecolor {steelblue2}          {rgb}{0.36,0.67,0.93}
\definecolor {steelblue3}          {rgb}{0.31,0.58,0.80}
\definecolor {steelblue4}          {rgb}{0.21,0.39,0.55}
\definecolor {deepskyblue2}        {rgb}{0.00,0.70,0.93}
\definecolor {deepskyblue3}        {rgb}{0.00,0.60,0.80}
\definecolor {deepskyblue4}        {rgb}{0.00,0.41,0.55}
\definecolor {skyblue1}            {rgb}{0.53,0.81,1.00}
\definecolor {skyblue2}            {rgb}{0.49,0.75,0.93}
\definecolor {skyblue3}            {rgb}{0.42,0.65,0.80}
\definecolor {skyblue4}            {rgb}{0.29,0.44,0.55}
\definecolor {lightskyblue1}       {rgb}{0.69,0.89,1.00}
\definecolor {lightskyblue2}       {rgb}{0.64,0.83,0.93}
\definecolor {lightskyblue3}       {rgb}{0.55,0.71,0.80}
\definecolor {lightskyblue4}       {rgb}{0.38,0.48,0.55}
\definecolor {slategray1}          {rgb}{0.78,0.89,1.00}
\definecolor {slategray2}          {rgb}{0.73,0.83,0.93}
\definecolor {slategray3}          {rgb}{0.62,0.71,0.80}
\definecolor {slategray4}          {rgb}{0.42,0.48,0.55}
\definecolor {lightsteelblue1}     {rgb}{0.79,0.88,1.00}
\definecolor {lightsteelblue2}     {rgb}{0.74,0.82,0.93}
\definecolor {lightsteelblue3}     {rgb}{0.64,0.71,0.80}
\definecolor {lightsteelblue4}     {rgb}{0.43,0.48,0.55}
\definecolor {lightblue1}          {rgb}{0.75,0.94,1.00}
\definecolor {lightblue2}          {rgb}{0.70,0.87,0.93}
\definecolor {lightblue3}          {rgb}{0.60,0.75,0.80}
\definecolor {lightblue4}          {rgb}{0.41,0.51,0.55}
\definecolor {lightcyan2}          {rgb}{0.82,0.93,0.93}
\definecolor {lightcyan3}          {rgb}{0.71,0.80,0.80}
\definecolor {lightcyan4}          {rgb}{0.48,0.55,0.55}
\definecolor {paleturquoise1}      {rgb}{0.73,1.00,1.00}
\definecolor {paleturquoise2}      {rgb}{0.68,0.93,0.93}
\definecolor {paleturquoise3}      {rgb}{0.59,0.80,0.80}
\definecolor {paleturquoise4}      {rgb}{0.40,0.55,0.55}
\definecolor {cadetblue1}          {rgb}{0.60,0.96,1.00}
\definecolor {cadetblue2}          {rgb}{0.56,0.90,0.93}
\definecolor {cadetblue3}          {rgb}{0.48,0.77,0.80}
\definecolor {cadetblue4}          {rgb}{0.33,0.53,0.55}
\definecolor {turquoise1}          {rgb}{0.00,0.96,1.00}
\definecolor {turquoise2}          {rgb}{0.00,0.90,0.93}
\definecolor {turquoise3}          {rgb}{0.00,0.77,0.80}
\definecolor {turquoise4}          {rgb}{0.00,0.53,0.55}
\definecolor {cyan2}               {rgb}{0.00,0.93,0.93}
\definecolor {cyan3}               {rgb}{0.00,0.80,0.80}
\definecolor {cyan4}               {rgb}{0.00,0.55,0.55}
\definecolor {darkslategray1}      {rgb}{0.59,1.00,1.00}
\definecolor {darkslategray2}      {rgb}{0.55,0.93,0.93}
\definecolor {darkslategray3}      {rgb}{0.47,0.80,0.80}
\definecolor {darkslategray4}      {rgb}{0.32,0.55,0.55}
\definecolor {aquamarine2}         {rgb}{0.46,0.93,0.78}
\definecolor {aquamarine4}         {rgb}{0.27,0.55,0.45}
\definecolor {darkseagreen1}       {rgb}{0.76,1.00,0.76}
\definecolor {darkseagreen2}       {rgb}{0.71,0.93,0.71}
\definecolor {darkseagreen3}       {rgb}{0.61,0.80,0.61}
\definecolor {darkseagreen4}       {rgb}{0.41,0.55,0.41}
\definecolor {seagreen1}           {rgb}{0.33,1.00,0.62}
\definecolor {seagreen2}           {rgb}{0.31,0.93,0.58}
\definecolor {seagreen3}           {rgb}{0.26,0.80,0.50}
\definecolor {palegreen1}          {rgb}{0.60,1.00,0.60}
\definecolor {palegreen2}          {rgb}{0.56,0.93,0.56}
\definecolor {palegreen3}          {rgb}{0.49,0.80,0.49}
\definecolor {palegreen4}          {rgb}{0.33,0.55,0.33}
\definecolor {springgreen2}        {rgb}{0.00,0.93,0.46}
\definecolor {springgreen3}        {rgb}{0.00,0.80,0.40}
\definecolor {springgreen4}        {rgb}{0.00,0.55,0.27}
\definecolor {green2}              {rgb}{0.00,0.93,0.00}
\definecolor {green3}              {rgb}{0.00,0.80,0.00}
\definecolor {green4}              {rgb}{0.00,0.55,0.00}
\definecolor {chartreuse2}         {rgb}{0.46,0.93,0.00}
\definecolor {chartreuse3}         {rgb}{0.40,0.80,0.00}
\definecolor {chartreuse4}         {rgb}{0.27,0.55,0.00}
\definecolor {olivedrab1}          {rgb}{0.75,1.00,0.24}
\definecolor {olivedrab2}          {rgb}{0.70,0.93,0.23}
\definecolor {olivedrab4}          {rgb}{0.41,0.55,0.13}
\definecolor {darkolivegreen1}     {rgb}{0.79,1.00,0.44}
\definecolor {darkolivegreen2}     {rgb}{0.74,0.93,0.41}
\definecolor {darkolivegreen3}     {rgb}{0.64,0.80,0.35}
\definecolor {darkolivegreen4}     {rgb}{0.43,0.55,0.24}
\definecolor {khaki1}              {rgb}{1.00,0.96,0.56}
\definecolor {khaki2}              {rgb}{0.93,0.90,0.52}
\definecolor {khaki3}              {rgb}{0.80,0.78,0.45}
\definecolor {khaki4}              {rgb}{0.55,0.53,0.31}
\definecolor {lightgoldenrod1}     {rgb}{1.00,0.93,0.55}
\definecolor {lightgoldenrod2}     {rgb}{0.93,0.86,0.51}
\definecolor {lightgoldenrod3}     {rgb}{0.80,0.75,0.44}
\definecolor {lightgoldenrod4}     {rgb}{0.55,0.51,0.30}
\definecolor {lightyellow2}        {rgb}{0.93,0.93,0.82}
\definecolor {lightyellow3}        {rgb}{0.80,0.80,0.71}
\definecolor {lightyellow4}        {rgb}{0.55,0.55,0.48}
\definecolor {yellow2}             {rgb}{0.93,0.93,0.00}
\definecolor {yellow3}             {rgb}{0.80,0.80,0.00}
\definecolor {yellow4}             {rgb}{0.55,0.55,0.00}
\definecolor {gold2}               {rgb}{0.93,0.79,0.00}
\definecolor {gold3}               {rgb}{0.80,0.68,0.00}
\definecolor {gold4}               {rgb}{0.55,0.46,0.00}
\definecolor {goldenrod1}          {rgb}{1.00,0.76,0.15}
\definecolor {goldenrod2}          {rgb}{0.93,0.71,0.13}
\definecolor {goldenrod3}          {rgb}{0.80,0.61,0.11}
\definecolor {goldenrod4}          {rgb}{0.55,0.41,0.08}
\definecolor {darkgoldenrod1}      {rgb}{1.00,0.73,0.06}
\definecolor {darkgoldenrod2}      {rgb}{0.93,0.68,0.05}
\definecolor {darkgoldenrod3}      {rgb}{0.80,0.58,0.05}
\definecolor {darkgoldenrod4}      {rgb}{0.55,0.40,0.03}
\definecolor {rosybrown1}          {rgb}{1.00,0.76,0.76}
\definecolor {rosybrown2}          {rgb}{0.93,0.71,0.71}
\definecolor {rosybrown3}          {rgb}{0.80,0.61,0.61}
\definecolor {rosybrown4}          {rgb}{0.55,0.41,0.41}
\definecolor {indianred1}          {rgb}{1.00,0.42,0.42}
\definecolor {indianred2}          {rgb}{0.93,0.39,0.39}
\definecolor {indianred3}          {rgb}{0.80,0.33,0.33}
\definecolor {indianred4}          {rgb}{0.55,0.23,0.23}
\definecolor {sienna1}             {rgb}{1.00,0.51,0.28}
\definecolor {sienna2}             {rgb}{0.93,0.47,0.26}
\definecolor {sienna3}             {rgb}{0.80,0.41,0.22}
\definecolor {sienna4}             {rgb}{0.55,0.28,0.15}
\definecolor {burlywood1}          {rgb}{1.00,0.83,0.61}
\definecolor {burlywood2}          {rgb}{0.93,0.77,0.57}
\definecolor {burlywood3}          {rgb}{0.80,0.67,0.49}
\definecolor {burlywood4}          {rgb}{0.55,0.45,0.33}
\definecolor {wheat1}              {rgb}{1.00,0.91,0.73}
\definecolor {wheat2}              {rgb}{0.93,0.85,0.68}
\definecolor {wheat3}              {rgb}{0.80,0.73,0.59}
\definecolor {wheat4}              {rgb}{0.55,0.49,0.40}
\definecolor {tan1}                {rgb}{1.00,0.65,0.31}
\definecolor {tan2}                {rgb}{0.93,0.60,0.29}
\definecolor {tan4}                {rgb}{0.55,0.35,0.17}
\definecolor {chocolate1}          {rgb}{1.00,0.50,0.14}
\definecolor {chocolate2}          {rgb}{0.93,0.46,0.13}
\definecolor {chocolate3}          {rgb}{0.80,0.40,0.11}
\definecolor {firebrick1}          {rgb}{1.00,0.19,0.19}
\definecolor {firebrick2}          {rgb}{0.93,0.17,0.17}
\definecolor {firebrick3}          {rgb}{0.80,0.15,0.15}
\definecolor {firebrick4}          {rgb}{0.55,0.10,0.10}
\definecolor {brown1}              {rgb}{1.00,0.25,0.25}
\definecolor {brown2}              {rgb}{0.93,0.23,0.23}
\definecolor {brown3}              {rgb}{0.80,0.20,0.20}
\definecolor {brown4}              {rgb}{0.55,0.14,0.14}
\definecolor {salmon1}             {rgb}{1.00,0.55,0.41}
\definecolor {salmon2}             {rgb}{0.93,0.51,0.38}
\definecolor {salmon3}             {rgb}{0.80,0.44,0.33}
\definecolor {salmon4}             {rgb}{0.55,0.30,0.22}
\definecolor {lightsalmon2}        {rgb}{0.93,0.58,0.45}
\definecolor {lightsalmon3}        {rgb}{0.80,0.51,0.38}
\definecolor {lightsalmon4}        {rgb}{0.55,0.34,0.26}
\definecolor {orange2}             {rgb}{0.93,0.60,0.00}
\definecolor {orange3}             {rgb}{0.80,0.52,0.00}
\definecolor {orange4}             {rgb}{0.55,0.35,0.00}
\definecolor {darkorange1}         {rgb}{1.00,0.50,0.00}
\definecolor {darkorange2}         {rgb}{0.93,0.46,0.00}
\definecolor {darkorange3}         {rgb}{0.80,0.40,0.00}
\definecolor {darkorange4}         {rgb}{0.55,0.27,0.00}
\definecolor {coral1}              {rgb}{1.00,0.45,0.34}
\definecolor {coral2}              {rgb}{0.93,0.42,0.31}
\definecolor {coral3}              {rgb}{0.80,0.36,0.27}
\definecolor {coral4}              {rgb}{0.55,0.24,0.18}
\definecolor {tomato2}             {rgb}{0.93,0.36,0.26}
\definecolor {tomato3}             {rgb}{0.80,0.31,0.22}
\definecolor {tomato4}             {rgb}{0.55,0.21,0.15}
\definecolor {orangered2}          {rgb}{0.93,0.25,0.00}
\definecolor {orangered3}          {rgb}{0.80,0.22,0.00}
\definecolor {orangered4}          {rgb}{0.55,0.15,0.00}
\definecolor {red2}                {rgb}{0.93,0.00,0.00}
\definecolor {red3}                {rgb}{0.80,0.00,0.00}
\definecolor {red4}                {rgb}{0.55,0.00,0.00}
\definecolor {deeppink2}           {rgb}{0.93,0.07,0.54}
\definecolor {deeppink3}           {rgb}{0.80,0.06,0.46}
\definecolor {deeppink4}           {rgb}{0.55,0.04,0.31}
\definecolor {hotpink1}            {rgb}{1.00,0.43,0.71}
\definecolor {hotpink2}            {rgb}{0.93,0.42,0.65}
\definecolor {hotpink3}            {rgb}{0.80,0.38,0.56}
\definecolor {hotpink4}            {rgb}{0.55,0.23,0.38}
\definecolor {pink1}               {rgb}{1.00,0.71,0.77}
\definecolor {pink2}               {rgb}{0.93,0.66,0.72}
\definecolor {pink3}               {rgb}{0.80,0.57,0.62}
\definecolor {pink4}               {rgb}{0.55,0.39,0.42}
\definecolor {lightpink1}          {rgb}{1.00,0.68,0.73}
\definecolor {lightpink2}          {rgb}{0.93,0.64,0.68}
\definecolor {lightpink3}          {rgb}{0.80,0.55,0.58}
\definecolor {lightpink4}          {rgb}{0.55,0.37,0.40}
\definecolor {palevioletred1}      {rgb}{1.00,0.51,0.67}
\definecolor {palevioletred2}      {rgb}{0.93,0.47,0.62}
\definecolor {palevioletred3}      {rgb}{0.80,0.41,0.54}
\definecolor {palevioletred4}      {rgb}{0.55,0.28,0.36}
\definecolor {maroon1}             {rgb}{1.00,0.20,0.70}
\definecolor {maroon2}             {rgb}{0.93,0.19,0.65}
\definecolor {maroon3}             {rgb}{0.80,0.16,0.56}
\definecolor {maroon4}             {rgb}{0.55,0.11,0.38}
\definecolor {violetred1}          {rgb}{1.00,0.24,0.59}
\definecolor {violetred2}          {rgb}{0.93,0.23,0.55}
\definecolor {violetred3}          {rgb}{0.80,0.20,0.47}
\definecolor {violetred4}          {rgb}{0.55,0.13,0.32}
\definecolor {magenta2}            {rgb}{0.93,0.00,0.93}
\definecolor {magenta3}            {rgb}{0.80,0.00,0.80}
\definecolor {magenta4}            {rgb}{0.55,0.00,0.55}
\definecolor {orchid1}             {rgb}{1.00,0.51,0.98}
\definecolor {orchid2}             {rgb}{0.93,0.48,0.91}
\definecolor {orchid3}             {rgb}{0.80,0.41,0.79}
\definecolor {orchid4}             {rgb}{0.55,0.28,0.54}
\definecolor {plum1}               {rgb}{1.00,0.73,1.00}
\definecolor {plum2}               {rgb}{0.93,0.68,0.93}
\definecolor {plum3}               {rgb}{0.80,0.59,0.80}
\definecolor {plum4}               {rgb}{0.55,0.40,0.55}
\definecolor {mediumorchid1}       {rgb}{0.88,0.40,1.00}
\definecolor {mediumorchid2}       {rgb}{0.82,0.37,0.93}
\definecolor {mediumorchid3}       {rgb}{0.71,0.32,0.80}
\definecolor {mediumorchid4}       {rgb}{0.48,0.22,0.55}
\definecolor {darkorchid1}         {rgb}{0.75,0.24,1.00}
\definecolor {darkorchid2}         {rgb}{0.70,0.23,0.93}
\definecolor {darkorchid3}         {rgb}{0.60,0.20,0.80}
\definecolor {darkorchid4}         {rgb}{0.41,0.13,0.55}
\definecolor {purple1}             {rgb}{0.61,0.19,1.00}
\definecolor {purple2}             {rgb}{0.57,0.17,0.93}
\definecolor {purple3}             {rgb}{0.49,0.15,0.80}
\definecolor {purple4}             {rgb}{0.33,0.10,0.55}
\definecolor {mediumpurple1}       {rgb}{0.67,0.51,1.00}
\definecolor {mediumpurple2}       {rgb}{0.62,0.47,0.93}
\definecolor {mediumpurple3}       {rgb}{0.54,0.41,0.80}
\definecolor {mediumpurple4}       {rgb}{0.36,0.28,0.55}
\definecolor {thistle1}            {rgb}{1.00,0.88,1.00}
\definecolor {thistle2}            {rgb}{0.93,0.82,0.93}
\definecolor {thistle3}            {rgb}{0.80,0.71,0.80}
\definecolor {thistle4}            {rgb}{0.55,0.48,0.55}
\definecolor {gray1}               {rgb}{0.01,0.01,0.01}
\definecolor {gray2}               {rgb}{0.02,0.02,0.02}
\definecolor {gray3}               {rgb}{0.03,0.03,0.03}
\definecolor {gray4}               {rgb}{0.04,0.04,0.04}
\definecolor {gray5}               {rgb}{0.05,0.05,0.05}
\definecolor {gray6}               {rgb}{0.06,0.06,0.06}
\definecolor {gray7}               {rgb}{0.07,0.07,0.07}
\definecolor {gray8}               {rgb}{0.08,0.08,0.08}
\definecolor {gray9}               {rgb}{0.09,0.09,0.09}
\definecolor {gray10}              {rgb}{0.10,0.10,0.10}
\definecolor {gray11}              {rgb}{0.11,0.11,0.11}
\definecolor {gray12}              {rgb}{0.12,0.12,0.12}
\definecolor {gray13}              {rgb}{0.13,0.13,0.13}
\definecolor {gray14}              {rgb}{0.14,0.14,0.14}
\definecolor {gray15}              {rgb}{0.15,0.15,0.15}
\definecolor {gray16}              {rgb}{0.16,0.16,0.16}
\definecolor {gray17}              {rgb}{0.17,0.17,0.17}
\definecolor {gray18}              {rgb}{0.18,0.18,0.18}
\definecolor {gray19}              {rgb}{0.19,0.19,0.19}
\definecolor {gray20}              {rgb}{0.20,0.20,0.20}
\definecolor {gray21}              {rgb}{0.21,0.21,0.21}
\definecolor {gray22}              {rgb}{0.22,0.22,0.22}
\definecolor {gray23}              {rgb}{0.23,0.23,0.23}
\definecolor {gray24}              {rgb}{0.24,0.24,0.24}
\definecolor {gray25}              {rgb}{0.25,0.25,0.25}
\definecolor {gray26}              {rgb}{0.26,0.26,0.26}
\definecolor {gray27}              {rgb}{0.27,0.27,0.27}
\definecolor {gray28}              {rgb}{0.28,0.28,0.28}
\definecolor {gray29}              {rgb}{0.29,0.29,0.29}
\definecolor {gray30}              {rgb}{0.30,0.30,0.30}
\definecolor {gray31}              {rgb}{0.31,0.31,0.31}
\definecolor {gray32}              {rgb}{0.32,0.32,0.32}
\definecolor {gray33}              {rgb}{0.33,0.33,0.33}
\definecolor {gray34}              {rgb}{0.34,0.34,0.34}
\definecolor {gray35}              {rgb}{0.35,0.35,0.35}
\definecolor {gray36}              {rgb}{0.36,0.36,0.36}
\definecolor {gray37}              {rgb}{0.37,0.37,0.37}
\definecolor {gray38}              {rgb}{0.38,0.38,0.38}
\definecolor {gray39}              {rgb}{0.39,0.39,0.39}
\definecolor {gray40}              {rgb}{0.40,0.40,0.40}
\definecolor {gray42}              {rgb}{0.42,0.42,0.42}
\definecolor {gray43}              {rgb}{0.43,0.43,0.43}
\definecolor {gray44}              {rgb}{0.44,0.44,0.44}
\definecolor {gray45}              {rgb}{0.45,0.45,0.45}
\definecolor {gray46}              {rgb}{0.46,0.46,0.46}
\definecolor {gray47}              {rgb}{0.47,0.47,0.47}
\definecolor {gray48}              {rgb}{0.48,0.48,0.48}
\definecolor {gray49}              {rgb}{0.49,0.49,0.49}
\definecolor {gray50}              {rgb}{0.50,0.50,0.50}
\definecolor {gray51}              {rgb}{0.51,0.51,0.51}
\definecolor {gray52}              {rgb}{0.52,0.52,0.52}
\definecolor {gray53}              {rgb}{0.53,0.53,0.53}
\definecolor {gray54}              {rgb}{0.54,0.54,0.54}
\definecolor {gray55}              {rgb}{0.55,0.55,0.55}
\definecolor {gray56}              {rgb}{0.56,0.56,0.56}
\definecolor {gray57}              {rgb}{0.57,0.57,0.57}
\definecolor {gray58}              {rgb}{0.58,0.58,0.58}
\definecolor {gray59}              {rgb}{0.59,0.59,0.59}
\definecolor {gray60}              {rgb}{0.60,0.60,0.60}
\definecolor {gray61}              {rgb}{0.61,0.61,0.61}
\definecolor {gray62}              {rgb}{0.62,0.62,0.62}
\definecolor {gray63}              {rgb}{0.63,0.63,0.63}
\definecolor {gray64}              {rgb}{0.64,0.64,0.64}
\definecolor {gray65}              {rgb}{0.65,0.65,0.65}
\definecolor {gray66}              {rgb}{0.66,0.66,0.66}
\definecolor {gray67}              {rgb}{0.67,0.67,0.67}
\definecolor {gray68}              {rgb}{0.68,0.68,0.68}
\definecolor {gray69}              {rgb}{0.69,0.69,0.69}
\definecolor {gray70}              {rgb}{0.70,0.70,0.70}
\definecolor {gray71}              {rgb}{0.71,0.71,0.71}
\definecolor {gray72}              {rgb}{0.72,0.72,0.72}
\definecolor {gray73}              {rgb}{0.73,0.73,0.73}
\definecolor {gray74}              {rgb}{0.74,0.74,0.74}
\definecolor {gray75}              {rgb}{0.75,0.75,0.75}
\definecolor {gray76}              {rgb}{0.76,0.76,0.76}
\definecolor {gray77}              {rgb}{0.77,0.77,0.77}
\definecolor {gray78}              {rgb}{0.78,0.78,0.78}
\definecolor {gray79}              {rgb}{0.79,0.79,0.79}
\definecolor {gray80}              {rgb}{0.80,0.80,0.80}
\definecolor {gray81}              {rgb}{0.81,0.81,0.81}
\definecolor {gray82}              {rgb}{0.82,0.82,0.82}
\definecolor {gray83}              {rgb}{0.83,0.83,0.83}
\definecolor {gray84}              {rgb}{0.84,0.84,0.84}
\definecolor {gray85}              {rgb}{0.85,0.85,0.85}
\definecolor {gray86}              {rgb}{0.86,0.86,0.86}
\definecolor {gray87}              {rgb}{0.87,0.87,0.87}
\definecolor {gray88}              {rgb}{0.88,0.88,0.88}
\definecolor {gray89}              {rgb}{0.89,0.89,0.89}
\definecolor {gray90}              {rgb}{0.90,0.90,0.90}
\definecolor {gray91}              {rgb}{0.91,0.91,0.91}
\definecolor {gray92}              {rgb}{0.92,0.92,0.92}
\definecolor {gray93}              {rgb}{0.93,0.93,0.93}
\definecolor {gray94}              {rgb}{0.94,0.94,0.94}
\definecolor {gray95}              {rgb}{0.95,0.95,0.95}
\definecolor {gray97}              {rgb}{0.97,0.97,0.97}
\definecolor {gray98}              {rgb}{0.98,0.98,0.98}
\definecolor {gray99}              {rgb}{0.99,0.99,0.99}
\definecolor {darkgrey}            {rgb}{0.66,0.66,0.66}

\newcommand{\Tone}{\ensuremath{\T_1}\xspace}
\newcommand{\Ttwo}{\ensuremath{\T_2}\xspace}
\newcommand{\Tonetwo}{\ensuremath{\Tone\cup \Ttwo}\xspace}
\newcommand{\smt}{SMT\xspace}
\newcommand{\smtt}{\ensuremath{\text{SMT}(\T)}\xspace}

\newcommand{\utvpi}{\ensuremath{\mathcal{UTVPI}}\xspace}
\newcommand{\utvpiint}{\ensuremath{\mathcal{UTVPI(\mathbb{Z})}}\xspace}
\newcommand{\utvpirat}{\ensuremath{\mathcal{UTVPI(\mathbb{Q})}}\xspace}

\newcommand{\dlrat}{\ensuremath{\mathcal{DL(\mathbb{Q})}}\xspace}
\newcommand{\dlint}{\ensuremath{\mathcal{DL(\mathbb{Z})}}\xspace}
\newcommand{\la}{\ensuremath{\mathcal{LA}}\xspace}
\newcommand{\larat}{\ensuremath{\mathcal{LA}(\mathbb{Q})}\xspace}
\newcommand{\laint}{\ensuremath{\mathcal{LA}(\mathbb{Z})}\xspace}

\newcommand{\tonetwo}{\ensuremath{\T_1\cup\T_2}\xspace}

\newcommand{\Tmodels}{\models_{\T}}
\newcommand{\laratmodels}{\models_{\larat}}

\newcommand{\Ti}{\ensuremath{\mathcal{T}_i}\xspace}
\newcommand{\TsolverGen}[1]{\ensuremath{{#1}\textit{-solver}}\xspace}
\newcommand{\TsolversGen}[1]{\ensuremath{{#1}\textit{-solvers}}\xspace}
\newcommand{\Tsolver}{\TsolverGen{\T}}
\newcommand{\Tsolvers}{\TsolversGen{\T}}

\newcommand{\Tisolver}{\TsolverGen{\ensuremath{\Ti}}}

\newcommand{\Tlemma}{\T-lemma\xspace}
\newcommand{\Tlemmas}{\T-lemmas\xspace}

\def\makenewenumerate#1#2{%
\newcounter{cnt#1}
\newenvironment{#1}%
{\begin{list}{\makebox[0pt][r]{#2}}%
{\setlength{\itemsep}{0pt}%
 \setlength{\parsep}{.2em}%
 \setlength{\leftmargin}{2em}%
 \setlength{\labelwidth}{.4em}%
 \usecounter{cnt#1}}}%
{\end{list}}}
\makenewenumerate{renumerate}{\rm(\roman{cntrenumerate})}
\makenewenumerate{renumerateprime}{\rm(\roman{cntrenumerateprime}$'$)}
\makenewenumerate{renumeratesecond}{\rm(\roman{cntrenumeratesecond}$''$)}

\makenewenumerate{aenumerate}{({\it\alph{cntaenumerate}})}
\makenewenumerate{aenumerateprime}{({\it\alph{cntaenumerateprime}$'$})}
\makenewenumerate{aenumeratesecond}{({\it\alph{cntaenumeratesecond}$''$})}

\newtheorem{theorem}{Theorem}[section]

\newtheorem{corollary}[theorem]{Corollary}

\newtheorem{lemma}[theorem]{Lemma}
\newdef{defn}[theorem]{Definition}
\newdef{remark}[theorem]{Remark}
\newtheorem{example}{Example}[section]

\newcommand{\dl}{\ensuremath{{\cal DL}}\xspace}
\newcommand{\laR}{\larat}
\newcommand{\euf}{\ensuremath{{\cal EUF}}\xspace}
\newcommand{\T}{\ensuremath{{\cal T}}\xspace}

\newcommand{\hyp}{\textsc{Hyp}\xspace}

\newcommand{\comb}{\textsc{Comb}\xspace}
\newcommand{\leqeq}{\textsc{LeqEq}\xspace}

\newcommand{\mathsat}{\textsc{MathSAT}\xspace}
\newcommand{\yices}{\textsc{Yices}\xspace}
\newcommand{\zthree}{\textsc{Z3}\xspace}
\newcommand{\cvcthree}{\textsc{CVC3}\xspace}
\newcommand{\argolib}{\textsc{ArgoLib}\xspace}

\newcommand{\foci}{\textsc{Foci}\xspace}
\newcommand{\clp}{\textsc{clp-prover}\xspace}
\newcommand{\blast}{\textsc{Blast}\xspace}
\newcommand{\zap}{\textsc{Zap}\xspace}
\newcommand{\lifter}{\textsc{Lifter}\xspace}
\newcommand{\csisat}{\textsc{CSIsat}\xspace}
\newcommand{\intjain}{\textsc{INT2}\xspace}

\newcommand{\infrule}[3]{\ensuremath{\infer{#3}{#2}}}
\newcommand{\hyprule}[1]{\ensuremath{#1}}

\newcommand{\ieproof}[1]{\ensuremath{\Pi^{\mathsf{ie}}_{#1}}\xspace}
\newcommand{\ielocal}{\textsf{ie\,}-local\xspace}

\newenvironment{agproof}{\begin{proof}}{\end{proof}}

\newcommand{\defas}{\ensuremath{\stackrel{\text{\tiny def}}{=}}\xspace}

\newenvironment{small2}{\fontsize{8}{10}\selectfont}{\normalsize}
\newcommand{\PP}{\ensuremath{\Pi}\xspace}
\newcommand{\PPie}{\ensuremath{\PP^{\mathsf{ie}}}\xspace}
\newcommand{\ie}{\textsf{ie}\xspace}

\newcommand{\proptofol}{\ensuremath{{\cal P}2{\cal T}}\xspace}
\newcommand{\foltoprop}{\ensuremath{{\cal T}2{\cal P}}\xspace}
\newcommand{\btot}{\proptofol}
\newcommand{\ttob}{\foltoprop}
\newcommand{\satres}{\ensuremath{\mathsf{sat}}\xspace}
\newcommand{\unsatres}{\ensuremath{\mathsf{unsat}}\xspace}
\newcommand{\dpll}{\textsc{DPLL}\xspace}

\newcommand{\agsubsubsection}[1]{%
\subsubsection{#1}%
}

\newcommand{\rdiff}[3]{\ensuremath{(0 \le #2-#1 #3)}} 
 
\newcommand{\pv}[1]{\ensuremath{#1^+}} 
\newcommand{\nv}[1]{\ensuremath{#1^-}}

\newcommand{\pp}[3]{\ensuremath{(0 \le -#1 -#2  #3)}} 
\newcommand{\pn}[3]{\ensuremath{(0 \le #2 - #1  #3)}} 
 
\newcommand{\nn}[3]{\ensuremath{(0 \le #1 + #2  #3)}} 
\newcommand{\ub}[2]{\ensuremath{(0 \le - #1 + #2)}}
\newcommand{\lb}[2]{\ensuremath{(0 \le #1 + #2)}} 
 
\newcommand{\ppdiff}[3]{\rdiff{\pv{#1}}{\pv{#2}}{#3}}
\newcommand{\pndiff}[3]{\rdiff{\pv{#1}}{\nv{#2}}{#3}}
\newcommand{\npdiff}[3]{\rdiff{\nv{#1}}{\pv{#2}}{#3}}
\newcommand{\nndiff}[3]{\rdiff{\nv{#1}}{\nv{#2}}{#3}}

\newcommand{\ppenc}[3]{\pndiff{#1}{#2}{#3}, \pndiff{#2}{#1}{#3}} 
\newcommand{\pnenc}[3]{\ppdiff{#1}{#2}{#3}, \nndiff{#2}{#1}{#3}} 
\newcommand{\nnenc}[3]{\npdiff{#1}{#2}{#3}, \npdiff{#2}{#1}{#3}} 
\newcommand{\ubenc}[2]{\pndiff{#1}{#1}{#2}} 
\newcommand{\lbenc}[2]{\npdiff{#1}{#1}{#2}}

\title{
Efficient Generation of Craig Interpolants \\in Satisfiability Modulo Theories
}
\author{ALESSANDRO CIMATTI\\
\textsc{fbk-irst}\\
ALBERTO GRIGGIO and ROBERTO SEBASTIANI\\
\textsc{disi}, Universit\`a di Trento}
\begin{abstract}
The problem of computing Craig Interpolants has recently received a
lot of interest. In this paper, we address the problem of efficient
generation of interpolants for some important fragments of first order
logic, which are amenable for effective decision procedures, called
Satisfiability Modulo Theory solvers.

We make the following contributions.
First, we provide interpolation procedures for several basic theories
of interest: the theories of linear arithmetic over the rationals,
difference logic over rationals and integers, and UTVPI over rationals
and integers.
Second, we define a novel approach to interpolate combinations of
theories, that applies to the Delayed Theory Combination approach.

Efficiency is ensured by the fact that the proposed interpolation
algorithms extend state of the art algorithms for Satisfiability
Modulo Theories.  Our experimental evaluation shows that the MathSAT
SMT solver can produce interpolants with minor overhead in search, and
much more efficiently than other competitor solvers.
\end{abstract}

\category{F.4.1}{Mathematical Logic and Formal Languages}{Mathematical Logic}[Mechanical theorem proving]
\terms{Theory, Algorithms} 
\keywords{Craig Interpolation, Decision Procedures, SMT}

\begin{document}
\begin{bottomstuff}
Authors addresses: 
Alessandro Cimatti ({\tt cimatti@fbk.eu}), 
\textsc{FBK-IRST}, Via Sommarive 18, 38050 Povo, Trento, Italy.
Alberto Griggio ({\tt griggio@disi.unitn.it}) and 
Roberto Sebastiani ({\tt rseba@disi.unitn.it}),
\textsc{DISI}, Universit\`a di Trento, Via Sommarive 14, 38050 Povo, Trento, Italy.
\end{bottomstuff}
\maketitle

 \newpage
\noindent
{\bf Note for reviewers.}
The table of contents is added only for  the sake of reviewer's
 convenience, and will be removed in the final version 
if the paper is accepted. 
 \tableofcontents
 \newpage

\section{Introduction}
\label{sec:intro}

One of the most successful applications of computational logic is Formal Verification,
It that aims at proving (or disproving) certain
properties of the behaviours of a reactive system.
In recent years, also thanks to the impressive improvements of SAT
solvers, a wide variety of verification methods based on SAT solving
have been proposed. These methods proved effective for discrete state
systems, most notably hardware components.
The approach is made practical by the fact that SAT solvers, in
addition to proving efficiently the satisfiability of huge
propositional formulas, provide several functionalities, such as model
generation, proof production, extraction of unsatisfiable cores, and
generation of Craig interpolants  (interpolation).
In particular, since the seminal paper of McMillan~\cite{interpolation-mc-sat},
interpolation has been recognized to be a substantial tool for
verification in the case of Boolean
systems~\cite{cabodi06,somenzi06,silva07}.

One of the main limitations of SAT-based approaches,
is in their
expressive power. Many systems of practical interest, containing
integer or real valued variables, such as software, and timed and
hybrid systems, can not be represented directly within propositional
logic.
This has prompted research in the analysis of fragments of first order
logic: given a formula referring to variables, the problem is to find
a satisfying assignment in a theory of interest (e.g. linear
arithmetic).
This field, referred to as Satisfiability Modulo Theory (SMT), has
resulted in substantial theoretical results, and in very effective
decision procedures, known as SMT solvers. State of the art SMT
solvers complement the Boolean SAT algorithms with specialized 
decision procedures for conjunctions of literals in some given theory ({\em theory solvers}). 
In addition to checking satisfiability, SMT solvers are able
to generate models, produce proofs, and extract unsatisfiable
cores. This has allowed, to lift many SAT-based verification
algorithms to SMT-based verification, as well as to to open up the way
to abstraction-refinement with SMT.

Quite surprisingly, however, the research on interpolation for SMT has
not kept the pace of SMT solving. In fact, the current approaches to
producing interpolants for fragments of first order
theories~\cite{mcmillan_interpolating_prover,musuvathi_interpolation,rybalchenko_interp,kroening_interp,interpolation_data_structures,jain_cav08}
all suffer from a number of problems. Some of the approaches are
severely limited in terms of their expressiveness. For instance, the
tool described in \cite{rybalchenko_interp} can only deal with
conjunctions of literals, whilst the recent work described
in~\cite{kroening_interp} can not deal with many useful theories.
Furthermore, very few tools are
available~\cite{rybalchenko_interp,mcmillan_interpolating_prover}, and
these tools do not seem to scale particularly well. More than to
na{\"i}ve implementation, this appears to be due to the underlying
algorithms, that substantially deviate from or ignore choices common
in state-of-the-art SMT.
For instance, in the domain of linear arithmetic over the rationals
(\larat), strict inequalities are encoded
in~\cite{mcmillan_interpolating_prover} as the conjunction of a weak
inequality and a disequality; although sound, this choice destroys the
structure of the constraints, {forces reasoning in the combination of
theories $\larat\cup\euf$,} requires additional splitting, and
ultimately results in a larger search space.
Similarly, the fragment of Difference Logic (\dlrat) is dealt with by
means of a general-purpose algorithm for full \larat, rather than
one of the well-known and much faster specialized algorithms.
An even more fundamental example is the fact that state-of-the-art SMT
reasoners use dedicated algorithms for Linear
Arithmetic~\cite{demoura_cav06}.

In this paper, we tackle the problem of generating interpolants for
SMT problems, fully leveraging the algorithms used in a state of the
art SMT solver. In particular, our main contributions are:
\begin{enumerate}
\item {An interpolation algorithm for \larat that exploits a
    variant of the algorithm presented in \cite{demoura_cav06}, and that is
    capable of handling the full \larat -- including strict inequalities and
    disequalities -- without the need of theory combination;}

\item {An algorithm for computing interpolants in \dl{} -- both over the
    rationals and over the integers -- that builds on top of the
    efficient graph-based decision algorithms given in
    \cite{cotton-maler,barcelogic-dl}, that ensures that the generated
    interpolants are still in the \dl fragment of linear arithmetic,
    and that allows for computing stronger interpolants than the
    existing algorithms for the full linear arithmetic;}

\item {An algorithm for computing interpolants in \utvpi -- both over
    the rationals and over the integers -- that builds on an encoding
    of \dl. The algorithm ensures that the generated interpolants are
    still in the \utvpi fragment of linear arithmetic, and that allows
    for computing stronger interpolants than the existing algorithms
    for the full linear arithmetic;}

\item {An algorithm for computing interpolants in a combination
    \tonetwo of theories based on the Delayed Theory Combination (DTC)
    method \cite{infocomp_dtc05,AMAI08_dtc} (as an alternative to the
    traditional Nelson-Oppen method), which does not require ad-hoc
    interpolant combination methods, but exploits the propositional
    interpolation algorithm for performing the combination of theories;}

\item {An efficient implementation of all the proposed techniques
    within the \mathsat 4 SMT solver \cite{mathsat4}, and an extensive
    experimental evaluation on a wide range of benchmarks.}
\end{enumerate}

This comprehensive approach advances the state of the art in two main
directions: on one side, we show how to extend efficient SMT solving
techniques to SMT interpolation, for a wide class of important
theories, without paying a substantial price in performance; on the
other side, we present an interpolating SMT solver that is able to
produce interpolants for a much wider class of problems than its
competitors, and, on problems that can be dealt with by other tools,
shows dramatic improvements in performance, often by orders of
magnitude.

\smallskip\noindent 
\textbf{Content.} The paper is structured as follows.
In~\sref{sec:stateoftheart} we present some background on
interpolation in SMT.
In~\sref{sec:la},~\sref{sec:dl} and~\sref{sec:utvpi} we show how to
efficiently interpolate \larat, \dl and \utvpi respectively.
In~\sref{sec:dtc} we discuss interpolation for combined theories.
\ignore{In~\sref{sec:dtc_multipleinterpolants} we discuss multiple interpolation.}
The proposed techniques are experimentally evaluated in~\sref{sec:expeval}.
In~\sref{sec:concl} we draw some conclusions, and outline directions for future work.
The discussion of related work is distributed in the technical sections (\sref{sec:la}-\sref{sec:dtc}).

\smallskip\noindent 
\textbf{Note to reviewers.} 
Some of the material contained in this paper, in a less
detailed form, has been published in two conference papers 
\cite{cgs_tacas08_interpolants,cgs_cade09_interpolants}.
%


\section{Background and state-of-the-art}
\label{sec:stateoftheart}

\subsection{Satisfiability Modulo Theory -- \smt}
\label{sec:smt}

Our setting is standard first order logic. A $0$-ary function symbol is
called a \emph{constant}. A \emph{term} is a first-order term built out of
function symbols and variables. 
We write $t_1 \equiv t_2$ when the
two terms $t_1$ and $t_2$ are syntactically identical.  If $t_1,\ldots,t_n$
are terms and $p$ is a predicate symbol, then $p(t_1,\ldots,t_n)$ is an
\emph{atom}. A \emph{literal} is either an atom or its negation.
A \emph{formula} $\phi$ is built in the usual way out of the universal and
existential quantifiers, Boolean connectives, and atoms. We call a formula
\emph{quantifier-free} if it does not contain quantifiers, and \emph{ground} if
it does not contain free variables.
A \emph{clause} is a disjunction of
literals. A formula is said to be in \emph{conjunctive normal
  form} (CNF) if it is a conjunction of clauses. 
For every non-CNF \T-formula $\varphi$, an equisatisfiable CNF
formula $\psi$ can be generated in polynomial time \cite{tseitin}.

We also assume the usual first-order notions of interpretation,
satisfiability, validity, logical consequence, and theory, as given,
e.g., in~\cite{enderton}. 
A {\em first-order theory}, \T, is a set of first-order sentences. 
In this paper, we consider only theories with equality.
A structure {\cal A} is a model of a theory \T if {\cal A} satisfies every
sentence in \T. A formula is {\em satisfiable in \T} (or
\emph{\T-satisfiable}) if it is satisfiable in a model of \T.

We call {\em Satisfiability Modulo (the) Theory \T}, \smtt, the
problem of deciding the satisfiability of quantifier-free 
formulas\footnote{The general definition of \smt deals also with quantified
  formulas. Nevertheless, in this paper we restrict our interest to
  quantifier-free formulas.} with respect to a background theory \T.
We denote formulas with $\phi, \psi, A, B, C, I$, {\T-}variables with
$x, y, z$, {Boolean variables with $p, q$} and 
numeric constants with $a, b, c, l, u$. 
Given a theory \T, we write $\phi \Tmodels \psi$ (or simply
$\phi \models \psi$) to denote that the formula $\psi$ is a logical
consequence of $\phi$ in the theory \T.
With $\phi \preceq \psi$ we denote
that all uninterpreted (in \T) symbols of $\phi$ appear in $\psi$. 
If $C$ is a
clause, $C \downarrow B$ is the clause obtained by removing all the literals
whose atoms do not occur in $B$, and $C \setminus B$ that obtained by
removing all the literals whose atoms do occur in $B$. With a little abuse of
notation, we might sometimes denote conjunctions of literals $l_1 \wedge
\ldots \wedge l_n$ as sets $\{l_1, \ldots, l_n\}$ and vice versa. If $\eta
\defas \{l_1, \ldots, l_n\}$, we might write $\neg\eta$ to mean $\neg l_1 \vee
\ldots \vee \neg l_n$.
A theory \T is \emph{stably-infinite} iff every quantifier-free
\T-satisfiable formula is satisfiable in an infinite model of \T.
A theory \T is \emph{convex} iff, for every collection $l_1,\ldots,
l_k, e_1,\ldots, e_n$ of literals in $\T$ s.t. $e_1,\ldots, e_n$ are
in the form $(x=y)$, $x,y$ being variables, we have that
$\{l_1,...,l_k\}\models_\T \bigvee_{i=1}^{n} e_i$ if and only if
$\{l_1,...,l_k\}\models_\T e_i \text{~for some~} 1 \leq i \leq n$.

Given a decidable first-order theory \T, we call a {\em theory solver
for \T}, \Tsolver, any tool able to decide the satisfiability in
\T of sets/conjunctions of ground atomic formulas and their
negations --- {\em theory literals} or {\em \T-literals} --- in the
language of \T.
If $S \defas \{l_1, \ldots, l_n\}$ is a set of literals
in \T, we call \emph{(\T)-conflict set} any subset $\eta$ of $S$ which is
inconsistent in \T.~\footnote{In the next sections, as we are in an \smtt
  context, we often omit specifying ``in the theory \T'' when speaking of
  consistency, validity, etc.  } We call 
$\neg\eta$ a \T-lemma. (Notice that $\neg\eta$ is a \T-valid clause.)

\begin{defn}[Resolution proof]\label{def:resolution_proof}
  Given a set of clauses $S \defas \{C_1,\ldots, C_n\}$ and a clause $C$, we
  call a \emph{resolution proof} {of the deduction} $\bigwedge_i C_i
  \Tmodels C$ a DAG ${\cal P}$ such that:
  \begin{enumerate}
  \item $C$ is the root of ${\cal P}$;
  \item the leaves of ${\cal P}$ are either elements of $S$ or \T-lemmas;
  \item each non-leaf node $C'$ has two {premises} 
    $C_{p_1}$ and $C_{p_2}$ such
    that $C_{p_1} \defas p \vee \phi_1$, $C_{p_2} \defas \neg p \vee \phi_2$,
    and $C' \defas \phi_1 \vee \phi_2$. 
    The atom $p$ is called the
    \emph{pivot} of $C_{p_1}$ and $C_{p_2}$.
  \end{enumerate}
  If $C$ is the empty clause (denoted with $\bot$), then ${\cal P}$ is a
  \emph{resolution proof of {(\T-)}unsatisfiability} for $\bigwedge_i C_i$.
\end{defn}

We consider the \smtt problem for some background theory \T.

\begin{defn}[Craig Interpolant]\label{def:interpolant}
  Given an ordered pair $(A, B)$ of formulas such that $A \wedge B
  \models_{\T} \bot$, a \emph{Craig interpolant} (simply ``interpolant''
  hereafter) is a formula $I$ s.t.:
  \begin{compactenum}[(i)]
  \item $A \Tmodels I$,
  \item $ I \wedge B \Tmodels \bot$,
  \item $I \preceq A$ and $I \preceq B$.
  \end{compactenum}
\end{defn}

\subsection{Algorithms for SMT}
\label{sec:algorithms_smt}

\begin{figure}[!t]
  \begin{center}
  \begin{minipage}{.45\textwidth}
{\tt
  \begin{tabbing}[c]
  fo \= fo \= fo \= fo \= fo \= \kill
1.\>   SatValue Lazy\_SMT\_Solver (\T-formula $\phi$) \{ \\
2.\>  \> $\phi'$ = \verb,convert_to_cnf,($\phi$)\\
3.\>  \> $\phi^p = \foltoprop(\phi')$\\
4.\>  \> {\bf while} (DPLL{}($\phi^p,\mu^p$) == \satres) \{\\
5. \> \>\> {$\langle\rho, \eta\rangle$} = \Tsolver{}$(\proptofol(\mu^p))$ \\
6.\>  \> \> {\bf if} ($\rho$ == \satres)  {\bf then return} \satres \\
7.\>  \> \> $\phi^p = \phi^p \land \foltoprop(\neg \eta)$ \\
8.\>  \> \}\\
9.\>  \> {\bf return} \unsatres \\
10.\>  \}
  \end{tabbing}
}
  \end{minipage}
  \end{center}
\caption{A simplified 
  schema for lazy \smtt procedures.}
\label{fig:offline-algo}
\end{figure}

A standard technique for solving the SMT(\T) problem is to integrate a
DPLL-based SAT solver and a \T-solver in a ``\emph{lazy}'' manner. 
The idea underlying every lazy \smtt procedure is that (a complete
set of) the truth assignments for the propositional abstraction of
$\phi$ are enumerated and checked for satisfiability in \T; the
procedure either returns \satres if one \T-satisfiable truth
assignment is found, or it returns \unsatres otherwise.

Figure \ref{fig:offline-algo} presents a simplified schema of a
lazy \smtt procedure, called the {\em off-line schema}. 
The bijective function \foltoprop (``Theory-to-Boolean''), called 
\emph{Boolean abstraction}, maps Boolean atoms into themselves and 
non-Boolean \T-atoms into fresh Boolean atoms 
--- so that two atom instances in $\phi$ are mapped into the same Boolean atom 
iff they are syntactically identical ---
and extends to \T-formulas and sets of \T-formulas in the obvious way 
--- i.e., $\foltoprop(\neg\phi_1) \defas \neg\foltoprop(\phi_1)$,
$\foltoprop(\phi_1\bowtie\phi_2) \defas \foltoprop(\phi_1)\bowtie\foltoprop(\phi_2)$ for each Boolean connective $\bowtie$, 
$\foltoprop(\{\phi_i\}_i) \defas \{\foltoprop(\phi_i)\}_i$.
The function \proptofol (``propositional-to-theory''), called {\em refinement},
is the inverse of \foltoprop.
The propositional abstraction $\phi^p$ of the input formula $\phi$ is given as
input to a SAT solver based on the DPLL algorithm
\cite{DBLP:journals/cacm/DavisLL62,DBLP:conf/cade/ZhangM02}, which either
decides that $\phi^p$ is unsatisfiable, and hence $\phi$ is
\T-unsatisfiable, or returns a satisfying assignment $\mu^p$; in the latter
case, $\btot(\mu^p)$ is given as input to \Tsolver{}.  If $\btot(\mu^p)$ is
found \T-consistent, then $\phi$ is \T-consistent. If not, \Tsolver{}
returns the conflict set $\eta$ which caused the \T-inconsistency of
$\btot(\mu^p)$; the abstraction of the \T-lemma $\neg\eta$, \ttob($\neg\eta$),
is then added as a clause to $\phi^p$. Then the \dpll solver is restarted from
scratch on the resulting formula.

Practical implementations follow a more elaborated schema, called the {\em
  on-line schema} (see \cite{seba_lazy_smt}).
As before, $\phi^p$ is given as input to a modified version of \dpll, and when
a satisfying assignment $\mu^p$ is found, the refinement $\mu$ of $\mu^p$ is
fed to the \Tsolver; if $\mu$ is found \T-consistent, then $\phi$ is
\T-consistent; otherwise, \Tsolver{} returns the conflict set $\eta$ which
caused the \T-inconsistency of $\btot(\mu^p)$.
Then the clause $\neg\eta^p$ is added in conjunction to $\phi^p$, either
temporarily or permanently (\emph{\T-learning}), and the algorithm backtracks
up to the highest point in the search where one of the literals in
$\neg\eta^p$ is unassigned (\emph{\T-backjumping}), and therefore its value is
(propositionally) implied by the others in $\neg\eta^p$.
Another important improvement is \emph{early pruning (EP)}: before
  every literal selection, intermediate assignments are checked for
  \T-satisfiability and, if not \T-satisfiable, they are pruned (since no
  refinement can be \T-satisfiable). Finally, \emph{theory propagation} can be
  used to reduce the search space by allowing the \T-solvers to explicitly
  return truth values for unassigned literals, which can be unit-propagated by
  the SAT solver.  %
The interested reader is pointed to, e.g., \cite{seba_lazy_smt} for
  details and further references.  

\smallskip
With a small modification of the embedded DPLL engine, a lazy SMT solver can
also be used to generate a resolution proof of unsatisfiability (see
e.g. \cite{avg_sat07}).

\subsection{Interpolation in SMT}
\label{sec:interpolation_smt}

The use of interpolation in formal verification has been introduced by
McMillan in \cite{interpolation-mc-sat} for purely-propositional formulas, and
it was subsequently extended to handle SMT($\euf\cup\larat$) formulas in
\cite{mcmillan_interpolating_prover}, \euf being the theory of equality and
uninterpreted functions.
The technique is based on earlier work by Pudl\'ak
\cite{pudlak_interpolation}, where two interpolant-generation algorithms are
described: one for computing interpolants for propositional formulas from
resolution proofs of unsatisfiability, and one for generating interpolants for
conjunctions of (weak) linear inequalities in \larat. An interpolant for a
pair $(A, B)$ of CNF formulas is constructed from a resolution proof of
unsatisfiability of $A \wedge B$, generated as outlined in \S\ref{sec:smt}.
The algorithm works by computing a formula $I_C$ for each clause 
in the resolution refutation, 
such that the formula $I_\bot$ associated to the empty root clause 
is the computed interpolant.%
The algorithm can be described as follows:
 
\noindent
\begin{minipage}[h]{\linewidth}
  \vspace{3mm} \hrule \vspace{2mm} \textbf{Algorithm 1: Interpolant generation
    for \smtt}
  \vspace{1mm} \hrule
  \vspace{2mm}
  \begin{enumerate}
  \item Generate a {resolution} proof of unsatisfiability ${\cal P}$
    for $A \wedge B$.
  \item For every \T-lemma $\neg\eta$ occurring in ${\cal P}$, generate an
    interpolant $I_{\neg\eta}$ for \mbox{$(\eta \setminus B, \eta \downarrow
      B)$}.
  \item For every input clause $C$ in ${\cal P}$, set $I_C \defas C \downarrow
    B$ if $C \in A$, and $I_C \defas \top$ if $C \in B$.
  \item For every inner node $C$ of ${\cal P}$ obtained by resolution from
    $C_1 \defas {p} \vee \phi_1$ and $C_2 \defas \neg {p} \vee \phi_2$, set
    $I_C \defas I_{C_1} \vee I_{C_2}$ if ${p}$ does not occur in $B$, and $I_C
    \defas I_{C_1} \wedge I_{C_2}$ otherwise.
  \item Output $I_\bot$ as an interpolant for $(A, B)$.
  \end{enumerate}
  \vspace{1mm}
  \hrule \vspace{3mm}
\end{minipage}

\begin{example}\label{ex:itp_example_proof}
  {Consider the following two formulas in \larat:}
  \begin{displaymath}
    \begin{split}
      A &\defas (p \vee (0 \leq x_1 - 3x_2 + 1)) \wedge (0 \leq x_1 + x_2)
      \wedge (\neg q \vee \neg(0 \leq x_1 + x_2)) \\
      B &\defas (\neg(0 \leq x_3 - 2x_1 - 3) \vee (0 \leq 1 - 2x_3)) \wedge
      (\neg p \vee q) \wedge (p \vee (0 \leq x_3 - 2x_1 - 3))
    \end{split}
  \end{displaymath}
  Figure~\ref{fig:itp_example_proof}(a) shows a resolution proof of
  unsatisfiability for $A \wedge B$, in which the clauses from $A$ have been
  underlined. The proof contains the following \larat-lemma (displayed in
  boldface):
  \begin{displaymath}
    \neg(0 \leq x_1 -3x_2+1) \vee \neg(0\leq
    x_1+x_2) \vee \neg(0\leq x_3 -2x_1-3) \vee \neg(0\leq 1 - 2x_3).
  \end{displaymath}
  Figure~\ref{fig:itp_example_proof}(b) shows, for each clause $\Theta_i$ in
  the proof, the formula $I_{\Theta_i}$ generated by Algorithm 1. For the
  \larat-lemma, it is easy to see that $(0 \leq 4x_1+1)$ is an interpolant for
  $((0 \leq x_1 - 3 x_2 + 1) \wedge (0 \leq x_1 + x_2), (0 \leq x_3 - 2 x_1 -
  3) \wedge (0 \leq 1 - 2 x_3))$ as required by Step 2 of the algorithm.
  {(We will show how to obtain this interpolant in
    Example~\ref{ex:la_proof_1}.)}  Therefore, $I_\bot\defas (p \vee (0 \leq
  4x_1+1)) \wedge \neg q$ is an interpolant for $(A, B)$.

\begin{figure}[t]
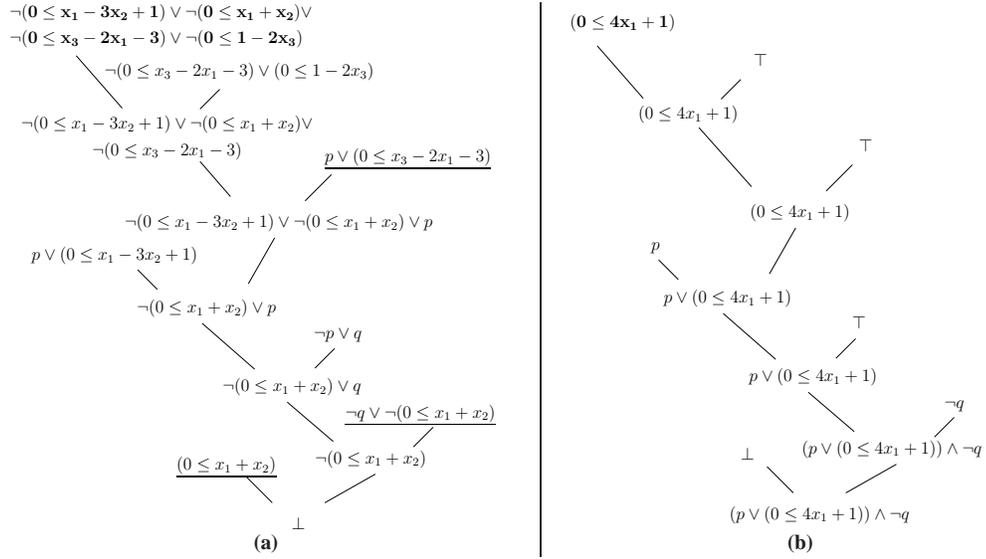

  \centering
  \begin{tabular}{c|c}
    \begin{minipage}{0.4\textwidth}
    \newcommand{\TlemmaOne}{$\mathbf{\neg(0 \leq x_1 -3x_2+1) \vee \neg(0\leq
        x_1+x_2) \vee}$}
    \newcommand{\TlemmaOneb}{$\mathbf{\neg(0\leq x_3 -2x_1-3) \vee \neg(0\leq 1 - 2x_3)}$}
    \newcommand{\HypOne}{$\neg(0 \leq x_3 -2x_1 -3) \vee (0 \leq 1 - 2x_3)$}
    \newcommand{\HypTwo}{$\underline{p \vee (0 \leq x_3 -2x_1 -3)}$}
    \newcommand{\HypThree}{$p \vee (0 \leq x_1 -3x_2 +1)$}
    \newcommand{\HypFour}{$\neg p \vee q$} 
    \newcommand{\HypFive}{$\underline{\neg q \vee
      \neg(0 \leq x_1 + x_2)}$} 
    \newcommand{\HypSix}{$\underline{(0 \leq x_1 + x_2)}$}

    \newcommand{\ResOne}{$\neg(0 \leq x_1 -3x_2+1) \vee \neg(0\leq x_1+x_2)
      \vee$} 
    \newcommand{\ResOneb}{$\neg(0\leq x_3 -2x_1-3)$} 
    \newcommand{\ResTwo}{$\neg(0 \leq x_1
      -3x_2+1) \vee \neg(0\leq x_1+x_2) \vee p$}
    \newcommand{\ResThree}{$\neg(0\leq x_1+x_2) \vee p$}
    \newcommand{\ResFour}{$\neg(0\leq x_1+x_2) \vee q$}
    \newcommand{\ResFive}{$\neg(0\leq x_1+x_2)$}
    \newcommand{\Bottom}{$\bot$}

    \scalebox{0.55}{\input{itp_example_proof.pstex_t}}
  \end{minipage}\hspace{1.8cm} &
  \hspace{1mm}
  \begin{minipage}{0.5\textwidth}
    \newcommand{\TlemmaOne}{$\mathbf{(0 \leq 4x_1 + 1)}$}
    \newcommand{\HypOne}{$\top$}
    \newcommand{\HypTwo}{$\top$}
    \newcommand{\HypThree}{$p$}
    \newcommand{\HypFour}{$\top$} 
    \newcommand{\HypFive}{$\neg q$}
    \newcommand{\HypSix}{$\bot$}

    \newcommand{\ResOne}{$(0 \leq 4x_1+1)$}
    \newcommand{\ResTwo}{$(0 \leq 4x_1+1)$}
    \newcommand{\ResThree}{$p \vee (0 \leq 4x_1+1)$}
    \newcommand{\ResFour}{$p \vee (0 \leq 4x_1+1)$}
    \newcommand{\ResFive}{$(p \vee (0 \leq 4x_1+1)) \wedge \neg q$}
    \newcommand{\Bottom}{$(p \vee (0 \leq 4x_1+1)) \wedge \neg q$}

    \scalebox{0.55}{\input{itp_example_proof2.pstex_t}}
  \end{minipage} \\[3mm]
  \textbf{(a)} & \textbf{(b)}
\end{tabular}
  \caption{Resolution proof of unsatisfiability (a) and interpolant (b) for
    the pair $(A, B)$ of formulas of Example~\ref{ex:itp_example_proof}. In
    the tree on the left, \T-lemmas are displayed in boldface, and clauses
    from $A$ are underlined.
    \label{fig:itp_example_proof}}
\end{figure}
\end{example}

Algorithm 1 can be applied also when $A$ and $B$ are not in CNF. In this case,
it suffices to pre-convert them into CNF by using disjoint sets of
auxiliary Boolean atoms in the usual way \cite{mcmillan_interpolating_prover}.

Notice that Step~2. of the algorithm is the only part which
  depends on the theory \T, so that the problem of
  interpolant generation in \smtt reduces to that of finding
  interpolants for \T-lemmas. 
To this extent,
 in \cite{mcmillan_interpolating_prover} McMillan gives a set of 
rules for constructing
interpolants for \T-lemmas in the theory of \euf, that of weak linear
inequalities $(0 \leq t)$ in \larat, %
and their combination. 
{Linear
  equalities $(0=t)$ can be reduced to conjunctions $(0 \leq t)\wedge
  (0 \leq -t)$ of inequalities.
Thanks to the combination of theories,
also strict linear inequalities $(0 < t)$ can be handled  in
  $\euf\cup\larat$ by replacing them 
with the conjunction 
$(0 \leq t) \wedge (0\neq t)$,%
\footnote{The details are not given in
    \cite{mcmillan_interpolating_prover}. One possible way of doing this is to
  rewrite $(0\neq t)$ as $(y=t) \wedge (z=0) \wedge (z\neq y)$, $z$ and $y$ being fresh variables.}
but this solution can be very inefficient.

The combination
$\euf\cup\larat$ can also be used to compute interpolants for other theories,
such as those of lists, arrays, sets and multisets
\cite{interpolation_data_structures}.

In \cite{mcmillan_interpolating_prover}, interpolants in the combined theory
$\euf\cup\larat$ are obtained by means of ad-hoc combination rules. The work in
\cite{musuvathi_interpolation}, instead, presents a method for generating
interpolants for \tonetwo using the interpolant-generation procedures of
$\T_1$ and $\T_2$ as black-boxes, using the Nelson-Oppen
  approach \cite{NO79}.  

Also the method of \cite{rybalchenko_interp} allows to compute interpolants
in $\euf\cup\larat$. Its peculiarity is that it is not based on unsatisfiability
proofs. Instead, it generates interpolants in \larat by solving a system of
constraints using an off-the-shelf Linear Programming (LP) solver. The method
allows both weak and strict inequalities. Extension to uninterpreted functions
is achieved by means of reduction to \larat using a hierarchical calculus
\cite{sofronie_interpolation_ijcar}.
The algorithm works only with conjunctions of atoms, although in principle it
could be integrated in Algorithm 1 to generate interpolants for \T-lemmas in
\larat. As an alternative, the authors show in \cite{rybalchenko_interp} how
to generate interpolants for formulas that are in Disjunctive Normal Form
(DNF).

Another different approach is explored in \cite{kroening_interp}. There, the
authors use the \emph{eager} SMT approach to encode the original SMT problem
into an equisatisfiable propositional problem, for which a propositional proof
of unsatisfiability is generated. This proof is later ``lifted'' to the
original theory, and used to generate an interpolant in a way similar to
Algorithm 1. At the moment, the approach is however limited to the theory of
equality only (without uninterpreted functions).

All the above techniques construct \emph{one} interpolant for $(A, B)$. In
general, however, interpolants are not unique. In particular, some of them can
be better than others, depending on the particular application domain. In
\cite{mcmillan_trans_itp}, it is shown how to manipulate proofs in order to
obtain stronger interpolants. In \cite{mcmillan_tacas06,mcmillan_cav07},
instead, a technique to restrict the language used in interpolants
is presented and shown to be useful in preventing divergence of techniques
based on predicate abstraction.

One of the most important applications of interpolation in Formal Verification
is abstraction refinement
\cite{DBLP:conf/popl/HenzingerJMM04,mcmillan06}. In such
setting, every input problem $\phi$ has the form $\phi \defas \phi_1 \wedge
\ldots \wedge \phi_n$, and the interpolating solver is asked to compute
several interpolants $I_1,\ldots,I_{n-1}$ corresponding to different
partitions of $\phi$ into $A_i$ and $B_i$, such that
\begin{equation}\label{eq:phi_ai_bi}
  \forall i,~~ A_i \defas \phi_1 \wedge \ldots \wedge \phi_i, \text{~~and~~} 
  B_i \defas \phi_{i+1} \wedge \ldots \wedge \phi_n.
\end{equation}
Moreover, $I_1,\ldots,I_{n-1}$ should be related by the following:
\begin{equation}
  \label{eq:multiple_interpolants}
  I_i \wedge \phi_{i+1} \models I_{i+1}
\end{equation}
A sufficient condition for \eqref{eq:multiple_interpolants} to hold is that
all the $I_i$'s are computed from the same proof of unsatisfiability \PP for
$\phi$ \cite{DBLP:conf/popl/HenzingerJMM04}. 

\begin{figure}[t]
  \centering
  \begin{small}
    \begin{tabular}{c}
      \begin{minipage}{1.0\linewidth}
        \begin{displaymath}
          \leqeq~\frac{\ignore{\Gamma \vdash }0 = t}{\ignore{\Gamma \vdash }0 \leq t}
          \hspace{13mm}
          \comb~\frac{\ignore{\Gamma \vdash }0 \leq t_1 \quad\quad 
            \ignore{\Gamma \vdash }0 \leq t_2}%
          {\ignore{\Gamma \vdash }0 \leq c_1t_1 + c_2t_2}~c_1,c_2 > 0
        \end{displaymath} 
      \end{minipage} \\
    \end{tabular}
  \end{small}
  \caption{%
    {\larat-proof rules for a conjunction $\Gamma$ of equalities and
      weak inequalities.}
    \label{fig:la_proof_rules}}
\end{figure}

\subsubsection{Interpolants for conjunctions of \larat-literals}
\label{sec:la_proofs}

We recall the algorithm of
\cite{mcmillan_interpolating_prover} for computing interpolants from
\laR-proofs of unsatisfiability, {for
  conjunctions of equalities and weak inequalities in \larat}.

An \emph{\laR-proof rule} $R$ for a conjunction $\Gamma$ of
  equalities and weak inequalities 
is either an element of $\Gamma$, or it
has the form $\dfrac{P}{\phi}$,
where $\phi$ is an equality or a weak inequality and $P$ is a
sequence of
proof rules, called the \emph{premises} of $R$. 
An \emph{\laR-proof of unsatisfiability}
for a conjunction of equalities and weak inequalities $\Gamma$ is simply a
rule in which $\phi\equiv 0 \leq c$ and where $c$ is a negative numerical
constant.\footnote{In the following, we might sometimes write $\bot$ as a
  synonym of an atom ``$0\le c$'' when $c$ is a negative numerical constant.}

Similarly to \cite{mcmillan_interpolating_prover}, we use the proof rules of
Figure~\ref{fig:la_proof_rules}: %
\leqeq for
deriving inequalities from equalities, and \comb for performing linear
combinations.\footnote{In \cite{mcmillan_interpolating_prover} the \leqeq rule
  is not used in \laR, because the input is assumed to consist only of
  inequalities.}

\smallskip Given an \laR-proof of unsatisfiability $P$ for a conjunction
$\Gamma$ of equalities and weak inequalities partitioned into $(A, B)$, an
interpolant $I$ can be computed simply by replacing every {atom} $0
\leq t$ occurring in B (resp. $0 = t$) with $0 \leq 0$ (resp. $0 = 0$) in each
leaf sub-rule of $P$, and propagating the results: the interpolant is
then the single weak inequality $0 \leq t$ at the root of $P$
\cite{mcmillan_interpolating_prover}.

\begin{example}\label{ex:la_proof_1}
  Consider the following sets of \laR atoms:
  \begin{displaymath}
    \begin{split}
      A &\defas \{ (0 \leq x_1 - 3 x_2 + 1), (0 \leq x_1 + x_2) \} \\
      B &\defas \{ (0 \leq x_3 - 2 x_1 - 3), (0 \leq 1 - 2 x_3) \}.
    \end{split}
  \end{displaymath}
  An \larat-proof of unsatisfiability $P$ for $A \wedge B$ is the following:
  \begin{small}
    \begin{displaymath}
      \infrule{Comb}{
      \infrule{Comb}{
        \hyprule{1 * (0\leq x_1-3x_2+1)} &
        \hyprule{4 * (0\leq x_1+x_2)}
      }{1 * (0\leq 4x_1 + 1)}
      \quad&\quad
      \infrule{Comb}{
        \hyprule{2 * (0\leq x_3-2x_1-3)} &
        \hyprule{1 * (0\leq 1-2x_3)}
      }{1 * (0\leq -4x_1 - 5)}
    }{(0\leq -4)}
    \end{displaymath}
  \end{small}
  By replacing inequalities in $B$ with $(0 \leq 0)$, we obtain the proof
  $P'$:
  \begin{small}
    \begin{displaymath}
      \infrule{Comb}{
      \infrule{Comb}{
        \hyprule{1 * (0\leq x_1-3x_2+1)} &
        \hyprule{4 * (0\leq x_1+x_2)}
      }{1 * (0\leq 4x_1 + 1)}
      \quad&\quad 
      \infrule{Comb}{
        \hyprule{2 * (0\leq 0)} &
        \hyprule{1 * (0\leq 0)}
      }{1 * (0\leq 0)}
    }{(0\leq 4x_1+1)}
    \end{displaymath}
  \end{small}
  Thus, the interpolant obtained is $(0 \leq 4x_1+1)$.
\end{example}


\section{From SMT(\larat) solving to SMT(\larat) interpolation}
\label{sec:la}

Traditionally, SMT solvers used some kind of incremental simplex algorithm
\cite{lp_book} as \T-solver for the \laR theory. Recently, Dutertre and de
Moura \cite{demoura_cav06} have proposed a new simplex-based algorithm,
specifically designed for integration in a lazy SMT solver. The algorithm
is extremely suitable for SMT, and SMT solvers embedding
  it were shown to significantly
outperform (often by
orders of magnitude) the ones based on other simplex
  variants. 
It has now been integrated in
several SMT solvers, including \argolib, \cvcthree,
\mathsat, \yices, and \zthree.
Remarkably, this algorithm allows for handling also strict inequalities.

In this Section, we show how to exploit this algorithm to
efficiently generate interpolants for \laR formulas.
Combined with the interpolation for the \smtt problem described in is then obtained by combining the general
 In \S\ref{sec:la_basic} we begin by considering the case in which
  the input atoms are only equalities and non-strict inequalities. In this
  case, we only need to show how to generate a proof of unsatisfiability,
  since then we can use the interpolation rules defined in
  \cite{mcmillan_interpolating_prover}. Then, in \S\ref{sec:strict_ineq} we
  show how to generate interpolants for problems containing also strict
  inequalities and disequalities.

\subsection{Interpolation with non-strict inequalities}
\label{sec:la_basic}

\agsubsubsection{The original Dutertre-de Moura algorithm}

In its original formulation, the Dutertre-de Moura algorithm assumes that the
variables $x_i$ are partitioned a priori in two sets, hereafter denoted as
$\hat{{\cal B}}$ (``initially basic'' or ``dependent'') and
$\hat{{\cal N}}$ (``initially non-basic'' or ``independent''), and
that the algorithm receives as inputs two kinds of atomic formulas:~
\footnote{Notationally, we use the hat symbol ~$\hat{}$~ to denote the initial
  value of the generic symbol.}
\begin{itemize}
\item a set of  \emph{equations} $\mathsf{eq}_i$, {one
    for each $x_i \in \hat{{\cal B}}$,} of the 
form $\textstyle\sum_{x_j\in \hat{{\cal N}}}\hat{a}_{ij}x_j + \hat{a}_{ii}x_i=
  0$ s.t. all $\hat{a}_{ij}$'s  are numerical constants;
\item \emph{elementary atoms} of the form $x_j \ge l_j$ or $x_j \le u_j$
  s.t. {$x_j \in \hat{{\cal B}}\cup\hat{{\cal N}}$ and} $l_j$, $u_j$
  are numerical constants.
\end{itemize}

In order to handle problems that are not in the above form, a
satisfiability-preserving preprocessing step is applied upfront, before
invoking the algorithm.

The initial equations $\mathsf{eq}_i$ are then used to build a tableau $T$:
\begin{equation}
\label{eq:tableau}
 \{ x_i = \textstyle\sum_{x_j \in {\cal N}}a_{ij}x_j ~|~ x_i \in {\cal B}\},
\end{equation}
where ${\cal B}$ (``basic'' or ``dependent''), ${\cal N}$ (``non-basic'' or
``independent'') and $a_{ij}$ are such that initially ${\cal B} \equiv
\hat{{\cal B}}$, ${\cal N} \equiv \hat{{\cal N}}$ and $a_{ij} \equiv
-\hat{a}_{ij}/\hat{a}_{ii}.$

In order to decide the satisfiability of the input problem, the algorithm
performs manipulations of the tableau that change the sets ${\cal B}$ and
${\cal N}$ and the values of the coefficients $a_{ij}$, always keeping the
tableau $T$ in \eqref{eq:tableau} equivalent to its initial version.
{
In particular, the algorithm maintains a mapping $\beta: {\cal B}\cup{\cal
    N}\longmapsto \mathbb{Q}$ 
representing a candidate model which, at every step,
satisfies the following invariants: 
\begin{gather}
    \forall x_j \in {\cal N},\quad l_j \leq \beta(x_j) \leq u_j
    \label{eq:la_inv1}, \qquad
    \forall x_i \in {\cal B},\quad \beta(x_i) = \textstyle\sum_{j \in {\cal N}}
    a_{ij}\beta(x_j).
\end{gather}
The algorithm
tries to adjust the values of $\beta$ and the sets ${\cal B}$ and ${\cal
  N}$, {and hence the coefficients $a_{ij}$ of the tableau,} such that
$l_i \leq \beta(x_i) \leq u_i$ holds also for all the $x_i$'s in ${\cal
  B}$. Inconsistency is detected when
this is not possible without violating any constraint in (\ref{eq:la_inv1}):
as the bounds on the variables in ${\cal N}$ are always satisfied {by
$\beta$}, then
there is a variable $x_i \in {\cal B}$ such that the inconsistency is caused
either by the elementary atom $x_i \geq l_i$ or by the atom $x_i \leq u_i$
\cite{demoura_cav06}; in the first case,~%
\footnote{Here we do not consider the second case $x_i \leq u_i$ as it is
  analogous to the first one.}  a conflict set $\eta$ is generated as follows:
\begin{gather}
\label{eq:la_conflict_set}
  \eta = \{ x_j \leq u_j | x_j \in {\cal N}^+\} \cup \{ x_j \geq l_j | x_j \in
  {\cal N}^-\} \cup \{x_i \geq l_i \}, 
\end{gather}
where $(x_i = \textstyle\sum_{x_j \in {\cal N}}a_{ij}x_j)$ is the row of the
current version of the tableau $T$ \eqref{eq:tableau} corresponding to $x_i$,
${\cal N}^+$ is $\{ x_j \in {\cal N} | a_{ij} > 0 \}$ and ${\cal N}^-$ is $\{
x_j \in {\cal N} | a_{ij} < 0\}$.

Notice that $\eta$ is a conflict set in the sense that it is made inconsistent
by (some of) the equations in the tableau $T$ \eqref{eq:tableau}, i.e.
$T\cup\eta\laratmodels\bot$. {In general, however, $\eta\not\laratmodels\bot$.}

\agsubsubsection{Our proof-producing variant}\label{sec:la_proofs_demoura}

In order to make it suitable for interpolant generation, we have
  conceived the following variant of the Dutertre-de Moura algorithm.

We take as input an arbitrary set of
inequalities $l_k \leq \textstyle\sum_h \hat{a}_{kh}\,y_h$ or $u_k \geq
\textstyle\sum_h \hat{a}_{kh}\,y_h$, and 
{apply an internal preprocessing step to obtain a set of equations
  and a set of elementary bounds.}
In particular, we introduce a ``slack'' variable $s_k$ for each
distinct term $\textstyle\sum_h \hat{a}_{kh}\,y_h$ occurring in the input
inequalities. Then, we replace such term with $s_k$ (thus obtaining $l_k \leq
s_k$ or $u_k \geq s_k$) and add an equation $s_k = \textstyle\sum_h
\hat{a}_{kh}\,y_h$. Notice that we introduce a slack variable even for
``elementary'' inequalities $(l_k \leq y_k)$. With this transformation, the
initial tableau $T$ \eqref{eq:tableau} is:
\begin{equation}
  \{ s_k = \textstyle\sum_h \hat{a}_{kh}\,y_h \}_k,
\end{equation}
s.t.  $\hat{{\cal B}}$ is made of all the slack
variables $s_k$'s, 
$\hat{{\cal N}}$ is made of all the original variables $y_h$'s, and
the elementary atoms contain only slack variables $s_k$'s.

Then the algorithm proceeds as described above, producing a set $\eta$
  \eqref{eq:la_conflict_set} in case of inconsistency.
In our variant {of the algorithm}, we can use $\eta$ to generate a conflict set $\eta'$, thanks
to the following {theorem}.
\begin{theorem}\label{thm:la_lemma_1}
  In the set $\eta$ of \eqref{eq:la_conflict_set}, $x_i$ and all the $x_j$'s
  are slack variables introduced by our preprocessing step.
  Moreover, the set $\eta' \defas \eta_{{\cal N}^+} \cup \eta_{{\cal
      N}^-} \cup \eta_i$   is a conflict set, where\\
  \begin{displaymath}
    \begin{split}
      \eta_{{\cal N}^+} &\defas~ \{ u_k \geq \textstyle\sum_h
      \hat{a}_{kh}\,y_h | s_k\equiv x_j \ and\ x_j \in
      {\cal N}^+ \}, \\
      \eta_{{\cal N}^-} &\defas~ \{ l_k \leq \textstyle\sum_h
      \hat{a}_{kh}\,y_h | s_k\equiv x_j \ and\ x_j
      \in {\cal N}^- \}, \\
      \eta_i &\defas~ \{ l_k \leq \textstyle\sum_h \hat{a}_{kh}\,y_h | s_k
      \equiv x_i \}.
    \end{split}
\end{displaymath}
\end{theorem}

  \begin{agproof}
{ We consider the case in which $\eta$
  \eqref{eq:la_conflict_set} is generated from a row $x_i = 
  \textstyle\sum_{x_j\in {\cal N}}a_{ij}\,x_j$ in the tableau $T$
  \eqref{eq:tableau} 
such that $\beta(x_i) <
  l_i$. In \cite{demoura_cav06} it is shown that in this case the following
  facts hold:
    \begin{equation}
      \forall x_j \in {\cal N}^+, \beta(x_j) = u_j, \text{~~~and~~~} \forall
      x_j \in {\cal N}^-, \beta(x_j) = l_j.
    \end{equation}
    (We recall that ${\cal N}^+ = \{ x_j \in {\cal N} | a_{ij} > 0 \}$ and
    ${\cal N}^- = \{ x_j \in {\cal N} | a_{ij} < 0\}$.)  The bounds $u_j$ and
    $l_j$ can be introduced only by elementary atoms. Since in our variant the
    elementary atoms contain only slack variables, each $x_j$ must be
    a slack variable (namely $s_k$). The same holds for $x_i$ (since its value is bounded by $l_i$).  }

{
    Now consider $\eta$ again. In \cite{demoura_cav06} it is shown that when a
    conflict is detected because
  $\beta(x_i) < l_i$, then the following fact holds:
  \begin{equation}\label{eq:la_conflict}
    \beta(x_i) = \textstyle\sum_{x_j \in {\cal N}^+} a_{ij}u_j + \textstyle\sum_{x_j \in {\cal
        N}^-} a_{ij}l_j.
  \end{equation}
  From the $i$-th row of the tableau $T$ \eqref{eq:tableau} we can derive
  \begin{equation}\label{eq:tableau_row}
    0 \leq \textstyle\sum_{x_j \in {\cal N}} a_{ij}\,x_j - x_i.
  \end{equation}
  
If we take each inequality $0 \leq u_j - x_j$ multiplied by the
 coefficient $a_{ij}$ for
  all $x_j \in {\cal N}^+$, each inequality $0 \leq x_j - l_j$
 multiplied by
  coefficient $-a_{ij}$ for all $x_j \in {\cal N}^-$,
and the inequality $(0 \leq x_i - l_i)$ multiplied by 1, and we add them to
  (\ref{eq:tableau_row}), we obtain
  \begin{equation}
    0 \leq \textstyle\sum_{{\cal N}^+}a_{ij}\,u_j + \textstyle\sum_{{\cal N}^-}a_{ij}\,l_j - l_i,
  \end{equation}
  which by (\ref{eq:la_conflict}) is equivalent to $0 \leq \beta(x_i)
  - l_i$.
Thus we have obtained  $0 \leq
  c$ with $c \equiv \beta(x_i) - l_i$, which is strictly lower than
  zero. Therefore, $\eta$ is inconsistent under the definitions in $T$.  
  Since we know that $x_i$ and all the $x_j$'s in $\eta$ are slack
  variables, we can replace every $x_j$ (i.e., every $s_k$)
 with its corresponding term
  $\textstyle\sum_h \hat{a}_{kh}\,y_h$, thus obtaining $\eta'$, which
  is thus inconsistent.
\qed
}
  \end{agproof}


When our variant of the algorithm detects an inconsistency, 
we  construct a proof of unsatisfiability
as follows. From the set $\eta$ of
\eqref{eq:la_conflict_set} we build a conflict set $\eta'$ by replacing
each elementary atom in it with the corresponding original atom, as shown in
Theorem~\ref{thm:la_lemma_1}. Using the \hyp rule, we introduce all the atoms in
$\eta_{\cal N^+}$, and combine them with repeated applications of the \comb
rule: if $u_k \geq \textstyle\sum_h \hat{a}_{kh}\,y_h$ is the atom corresponding to
$s_k$, we use as coefficient for the \comb the $a_{ij}$ (in the $i$-th row of
the current tableau) such that $s_k \equiv x_j$. Then, we introduce each of
the atoms in $\eta_{\cal N^-}$ with \hyp, and add them to the previous
combination, again using \comb. In this case, the coefficient to use is
$-a_{ij}$. Finally, we introduce the atom in $\eta_i$ and add it to the
combination with coefficient $1$.

\begin{corollary}\label{thm:la_lemma_2}
  The result of the linear combination described above is the atom $0 \leq c$,
  such that  $c$ is
  a numerical constant strictly lower than zero.
\end{corollary}
{
  \begin{agproof}
    Follows immediately by the proof of Theorem~\ref{thm:la_lemma_1}.\qed
  \end{agproof}


}
{
Besides the case just described (and its dual {when the inconsistency
  is due to an elementary atom $x_i \leq u_i$}),
another case in which an inconsistency can be detected is when 
two contradictory atoms are asserted: $l_k \leq \textstyle\sum_h \hat{a}_{kh}\,y_h$ and
$u_k \geq \textstyle\sum_h \hat{a}_{kh}\,y_h$, with $l_k > u_k$.
 In this case, the proof is simply the combination of
the two atoms  with coefficient~1.%
}

The extension for handling also equalities like $b_k =
\textstyle\sum_h\hat{a}_{kh}\,y_h$ is 
straightforward: we simply introduce two elementary atoms $b_k \leq s_k$ and
$b_k \geq s_k$ and, in the construction of the proof, we use the \leqeq rule
to introduce the proper inequality.

Finally, notice that the current implementation in \mathsat (see
\sref{sec:expeval}) is slightly different from what presented here, and
significantly more efficient. In practice, $\eta$, $\eta'$ are not constructed
in sequence; rather, they are built simultaneously. Moreover, some
optimizations are applied to eliminate some slack variables when they are not
needed.

\begin{example}\label{ex:la_proof_2}
  Consider again the two sets of \laR atoms of Example~\ref{ex:la_proof_1}:
  \begin{displaymath}
    \begin{split}
      A &\defas \{ (0 \leq x_1 - 3 x_2 + 1), (0 \leq x_1 + x_2 \} \\
      B &\defas \{ (0 \leq x_3 - 2 x_1 - 3), (0 \leq 1 - 2 x_3) \}.
    \end{split}
  \end{displaymath}
  With our variant of the Dutertre-de Moura algorithm, four ``slack''
  variables are introduced, resulting in the following tableau and elementary
  constraints: 
  \begin{displaymath}
      T \defas\left\{
      \begin{array}{rl@{\quad\quad}rl}
        s_1 =& x_1 - 3x_2 & -1 &\leq s_1 \\
        s_2 =& x_1 + x_2  &  0 &\leq s_2 \\
        s_3 =& x_3 - 2x_1 &  3 &\leq s_3 \\
        s_4 =& -2x_3      & -1 &\leq s_4
      \end{array}\right.
  \end{displaymath}
  To detect the inconsistency, the algorithm performs some pivoting steps,
  resulting in the final tableau $T'$:
  \begin{displaymath}
      T' \defas\left\{
      \begin{array}{rl}
        x_2 =& -\frac{1}{12} s_4 -\frac{1}{6} s_3 -\frac{1}{3} s_1 \\
        s_2 =& -\frac{1}{3} s_4 -\frac{2}{3} s_3 -\frac{1}{3} s_1 \\
        x_1 =& -\frac{1}{4} s_4 -\frac{1}{2} s_3 \\
        x_3 =& -\frac{1}{2} s_4 
      \end{array}\right.    
  \end{displaymath}
  The final values of $\beta$ are as follows:
  \begin{displaymath}
    \begin{array}{c@{\quad\quad}c@{\quad\quad}c@{\quad\quad}c}
      \beta(x_1) = \frac{7}{4} & \beta(x_2) = -\frac{1}{12} & \beta(x_3) =
      \frac{1}{2} & \\
      \beta(s_1) = -1 & \beta(s_2) = -\frac{4}{3} & \beta(s_3) = 3 & 
      \beta(s_4) = -1
    \end{array}
  \end{displaymath}
  Therefore, the bound $(0 \leq s_2)$ is violated. From the second row
  of $T'$, the set $\eta$ and the conflict set $\eta'$ are computed:
  \begin{displaymath}
    \begin{split}
      \eta &\defas \emptyset \cup \{ (-1 \leq s_4), (3 \leq s_3), (-1 \leq s_1)
      \} \cup \{ (0 \leq s_2) \} \\
      \eta' &\defas \emptyset \cup \{ (0 \leq 1 - 2 x_3), (0 \leq x_3 -
      2x_1 -3), (0 \leq x_1 -3x_2 + 1) \} \cup \{ (0 \leq x_1 + x_2) \}
    \end{split}
  \end{displaymath}
  The generated proof of unsatisfiability $P$ is:
  \begin{small}
    \begin{displaymath}
      \infrule{Comb}{
        \infrule{Comb}{
          \infrule{Comb}{
            \hyprule{\frac{1}{3} * (0 \leq 1 - 2x_3)} &
            \hyprule{\frac{2}{3} * (0 \leq x_3 - 2x_1 -3)}
          }{1 * (0 \leq -\frac{4}{3}x_1-\frac{5}{3})} & 
          \hyprule{\frac{1}{3} * (0 \leq x_1-3x_2+1)}}
        {1 * (0\leq-x_1-x_2-\frac{4}{3})} &
        \hyprule{1 * (0\leq x_1 + x_2)}}
      {(0\leq-\frac{4}{3})}
    \end{displaymath}
  \end{small}
  After replacing the inequalities of $B$ with $(0 \leq 0)$ in $P$, the new
  proof $P'$ is:
  \begin{small}
    \begin{displaymath}
      \infrule{Comb}{
        \infrule{Comb}{
          \infrule{Comb}{
            \hyprule{\frac{1}{3} * (0 \leq 0)} &
            \hyprule{\frac{2}{3} * (0 \leq 0)}
          }{1 * (0 \leq 0)} & 
          \hyprule{\frac{1}{3} * (0 \leq x_1-3x_2+1)}}
        {1 * (0\leq \frac{1}{3}x_1-x_2+\frac{1}{3})} &
        \hyprule{1 * (0\leq x_1 + x_2)}}
      {(0\leq\frac{4}{3}x_1+\frac{1}{3})}
    \end{displaymath}
  \end{small}  
  Thus the computed interpolant is $(0\leq\frac{4}{3}x_1+\frac{1}{3})$ (which
  is equivalent to that of Example~\ref{ex:la_proof_1}).
\end{example}

\subsection{Interpolation with strict inequalities and disequalities}
\label{sec:strict_ineq}

Another benefit of the Dutertre-de Moura algorithm is that it can handle
strict inequalities directly. Its method is based on the following lemma.

\begin{lemma}[Lemma 1 in \cite{demoura_cav06}]\label{thm:la_strict_ineq_lemma}
  A set of linear arithmetic atoms $\Gamma$ containing strict
  inequalities $S = \{ 0 < t_1, \ldots, 0 < t_n\}$ is satisfiable iff there
  exists a rational number $\varepsilon > 0$ such that $\Gamma_\varepsilon \defas
  (\Gamma \cup S_\varepsilon) \setminus S$ is satisfiable, where
  $S_\varepsilon \defas \{\varepsilon \leq t_1, \ldots, \varepsilon \leq t_n \}$.
\end{lemma}

The idea of \cite{demoura_cav06} is that of treating the
\emph{infinitesimal parameter} $\varepsilon$
symbolically instead of explicitly computing its value. Strict bounds $(x <
b)$ are replaced with weak ones $(x \leq b - \varepsilon)$, and the operations
on bounds are adjusted to take  
$\varepsilon$ into account.
 
{We extend the same idea to the computation of interpolants.} We
transform every atom $(0 < t_i)$ occurring in the proof of unsatisfiability
into $(0 \leq t_i - \varepsilon)$.  Then we compute an interpolant
$I_{\varepsilon}$ in the usual way.  As a consequence of the rules of
\cite{mcmillan_interpolating_prover}, $I_{\varepsilon}$ is always a single
atom.  As shown by the following lemma, if $I_{\varepsilon}$ contains
$\varepsilon$, then it must be in the form $(0 \leq t -c\,\varepsilon)$ with
$c > 0$, and
we can  rewrite $I_\varepsilon$ into $(0 < t)$.

\begin{theorem}[Interpolation with strict inequalities]
\label{thm:int_strict_ineq_lemma}
  Let $\Gamma$, $S$, $\Gamma_\varepsilon$ and $S_\varepsilon$ be defined as in
  Lemma~\ref{thm:la_strict_ineq_lemma}. Let $\Gamma$ be partitioned into $A$
  and $B$, and let $A_\varepsilon$ and $B_\varepsilon$ be obtained from $A$
  and $B$ by replacing atoms in $S$ with the corresponding ones in
  $S_\varepsilon$. Let $I_\varepsilon$ be an interpolant for $(A_\varepsilon,
  B_\varepsilon)$. Then:
  \begin{itemize}
  \item If $\varepsilon \not\preceq I_\varepsilon$, then $I_\varepsilon$ is an
    interpolant for $(A, B)$.
  \item If $\varepsilon \preceq I_\varepsilon$, then $I_\varepsilon \equiv (0
    \leq t -c\,\varepsilon)$ for some $c > 0$, and $I \defas (0 < t)$ is an
    interpolant for $(A, B)$.
  \end{itemize}
\end{theorem}

\begin{agproof}
  Since the side condition of the \comb rule ensures that equations are
  combined only using positive coefficients, and since the atoms
  introduced in the proof either do not contain $\varepsilon$ or contain it
  with a negative coefficient, if $\varepsilon$ appears in $I_\varepsilon$, it
  must have a negative coefficient. 

If $\varepsilon$ does not appear in $I_\varepsilon$,
  then $I_\varepsilon$ has been obtained from atoms
  appearing in $A$ or $B$, 
  so that $I_\varepsilon$ is an interpolant for $(A, B)$. 

  If $\varepsilon$ appears in $I_\varepsilon$, since its value has not been
  explicitly computed, it can be arbitrarily small, so thanks to
  Lemma~\ref{thm:la_strict_ineq_lemma} we have that 
{$B_\varepsilon \wedge
  I_\varepsilon \laratmodels \bot$ implies $B \wedge I \laratmodels \bot$}.

We can prove that $A \laratmodels I$ as follows. 
We consider some interpretation $\mu$ which is a model for
  $A$. Since $\varepsilon$ does not occur in $A$, we can
  extend $\mu$ by setting $\mu(\varepsilon) = \delta$ for some $\delta
  > 0$ such that $\mu$ is
  a model also for $A_\varepsilon$. As $A_{\epsilon}\laratmodels
  I_{\epsilon}$,  $\mu$ is also a model for
  $I_\varepsilon$, and hence $\mu$ is also a model for $I$.
Thus, we have that $A\laratmodels I$. \qed
\end{agproof}


Notice that Theorem~\ref{thm:int_strict_ineq_lemma} can be extended
straightforwardly to the case in which the interpolant is a conjunction of
inequalities.

{Thus, in case of strict inequalities,
  Theorem~\ref{thm:int_strict_ineq_lemma} gives us a way for constructing
  interpolants with no need of expensive theory combination (as instead was
  the case in \cite{mcmillan_interpolating_prover}). Moreover, thanks to it }
we can handle also negated equalities $(0 \neq
  t)$ directly. Suppose our set $S$ of input atoms (partitioned into $A$ and
  $B$) is the union of a set $S'$ of equalities and inequalities (both weak
  and strict) and a set $S^{\neq}$ of disequalities, and suppose that $S'$ is
  consistent. (If not so, an interpolant can be computed
  from $S'$.) Since \laR is convex, $S$ is inconsistent iff exists
  $(0 \neq 
  t) \in S^{\neq}$ such that $S' \cup \{(0 \neq t)\}$ is inconsistent, that
  is, such that both $S' \cup \{(0 < t)\}$ and $S' \cup \{(0 > t)\}$ are
  inconsistent. 

Therefore, we pick one element $(0 \neq t)$ of $S^{\neq}$ at a
  time, and check the satisfiability of $S' \cup \{(0 < t)\}$ and $S' \cup
  \{(0 > t)\}$. If both are inconsistent, from the two proofs we can generate
  two interpolants $I^-$ and $I^+$. We combine $I^+$ and $I^-$ to obtain
  an interpolant $I$ for $(A, B)$: if $(0 \neq t) \in A$, then $I$ is $I^+ \vee
  I^-$; if $(0 \neq t) \in B$, then $I$ is $I^+ \wedge I^-$, as shown by the
  following lemma.

\begin{theorem}[Interpolation for negated equalities]\label{thm:la_diseq_lemma}
  Let $A$ and $B$ two conjunctions of \larat atoms, and let $n
  \defas (0 \neq t)$ be one such atom. Let $g \defas (0 < t)$ and $l
  \defas (0 > t)$. 
\\
If $n \in A$, then let 
$ A^+ \defas A \setminus \{n\} \cup \{g\},\ 
      A^-  \defas A \setminus \{n\} \cup \{l\},\
      \text{~and~} 
      B^+  \defas B^- \defas B$.
\\
If  $n \in B$, then let 
$A^+  \defas A^- \defas A,\
B^+ \defas B \setminus \{n\} \cup \{g\},\
      \text{~and~} 
B^-  \defas B \setminus \{n\} \cup \{l\}$.
\\
Assume that $A^+ \wedge B^+ \laratmodels \bot$ and that $A^-
  \wedge B^- \laratmodels \bot$, and let $I^+$ and $I^-$ be two interpolants for
  $(A^+, B^+)$ and $(A^-, B^-)$ respectively, and let 
  \begin{displaymath}
    I \defas \left\{ 
      \begin{array}{ll}
        I^+ \vee I^- & \text{if $n \in A$} \\
        I^+ \wedge I^- & \text{if $n \in B$}.
      \end{array}
    \right.
  \end{displaymath}
  Then $I$ is an interpolant for $(A, B)$.
\end{theorem}

\begin{agproof}
  We have to prove that:
  \begin{enumerate}[(i)]
  \item $A \laratmodels I$
  \item $B \wedge I \laratmodels \bot$
  \item $I \preceq A$ and $I \preceq B$.
  \end{enumerate}

  \begin{enumerate}[(i)]
  \item If $n \in A$, then $A \laratmodels g \vee l$. By 
    hypothesis, we know that $A^+ \laratmodels I^+$ and $A^- \laratmodels I^-$. Then
    trivially $A \cup \{g\} \laratmodels I^+$ and $A \cup \{l\} \laratmodels
    I^-$. Therefore $A \cup \{g\} \laratmodels I^+ \vee I^-$ and $A \cup \{l\}
    \laratmodels I^- \vee I^+$, so that $A \laratmodels I$. 

    If $n \in B$, then $A^+ \equiv A^- \equiv A$. By 
    hypothesis $A \laratmodels I^+$ and $A \laratmodels I^-$, so that $A \laratmodels I$.

  \item If $n \in A$, then $B^+ \equiv B^- \equiv B$. 
By hypothesis $B \wedge I^+ \laratmodels \bot$ and $B \wedge I^-
    \laratmodels \bot$,  so that $B \wedge I \laratmodels \bot$.

    If $n \in B$, then $B \laratmodels g \vee l$, so that either $B \rightarrow g$ or
    $B \rightarrow l$ must hold. 
By  hypothesis we have $B^+ \wedge I^+ \laratmodels \bot$, so that $B \cup
\{g\} \wedge I^+ \laratmodels \bot$. 
If $B \rightarrow g$ holds, then $B \wedge I^+ \laratmodels \bot$, and hence $B \wedge I \laratmodels
    \bot$. 
Similarly, if $B \rightarrow l$ holds, then $B \wedge I^- \laratmodels \bot$, and
    so again $B \wedge I \laratmodels \bot$.

  \item By the hypothesis, both $I^+$ and $I^-$ contain only symbols
    common to $A$ and $B$, so that $I \preceq A$ and $I \preceq B$.\qed
  \end{enumerate}
\end{agproof}


\begin{example}
  {Consider the following sets of \larat atoms:}
  \begin{displaymath}
    \begin{split}
      A &\defas \{ (0 \neq x_1 -3 x_2 + 1), (0 = x_1 + x_2) \}\\
      B &\defas \{ (0 = x_3 -2x_1 - 1), (0 = 1 - 2x_3) \}.
    \end{split}
  \end{displaymath}
  To compute an interpolant for $(A, B)$, we first split $n \defas (0 \neq x_1
  -3x_2 + 1)$ into $g \defas (0 < x_1 -3x_2 + 1)$ and $l \defas (0 < -x_1 +3x_2
  - 1)$, thus obtaining $A^+$ and $A^-$ defined as in
  Theorem~\ref{thm:la_diseq_lemma}. We then generate two \larat-proofs of
  unsatisfiability $P^+$ for $A^+ \wedge B$ and $P^-$ for $A^-
  \wedge B$, and replace $g$ in $P^+$ with $g_\varepsilon \defas (0
  \leq x_1 -3x_2 +1 -\varepsilon)$ and $l$ in $P^-$ with $l_\varepsilon
  \defas (0 \leq -x_1 +3x_2 -1 -\varepsilon)$, obtaining $P^+_\varepsilon$ and
  $P^-_\varepsilon$ (we omit the names of the inference rules):
  \begin{small2}
    \begin{displaymath}
      P^+_\varepsilon \defas
      \infrule{Comb}{
      \infrule{Comb}{
        \hyprule{(0\leq x_1-3x_2+1-\varepsilon)}\quad
        \infrule{Hyp}{\hyprule{(0 = x_1+x_2)}}{(0\leq x_1+x_2)}
      }{(0\leq 4x_1 + 1 -\varepsilon)}
      \quad\quad
      \infrule{Comb}{
        \infrule{Hyp}{\hyprule{(0 = x_3-2x_1-1)}}{(0\leq x_3-2x_1-1)}\quad
        \infrule{Hyp}{\hyprule{(0 = 1-2x_3)}}{(0\leq 1-2x_3)}
      }{(0\leq -4x_1 - 1)}
    }{(0\leq -\varepsilon)}
    \end{displaymath}
  \end{small2}
  \begin{small2}
    \begin{displaymath}
      P^-_\varepsilon \defas
      \infrule{Comb}{
      \infrule{Comb}{
        \hyprule{(0\leq -x_1+3x_2-1-\varepsilon)}\quad
        \infrule{Hyp}{\hyprule{(0 = x_1+x_2)}}{(0\leq -x_1-x_2)}
      }{(0\leq -4x_1 - 1 -\varepsilon)}
      \quad\quad
      \infrule{Comb}{
        \infrule{Hyp}{\hyprule{(0 = x_3-2x_1-1)}}{(0\leq -x_3+2x_1+1)}\quad
        \infrule{Hyp}{\hyprule{(0 = 1-2x_3)}}{(0\leq -1+2x_3)}
      }{(0\leq +4x_1 + 1)}
    }{(0\leq -\varepsilon)}
    \end{displaymath}
  \end{small2}
  We then compute the two interpolants $I^+_\varepsilon$ from
  $P^+_\varepsilon$ and $I^-_\varepsilon$ from $P^-_\varepsilon$:
  \begin{displaymath}
      I^+_\varepsilon \defas (0 \leq 4x_1 +1-\varepsilon) \hspace{2cm}
      I^-_\varepsilon \defas (0 \leq -4x_1 -1-\varepsilon).
  \end{displaymath}
  Therefore, according to Theorem~\ref{thm:int_strict_ineq_lemma} the two
  interpolants $I^+$ for $(A^+, B)$ and $I^-$ for $(A^-,B)$ are:
  \begin{displaymath}
      I^+ \defas (0 < 4x_1 +1) \hspace{2cm}
      I^- \defas (0 < -4x_1 -1).
  \end{displaymath}
  Finally, since $n \in B$, according to Theorem~\ref{thm:la_diseq_lemma}, the
  interpolant $I$ for $(A, B)$ is 
  \begin{displaymath}
    I \defas I^+ \vee I^- \equiv (0 < 4x_1 +1) \vee (0 < -4x_1 -1).
  \end{displaymath}
\end{example}

\subsection{Obtaining stronger interpolants}
\label{sec:la_stronger_interpolants}

We conclude this Section by illustrating a simple technique for improving the
strength of interpolants in \larat. 
The technique is orthogonal to our proof-generation algorithm
described in \S\ref{sec:la_proofs_demoura}, and it is therefore of
independent interest. It is an improvement of the general algorithm of
\cite{mcmillan_interpolating_prover} (and outlined in \S\ref{sec:la_proofs})
for generating interpolants from \larat-proofs of unsatisfiability.

\begin{defn}
  Given two interpolants $I_1$ and $I_2$ for the same pair $(A, B)$ of
  conjunctions of \larat-literals, we say that $I_1$ is \emph{stronger} than
  $I_2$ if and only if $I_1 \models_{\larat} I_2$ but $I_2
  \not\models_{\larat} I_1$.
\end{defn}

Our technique is based on the simple observation that the only purpose of the
summations performed during the traversal of proof trees for computing the
interpolant (as described in \S\ref{sec:la_proofs}) is that of eliminating
$A$-local variables. In fact, it is easy to see that the conjunction of the
constraints of $A$ occurring as leaves in an \larat-proof of unsatisfiability satisfies the first
two points of the definition of interpolant
(Definition~\ref{def:interpolant}): if such constraints do not contain
$A$-local variables, therefore, their conjunction is already an
interpolant;
{%
if not, it suffices to perform only the summations constraints of $A$ 
that are necessary to eliminate $A$-local variables.
}
Moreover, such interpolant is stronger than that obtained by
performing the summations with the coefficients found in the proof tree, since
for any set of constraints $\{s_1, \ldots, s_n\}$ and any set of positive
coefficients $\{c_1, \ldots, c_n\}$, $s_1 \wedge \ldots \wedge s_n
\models_{\larat} \sum_{i=1}^n c_i*s_i$ holds.

According to this observation, our proposal can be described as: 
{%
\emph{perform only those summations which are are necessary 
for eliminating $A$-local variables.}
}

\begin{example}
  Consider the following sets of \larat-atoms:
  \begin{displaymath}
    \begin{split}
      A & \defas \{ (0 \leq x_1 - 3 x_2 + 1), (0 \leq x_2 - \frac{1}{3} x_3),
      (0 \leq x_4
      - \frac{3}{2} x_5 - 1) \} \\
      B & \defas \{ (0 \leq 3 x_5 - x_1), (0 \leq x_3 - 2 x_4) \}
    \end{split}
  \end{displaymath}
  and the following \larat-proof of unsatisfiability of $A \wedge B$:
  \begin{small2}
    \begin{displaymath}
      \infrule{}{
        \infrule{}{
          \infrule{}{
            \infrule{}{
              \hyprule{(0 \leq x_1 -3x_2 +1)} &
              \hyprule{3*(0\leq x_2 -\frac{1}{3} x_3)}
            }{(0 \leq x_1 -x_3+1)} &
            \hyprule{2*(0 \leq x_4 - \frac{3}{2}x_5 -1)}
          }{(0 \leq x_1 -x_3 +2x_4 -3x_5 -1)} &
          \hyprule{(0 \leq 3x_5 -x_1)}
        }{
          (0 \leq -x_3 +2x_4 -1)
        } &
        \hyprule{(0 \leq x_3 - 2 x_4)}
      }{(0\leq -1)}
    \end{displaymath}
  \end{small2}
  Here, the variable $x_2$ is $A$-local, whereas all the others are
  $AB$-common. The interpolant computed with the algorithm of
  \S\ref{sec:la_proofs} is
  \begin{displaymath}
    (0 \leq x_1 -x_3 +2x_4 -3x_5 -1),
  \end{displaymath}
  which is the result of the linear combination of \emph{all} the atoms of $A$
  in the proof. However, in order to eliminate the $A$-local variable $x_2$, it is
  enough to combine $(0 \leq x_1 -3x_2 +1)$ (with coefficient 1) and $(0\leq
  x_2 -\frac{1}{3} x_3)$ (with coefficient 3), obtaining $(0 \leq x_1
  -x_3+1)$. Therefore, a stronger interpolant is
  \begin{displaymath}
    (0 \leq x_1 -x_3+1) \wedge (0 \leq x_4 - \frac{3}{2}x_5 -1).
  \end{displaymath}
\end{example}

The technique can be implemented with a small modification 
of the proof-based algorithm described in \S\ref{sec:la_proofs}.
We associate with each node in the proof $P'$ 
(which is obtained from the original proof $P$ by replacing 
inequalities from $B$ with $(0\leq 0)$)
a list of pairs $\langle$\emph{coefficient, inequality}$\rangle$.
For a leaf, this list is a singleton in which the coefficient is 1 
and the inequality is the atom in the leaf itself.
For an inner node (which corresponds to an application of the \comb rule), 
the list $l$ is generated from the two lists $l_1$ and $l_2$ of the premises 
as follows:
\begin{enumerate}
\item Set $l$ as the concatenation of $l_1$ and $l_2$;

\item Let $c_1$ and $c_2$ be the coefficients used in the \comb rule. 
  Multiply each coefficient $c'_i$ occurring in a pair
  $\langle c'_i, 0 \leq t_i\rangle$ of $l$ by $c_1$ 
  if the pair comes from $l_1$, and by $c_2$ otherwise;

\item While there is an $A$-local variable $x$ occurring in more than one pair
  $\langle c', 0\leq t\rangle$ of $l$:%
  \footnote{That is, $x$ occurs in $t$.}
  \begin{enumerate}
  \item Collect all the pairs $\langle c'_i, 0\leq t_i\rangle$ 
    in which $x$ occurs;

  \item Generate a new pair $p \defas \langle 1, 0 \leq \sum_i c'_i*t_i\rangle$;

  \item Add $p$ to $l$, and remove all the pairs 
    $\langle c'_i, 0\leq t_i\rangle$.
  \end{enumerate}
\end{enumerate}
{%
After having applied the above algorithm, 
we can take the conjunction of the inequalities in the list associated with
the root of $P'$ as an interpolant.
}

\begin{theorem}
{%
  Let $P$ be a \larat-proof of unsatisfiability for a conjunction 
  $A \land B$ of inequalities, and $P'$ be obtained from $P$ by replacing
  each inequality of $B$ with $(0 \leq 0)$.
  Let $l \defas \langle c_1, 0 \leq t_1\rangle, \ldots, \langle c_n, 0 \leq t_n\rangle$
  be the list associated with the root of $P'$, computed as described above.
  Then $I \defas \bigwedge_{i=1}^n (0 \leq t_i)$ is an interpolant for $(A, B)$.
  Moreover, $I$ is always stronger than or equal to the interpolant 
  obtained with the algorithm of \S\ref{sec:la_proofs} for the same proof $P'$.
}
\end{theorem}
\begin{proof}
{%
  By induction on the structure of $P'$, it is easy to prove that, 
  for each constraint $(0 \leq t)$ in $P'$ 
  with its associated list 
  $l \defas \langle c_1, 0 \leq t_1\rangle, \ldots, \langle c_n, 0 \leq t_n\rangle$:
  \begin{enumerate}
  \item $A \models \bigwedge_{i=1}^n (0 \leq t_i)$; and
  \item $(0 \leq t) \equiv \sum_{i=1}^n c_i \cdot (0 \leq t_i)$
  \end{enumerate}
  Since the root of $P'$ is an interpolant for $(A, B)$, 
  this immediately proves the theorem.
}
\end{proof}


\section{From SMT(\dl) solving to SMT(\dl) interpolation}
\label{sec:dl}

Several interesting verification problems can be encoded using only a
subset of \la, the theory of Difference Logic (\dl),
either over the rationals (\dlrat) or over the integers (\dlint).
\dl is much simpler than \la{}, 
since 
in \dl{} 
all atoms are inequalities of the form $(0 \leq y - x + c)$, where $x$ and
$y$ are variables and $c$ is an integer constant.~%
\footnote{
Notice that we can assume w.l.o.g. that all constants are in $\mathbb{Z}$ because,
if this is not so,
then we can rewrite the whole formula into an equivalently-satisfiable
one by multiplying all constant symbols occurring in the formula 
by their greatest common denominator.}
Equalities can be handled as conjunctions of inequalities.
Here we do not consider the case when we also have strict
  inequalities $(0<y-x+c)$ {and disequalities $(0\neq y-x+c)$}, 
because 
in \dlrat{}
{they}  can be handled in a way which is similar 
to that described in \sref{sec:strict_ineq} for \larat,
whilst in \dlint{}
a strict inequality $(0<y-x+c)$ can be rewritten a priori 
into a weak one $(0\leq y-x+c-1)$,
{and a disequality can be replaced by a disjunction of strict inequalities.}

Very efficient solving algorithms have been
conceived for \dl~\cite{cotton-maler,barcelogic-dl}. 
In this section we present a specialized technique for
computing interpolants in \dl{} which exploits such state-of-the-art 
decision procedures.
Since a set of weak inequalities in \dl{} is consistent over the rationals 
if and only if it is consistent over the integers, 
 our algorithm is applicable without any modifications to both \dlrat{} and \dlint{} {(see e.g. \cite{barcelogic-dl})}.

Many SMT solvers use
dedicated, graph-based algorithms for checking the consistency of a
set of \dlrat atoms~\cite{cotton-maler,barcelogic-dl}. Intuitively, a set
$S$ of \dlrat atoms induces a graph whose vertexes are the variables of
the atoms, and there exists an edge $x \xrightarrow{c} y$ for every
$(0 \leq y - x + c) \in S$. $S$ is inconsistent if and only if the induced graph
has a cycle of negative weight.

We now extend the graph-based approach to generate interpolants.  
Consider the interpolation problem $(A,B)$ 
where $A$ and $B$ are sets of inequalities as above, and 
let $C$ be (the set of atoms in) a negative cycle
in the graph corresponding to $A\cup B$.

If $C\subseteq A$, then $A$ is inconsistent, in which case the
interpolant is $\bot$. Similarly, when $C\subseteq B$, the
interpolant is $\top$. 
If neither of these occurs, then the edges in the cycle can be
partitioned in subsets of $A$ and $B$.
We call maximal $A$-paths of $C$ a path
$x_1\xrightarrow{c_1}\ldots\xrightarrow{c_{n-1}}x_n$ such that ({\sc i})
$x_i\xrightarrow{c_i}x_{i+1}\in A$ {for $i \in [1, n-1]$}, and ({\sc ii}) $C$ contains
$x'\xrightarrow{c'} x_1$ and $x_n\xrightarrow{c''}x''$ that are in
$B$.
Clearly, the end-point variables $x_1,x_n$ of the maximal $A$-path are
such $x_1,x_n \preceq A$ and $x_1,x_n \preceq B$.
Let the {\em summary constraint} of a maximal $A$-path $x_1
\xrightarrow{c_1} \ldots \xrightarrow{c_{n-1}} x_n$ be the inequality
$0 \leq x_n - x_1 
+ \sum_{i=1}^{n-1} c_i$.

\begin{theorem}
The conjunction of summary constraints of the $A$-paths
of $C$ is an interpolant for $(A,B)$.
\end{theorem}
\begin{proof}
Using the rules for $\larat$ of
  Figure~\ref{fig:la_proof_rules},
we build a deduction of 
the summary constraint of an  maximal $A$-path 
from the conjunction of its corresponding set of constraints
  $\bigwedge_{i=1}^{n-1} (0\le x_{i+1} - x_i + c_i)$:
\begin{small2}
  \begin{displaymath}
      \infrule{Comb}{
       \infrule{}{
         \infrule{Comb}{
          \hyprule{(0 \leq x_2 - x_1 + c_1)} &
        \hyprule{(0 \leq x_3 - x_2 + c_2)}
      }{(0 \le x_3 - x_1 + c_1 + c_2)} & 
        \hyprule{(0 \leq x_4 - x_3 + c_3)}}{\ldots\phantom{X}}\qquad\ldots \qquad &
      \hyprule{(0 \leq x_n - x_{n-1} + c_{n-1})}
    }{(0 \leq x_n - x_1 + \sum_{i=1}^{n-1} c_i).}
  \end{displaymath}
\end{small2}
\noindent
Hence, $A$ entails the conjunction of the summary
constraints of all maximal $A$-paths.
Then, we notice that the conjunction of the summary constraints is
inconsistent with $B$. In fact, the weight of a maximal $A$-path and
the weight of its summary constraint are the same. Thus the cycle
obtained from $C$ by replacing each maximal $A$-path with the
corresponding summary constraint is also a negative cycle.
Finally, we notice that  every variable $x$ occurring in the
conjunction of the summary constraints is an end-point variable, and thus
$x\preceq A$ and $x\preceq B$.
\end{proof}

{
A final remark is in order. 
In principle, in order to generate a proof of
unsatisfiability for a conjunction of \dlrat atoms $A\wedge B$, 
the same rules used
for \laR \cite{mcmillan_interpolating_prover} could be used.
For instance, it is easy to build a proof which repeatedly applies the 
 \comb rule with $c_1=c_2=1$.
In general, however, 
the interpolants generated from such proofs are 
not \dlrat formulas anymore and, if computed starting from the same
inconsistent set $C$, they are either identical or weaker than
those generated with 
our method. 
In fact, %
it is easy to see that, unless our technique of
\sref{sec:la_stronger_interpolants} is adopted,
such interpolants are in the
form $(0 \le \textstyle\sum_i t_i)$ s.t. $\textstyle\bigwedge_i (0 \le
t_i)$ is the corresponding interpolant generated with our graph-based
method.
}

\begin{example}\label{ex:no_dl_itp}
  Consider the following sets of \dlrat atoms:\\
\begin{minipage}{.7\textwidth}
  \begin{displaymath}
    \begin{split}
      A &\defas \{ (0 \leq x_1 - x_2 + 1), (0 \leq x_2 - x_3), (0 \leq x_4 - x_5 - 1) \} \\
      B &\defas \{ (0 \leq x_5 - x_1), (0 \leq x_3 - x_4 - 1) \}.
    \end{split}
  \end{displaymath}
\end{minipage}
\begin{minipage}{.3\textwidth}
\vspace{-.4cm}
\hspace{.4cm}
\scalebox{0.55}{\input{dl_ex.pstex_t}}
\end{minipage}
\noindent
corresponding to the negative cycle on the right.
It is straightforward to see from the graph that the resulting interpolant is 
$(0 \leq x_1- x_3 + 1)\wedge (0 \leq x_4-x_5  - 1)$, because the first conjunct is
the summary constraint of the first two conjuncts in $A$.

Applying instead the rules of Figure~\ref{fig:la_proof_rules} 
{with coefficients 1,}
the proof of
unsatisfiability is:

\begin{small2}
  \begin{displaymath}
    \infrule{Comb}{
      \infrule{Comb}{
        \infrule{Comb}{
          \infrule{Comb}{
            \hyprule{(0 \leq x_1 - x_2 + 1)} &
            \hyprule{(0 \leq x_2 - x_3)}
          }{(0 \leq x_1 - x_3 + 1)} &
          \hyprule{(0 \leq x_4 - x_5 - 1)}
        }{(0 \leq x_1 - x_3 + x_4 - x_5)} &
        \hyprule{(0 \leq x_5 - x_1)}
      }{(0 \leq - x_3 + x_4)} &
      \hyprule{(0 \leq x_3 - x_4 - 1)}
    }{(0 \leq -1)}
  \end{displaymath}
\end{small2}
\noindent
  By using the interpolation rules for \laR, the interpolant we obtain is
  $  (0 \leq x_1 - x_3 + x_4 - x_5)$, which is not in \dlrat, and
  is weaker than that computed above:
\begin{small2}
  \begin{displaymath}
    \infrule{Comb}{
      \infrule{Comb}{
        \infrule{Comb}{
          \infrule{Comb}{
            \hyprule{(0 \leq x_1 - x_2 + 1)} &
            \hyprule{(0 \leq x_2 - x_3)}
          }{(0 \leq x_1 - x_3 + 1)} &
          \hyprule{(0 \leq x_4 - x_5 - 1)}
        }{(0 \leq x_1 - x_3 + x_4 - x_5)} &
        \hyprule{(0 \leq 0)}
      }{(0 \leq x_1 - x_3 + x_4 - x_5)} &
      \hyprule{(0 \leq 0)}
    }{(0 \leq x_1 - x_3 + x_4 - x_5)}
  \end{displaymath}
\end{small2}
{Notice that, if instead we apply our technique of
\sref{sec:la_stronger_interpolants}, 
then the \larat-interpolant generated from the above proof is identical
to the \dlrat one above.}

\end{example}


\section{From SMT(\utvpi) solving to SMT(\utvpi) interpolation}
\label{sec:utvpi}

{
The Unit-Two-Variables-Per-Inequality (\utvpi{}) theory 
is a subtheory of linear arithmetic, in which all constraints are in the form
$(0 \le ax_1+bx_2+ k)$, where $k$ is a numerical constant,
$a,b\in\{-1,0,1\}$, and variables $x_i$, $x_2$ range either over the rationals
(for \utvpirat) or over the integers (for \utvpiint{}). 
Consequently, \dlrat{} is a subtheory of \utvpirat, which is itself a subtheory of \larat{}, 
and \dlint{} is a subtheory of \utvpiint{}, which is itself a subtheory of \laint{}.
}
 
{
As for \dl{}, \utvpi{} can be treated more efficiently than the full \la,
and several specialized algorithms for \utvpi{} have been proposed in the literature. 
Traditional techniques are based on the  iterative computation of the transitive closure
of the constraints
\cite{harvey-utvpi,JaffarMSY94}; 
more recently \cite{lahiri_frocos2005} proposed a novel technique 
 based on a reduction to \dl{}, so that graph-based
techniques can be exploited, resulting into an asymptotically-faster
algorithm.
We adopt the latter approach and show how the graph-based interpolation
technique of \sref{sec:dl} can be extended to  
 \utvpi, for both the rationals (\sref{sec:utvpi_rat}) and the integers (\sref{sec:utvpi_int}).
}

\subsection{Graph-based interpolation for \utvpi on the Rationals}
\label{sec:utvpi_rat}

\begin{figure}[t]
$$
\begin{array}{|l|l|l|}\hline
\mbox{\utvpirat constraints\ } & \mbox{\dlrat{} constraints}  \\ 
\hline
\pn{x_2}{x_1}{+k} & \pnenc{x_2}{x_1}{+k} \\
\pp{x_1}{x_2}{+k} & \ppenc{x_2}{x_1}{+k} \\
\nn{x_1}{x_2}{+k} & \nnenc{x_2}{x_1}{+k} \\
\ub{x_1}{k} &  \ubenc{x_1}{+2\cdot k}  \\
\lb{x_1}{k} &  \lbenc{x_1}{+2\cdot k} \\ 
\hline
\end{array}
$$
\caption{\label{utvpi_fig}
The conversion map %
from \utvpirat to \dlrat.} %
\end{figure}
  
We analyze first the simpler case of \utvpirat.
Min\'e \cite{DBLP:conf/wcre/Mine01} showed that it is
possible to encode a set of \utvpirat constraints into a
\dlrat one in a satisfiability-preserving way.
The encoding works as follows. 
  We use $x_i$ to denote variables in the \utvpirat
 domain and $u$, $v$ for variables in the \dlrat domain.
For every variable $x_i$ in \utvpirat, 
we introduce two distinct variables \pv{x_i} and \nv{x_i} in \dlrat.
We introduce a mapping $\Upsilon$ from \dlrat variables to \utvpirat signed variables,
such that $\Upsilon(\pv{x_i}) = x_i$ and $\Upsilon(\nv{x_i}) = -x_i$.
$\Upsilon$ extends to (sets of) constraints in the natural way:
{%
$\Upsilon(0 \leq ax_1 + bx_2 + k) \defas (0 \leq a\Upsilon(x_1) + b\Upsilon(x_2)+c)$, and $\Upsilon(\{c_i\}_i) \defas \{\Upsilon(c_i)\}_i$.}
 We  say that $(\pv{x_i})^-= \nv{x_i}$ and $(\nv{x_i})^-=
  \pv{x_i}$.
We say that the constraints 
$(0\le u - v)$ and $(0\le (v)^- - (u)^-)$
s.t. $u,v\in \{\pv{x_i},\nv{x_i}\}_i$
are {\em dual}.
We encode each \utvpi constraint into the conjunction of  two dual
\dlrat{} constraints, as represented in  Figure~\ref{utvpi_fig}.
For each \dlrat constraint $(0 \leq v - u + k)$, 
$(0 \leq \Upsilon(v) - \Upsilon(u) + k)$ is the corresponding \utvpirat constraint.
Notice that the two dual \dlrat constraints 
{in the right column of Figure~\ref{utvpi_fig}}
are just different
representations of the original \utvpirat constraint.
 (The two dual constraints encoding a single-variable
   constraint are identical, so that their conjunction is collapsed
   into one constraint only.) 
The resulting set of constraints is satisfiable in \dlrat if and only
if the original one is satisfiable in \utvpirat \cite{DBLP:conf/wcre/Mine01,lahiri_frocos2005}.

Consider the pair $(A,B)$ where $A$ and $B$ are
sets of \utvpirat constraints. 
We apply the map of Figure~\ref{utvpi_fig} and 
we encode $(A,B)$ into a \dlrat{} pair $(A',B')$, and build the
constraint graph $G(A'\wedge B')$.
If $G(A'\wedge B')$ has no negative cycle, we can conclude that
$A'\wedge B'$ is \dlrat-consistent, and hence that $A\wedge B$ is
\utvpirat-consistent; otherwise, $A'\wedge B'$ is \dlrat-inconsistent,
and hence $A\wedge B$ is \utvpirat-inconsistent {\cite{DBLP:conf/wcre/Mine01,lahiri_frocos2005}}.
In fact, {it is straightforward to} observe that 
for any set of \dlrat constraints $\{C_1, \ldots, C_n, C\}$
resulting from the encoding of some \utvpirat constraints,
if $\bigwedge_{i=1}^n C_i \models_\dlrat C$ then 
$\bigwedge_{i=1}^n \Upsilon(C_i) \models_\utvpirat \Upsilon(C)$.

When $A\land B$ is inconsistent, we can generate an \utvpirat-interpolant
by extending the graph-based approach used for \dlrat.

\begin{theorem}
  Let $A \land B$ be an inconsistent conjunction of \utvpirat-constraints,
  and let $G(A'\land B')$ be the corresponding graph of \dlrat-constraints.
  Let $I'$ be a \dlrat-interpolant built from $G(A'\land B')$ 
  with the technique described in \sref{sec:dl}. 
  Then $I \defas \Upsilon(I')$ is an interpolant for $(A, B)$.
\end{theorem}
\begin{proof}
  (i) %
$I'$ is a conjunction of summary constraints, so it is in the form
$\bigwedge_i C_i$.  Therefore $A' \models_\dlrat C_i$ for all $i$, and so by
the observation above $A \models_\utvpirat \Upsilon(C_i)$. Hence, $A
\models_\utvpirat I$.
(ii) %
From the \dlrat-inconsistency of $I' \land B'$ we immediately derive that $I
\land B$ is \utvpirat-inconsistent.
(iii) %
$I \preceq A$ and $I \preceq B$ derive from $I' \preceq A'$ and $I'
\preceq B'$ by the definitions of $\Upsilon$ and the map of Figure~\ref{utvpi_fig}.  
\end{proof}

{
As with the \dlrat{} case, in principle, it is possible 
to generate a proof of
unsatisfiability for a conjunction of \utvpirat atoms $A\wedge B$
by repeatedly applying the \comb rule for \laR
\cite{mcmillan_interpolating_prover} with $c_1=c_2=1$.
As with \dlrat, however, 
the interpolants generated from such proofs may not be
 \utvpirat formulas anymore. Moreover, if computed starting from the same
inconsistent set $C$ and unless our technique of
\sref{sec:la_stronger_interpolants} is adopted, they are either
identical or weaker than those generated with 
our graph-based method, since they are in the form $(0 \le
\textstyle\sum_i t_i)$ s.t. $\textstyle\bigwedge_i (0 \le 
t_i)$ is the interpolant generated with our method. 
}

\newcommand{\cone}{{\ensuremath {\pp{x_3}{x_4}{-6}}}}
\newcommand{\ctwo}{{\ensuremath{\pn{x_2}{x_3}{+4}}}}
\newcommand{\ctwoone}{{\ensuremath{\pp{x_2}{x_1}{+3}}}}
\newcommand{\ctwotwo}{{\ensuremath{\nn{x_1}{x_3}{+1}}}}
\newcommand{\cthree}{{\ensuremath{\nn{x_5}{x_4}{+1}}}}
\newcommand{\cfour}{{\ensuremath{\nn{x_2}{x_3}{+3}}}}
\newcommand{\cfive}{{\ensuremath{\pn{x_5}{x_4}{+2}}}}
\newcommand{\cfiveone}{{\ensuremath{\pn{x_5}{x_6}{-1}}}}
\newcommand{\cfivetwo}{{\ensuremath{\pn{x_6}{x_4}{+4}}}}

\newcommand{\coneenc}{{\ensuremath {\ppenc{x_3}{x_4}{-6}}}}
\newcommand{\ctwoenc}{{\ensuremath{\pnenc{x_2}{x_3}{+4}}}}
\newcommand{\ctwooneenc}{{\ensuremath{\ppenc{x_2}{x_1}{+3}}}}
\newcommand{\ctwotwoenc}{{\ensuremath{\nnenc{x_1}{x_3}{+1}}}}
\newcommand{\cthreeenc}{{\ensuremath{\nnenc{x_5}{x_4}{+1}}}}
\newcommand{\cfourenc}{{\ensuremath{\nnenc{x_2}{x_3}{+3}}}}
\newcommand{\cfiveenc}{{\ensuremath{\pnenc{x_5}{x_4}{+2}}}}
\newcommand{\cfiveoneenc}{{\ensuremath{\pnenc{x_5}{x_6}{-1}}}}
\newcommand{\cfivetwoenc}{{\ensuremath{\pnenc{x_6}{x_4}{+4}}}}

\begin{figure}[t]
\begin{center} 
\scalebox{0.6}{\input{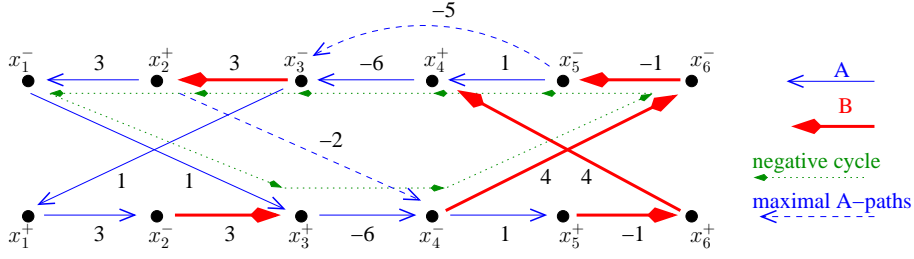}}
\end{center}
\caption{\label{utvpirat_ex_fig}
The constraint graph of Example~\ref{utvpirat_ex}.
 (We represent only one negative cycle with its corresponding
 A-paths, because the other is dual.)
}
\end{figure}
\begin{example}
\label{utvpirat_ex}
Consider the following sets of\ \utvpirat constraints:
\begin{displaymath}
  \begin{split}
    A  = \{ 
            &\ctwoone, \ctwotwo, \\
            &\cone, \cthree \}\\[1ex]
    B  = \{ &\cfour, 
            \cfiveone, \cfivetwo \}
  \end{split}
\end{displaymath}

\noindent
By the map of Figure~\ref{utvpi_fig}, they are converted
into the following sets of \dlrat constraints:
\begin{displaymath}
  \begin{split}
    A' = \{ 
            & \ctwooneenc,  \\
            & \ctwotwoenc,  \\
            & \coneenc,  \\
            & \cthreeenc \} 
  \end{split}
\end{displaymath}
\begin{displaymath}
  \begin{split}
    B' = \{ & \cfourenc,  \\
            & \cfiveoneenc, \\
            & \cfivetwoenc \}
  \end{split}
\end{displaymath}
\noindent
whose conjunction corresponds to the constraint graph of Figure~\ref{utvpirat_ex_fig}.
This graph has a negative cycle
$$%
\label{negcycle1_eq}
C' \defas  
\pv{x_2} \xrightarrow{3} 
\nv{x_1} \xrightarrow{1} 
\pv{x_3} \xrightarrow{-6} 
\nv{x_4} \xrightarrow{4} 
\nv{x_6} \xrightarrow{-1} 
\nv{x_5} \xrightarrow{1} 
\pv{x_4} \xrightarrow{-6} 
\nv{x_3} \xrightarrow{3} 
\pv{x_2}.
$$%
\noindent
 Thus, %
 $A\wedge B$ is inconsistent in \utvpirat.
From the negative cycle $C'$ %
we can extract the set of $A'$-paths
$
\{\pv{x_2} \xrightarrow{-2} \nv{x_4}, 
\ 
\nv{x_5} \xrightarrow{-5} \nv{x_3} \}, 
$
corresponding to the formula %
$
I'  \defas \pndiff{x_2}{x_4}{-2} \wedge \nndiff{x_5}{x_3}{-5},
$ %
which is an interpolant for $(A',B')$. $I'$
is thus mapped back into
$  
I  \defas \Upsilon(I') \defas \pp{x_2}{x_4}{-2} \wedge \pn{x_3}{x_5}{-5},
$
which is an interpolant for $(A,B)$.

Applying instead the \larat{} interpolation technique of \cite{mcmillan_interpolating_prover},
we find the interpolant $(0 \leq -x_2 -x_4 +x_5 -x_3 -7)$,
which is not in \utvpirat and is strictly weaker than that computed with our method.

\end{example}

\subsection{Graph-based interpolation for \utvpi on the Integers}
\label{sec:utvpi_int}

In order to deal with the more complex case of \utvpiint,
we adopt a layered approach~\cite{seba_lazy_smt}.
First, we check the consistency in \utvpirat using the technique of \cite{DBLP:conf/wcre/Mine01}.
If this results in an inconsistency, we compute an \utvpirat-interpolant 
as described in \sref{sec:utvpi_rat}. 
If the \utvpirat-procedure does not detect an inconsistency, 
we check the consistency in \utvpiint using the algorithm proposed by
Lahiri and Musuvathi in \cite{lahiri_frocos2005}, 
which extends the ideas of
\cite{DBLP:conf/wcre/Mine01} to the integer domain. 
In particular, it gives necessary and sufficient conditions 
to decide unsatisfiability by detecting particular kinds 
of zero-weight cycles in the induced \dl constraint graph.
This procedure works in $O(n\cdot m)$ time and $O(n+ m)$ space, $m$ and
$n$ being the number of constraints and variables respectively, which 
improves the previous  $O(n^2\cdot m)$ time and $O(n^2)$ space complexity
of the previous procedure of \cite{JaffarMSY94}.

We build on top of this algorithm and
we extend  the graph-based approach of \sref{sec:utvpi_rat} 
for producing interpolants also in \utvpiint{}. 
In particular, 
we use the following reformulation of a result of \cite{lahiri_frocos2005}.

\begin{theorem}  \label{teo:utvpi_int}
Let $\phi$ be a conjunction of \utvpiint constraints s.t. $\phi$ is 
satisfiable in \utvpirat. Then $\phi$ is unsatisfiable in 
\utvpiint iff the constraint graph $G(\phi)$ generated from $\phi$ has a cycle $C$
of weight 0 containing two vertices $\pv{x_i}$ and $\nv{x_i}$ s.t. the weight of
the path $\nv{x_i} \leadsto \pv{x_i}$ along $C$
is odd.~%
\end{theorem}
  
\begin{proof}
The ``only if'' part is a corollary of lemmas 1, 2 and 4 in
 \cite{lahiri_frocos2005}. 
The ``if'' comes straightforwardly from the analysis done in 
 \cite{lahiri_frocos2005}, whose main intuitions we recall in what follows.
\label{utvpi_tightening}
Assume the constraint graph $G(\phi)$ generated from $\phi$ has one cycle $C$
of weight 0 containing two vertices $\pv{x_i}$ and $\nv{x_i}$ s.t. the weight of
the path $\nv{x_i} \leadsto \pv{x_i}$ along $C$ is $2k+1$ for some
integer value $k$.
(Since $C$ has weight $0$, the weight
of the other path $\pv{x_i} \leadsto \nv{x_i}$ along $C$ is $-2k-1$.)
Then, the paths $\nv{x_i} \leadsto \pv{x_i}$ and $\pv{x_i} \leadsto \nv{x_i}$
contain at least two constraints, 
because otherwise their weight would be even (see the last two lines of Figure~\ref{utvpi_fig}).
Then, $\nv{x_i} \leadsto \pv{x_i}$ is in the form
$\nv{x_i} \leadsto v \xrightarrow{n} \pv{x_i}$, for some $v$ and $n$. 
From $\nv{x_i} \leadsto v$, we can derive the summary constraint
$(0 \le v - \nv{x_i} + (2k+1-n))$, which corresponds to the \utvpiint constraint
$(0 \le \Upsilon(v) + x_i + (2k+1-n))$. 
(This corresponds to $l-2$ applications of the \textsc{Transitive} rule of \cite{lahiri_frocos2005}, 
$l$ being the number of constraints in $\nv{x_i} \leadsto \pv{x_i}$.)
Then, by observing that the \utvpiint constraint corresponding to 
$v \xrightarrow{n} \pv{x_i}$ is $(0 \leq x_i - \Upsilon(v) + n)$,
we can apply the \textsc{Tightening} rule of \cite{lahiri_frocos2005} to
obtain $(0 \leq x_i + \lfloor (2k+1-n + n)/2 \rfloor)$, which is equivalent to 
$(0 \leq x_i + k)$.
Similarly, from $\pv{x_i} \leadsto \nv{x_i}$ we can obtain $(0 \leq -x_i -k-1)$,
and thus an inconsistency using the \textsc{Contradiction} rule of \cite{lahiri_frocos2005}.
\end{proof}

Consider a pair $(A, B)$ of \utvpiint constraints such that 
$A\land B$ is consistent in \utvpirat but inconsistent in \utvpiint.
By Theorem 1, %
the constraint graph $G(A'\land B')$ has a cycle $C$
of weight 0 containing two vertices $\pv{x_i}$ and $\nv{x_i}$ s.t. 
the weight of the paths  $\nv{x_i} \leadsto \pv{x_i}$ and $\pv{x_i}
\leadsto \nv{x_i}$ along $C$ are $2k+1$ and $-2k-1$ respectively, for
some value $k\in\mathbb{Z}$.
Our algorithm computes an interpolant for $(A, B)$ from the cycle $C$.
Let $C_A$ and $C_B$ be the subsets of the edges in $C$ corresponding to 
constraints in $A'$ and $B'$ respectively. 
We have to distinguish four distinct sub-cases. 

\newcommand{\utvpicase}[1]{\medskip\noindent\textbf{#1~}}

\utvpicase{Case 1:} {\em $x_i$ occurs in $B$ but not in $A$.}
Consequently, $\pv{x_i}$ and $\nv{x_i}$ occur in $B'$ but not in $A'$,
and hence they occur in  $C_B$ but not in $C_A$.
Let $I'$ be the conjunction of the summary constraints of the
maximal $C_A$-paths, and let $I$ be the conjunction of the corresponding 
\utvpiint constraints. 
\begin{theorem}
$I$ is an interpolant for $(A,B)$.  
\end{theorem}
\begin{proof}
(i)
By construction, $A\models_{\utvpiint}I$, as in \sref{sec:utvpi_rat}.
(ii)
The constraints in $I'$
and $C_B$ form a cycle matching the hypotheses of
Theorem~\ref{teo:utvpi_int}, from which $I\wedge B$ is \utvpiint-inconsistent. 
(iii)
We notice that  every variable $\pv{x_j},\nv{x_j}$ occurring in the
conjunction of the summary constraints is an end-point variable, 
so that 
$I'\preceq C_A$ and $I'\preceq C_B$, and thus 
$I\preceq A$ and $I\preceq B$.
\end{proof}

\begin{figure}[t]
\vspace{-1mm}
  \begin{minipage}[t]{0.5\linewidth}
    \centering
    \scalebox{0.45}{\input{utvpiint_case1.pstex_t}}
    \caption{\utvpiint interpolation, Case 1.
      \label{fig:case1}}    
  \end{minipage}
  \begin{minipage}[t]{0.5\linewidth}
    \centering
    \scalebox{0.45}{\input{utvpiint_case2.pstex_t}}
    \caption{\utvpiint interpolation, Case 2.
      \label{fig:case2}}    
  \end{minipage}
\end{figure}

\begin{example}\label{ex:case1}
Consider the following set of constraints:
\begin{displaymath}
  \begin{split}
    S = \{ & 
    (0 \leq x_1 - x_2 + 4),
    (0 \leq -x_2 - x_3 -5),
    (0 \leq x_2 + x_6 -4),
    (0 \leq x_5 + x_2 + 3), \\
    & (0 \leq -x_1 + x_3 + 2),
    (0 \leq -x_6 - x_4),
    (0 \leq x_4 - x_5)
    \},
  \end{split}
\end{displaymath}
partitioned into $A$ and $B$ as follows:
\begin{displaymath}
  A \left\{
    \begin{array}{l}
      (0 \leq x_3 - x_1 + 2) \\
      (0 \leq -x_6 - x_4) \\
      (0 \leq x_4 - x_5)
    \end{array}
  \right.
  \hspace{3cm}
  B \left\{
    \begin{array}{l}
      (0 \leq x_1 - x_2 + 4) \\
      (0 \leq -x_2 - x_3  -5) \\
      (0 \leq x_2 + x_6 -4) \\
      (0 \leq x_5 + x_2 + 3)
    \end{array}
  \right.
\end{displaymath}

Figure~\ref{fig:case1} shows a zero-weight cycle $C$ in $G(A'\land B')$ 
such that the paths $\nv{x_2} \leadsto \pv{x_2}$ and $\pv{x_2} \leadsto \nv{x_2}$ have an odd weight ($-1$ and $1$ resp.)
Therefore, by Theorem~\ref{teo:utvpi_int} $A\wedge B$ is \utvpiint-inconsistent.
The two summary constraints of the maximal $C_A$ paths are 
$(0 \leq \nv{x_6} - \pv{x_5})$ and $(0 \leq \pv{x_3} - \pv{x_1} + 2)$.
It is easy to see that
$I = (0 \leq -x_6 - x_5) \land (0 \leq x_3 - x_1 + 2)$ is an $\utvpiint$-interpolant for $(A, B)$.
\end{example}

\utvpicase{Case 2:} {\em $x_i$ occurs in both $A$ and $B$.}
Consequently, $\pv{x_i}$ and $\nv{x_i}$ occur in both $A'$ and $B'$.
If neither $\pv{x_i}$ nor $\nv{x_i}$ is such that both the incoming
and outgoing edges belong to $C_A$, 
then the cycle obtained by replacing each maximal $C_A$-path 
with its summary constraint still contains both $\pv{x_i}$ and $\nv{x_i}$, so
we can apply the
same process of Case 1.
Otherwise, if both the incoming and outgoing edges of $\pv{x_i}$ belong
to $C_A$, then 
we split the maximal $C_A$-path 
$u_1\xrightarrow{c_1}
\ldots
\xrightarrow{c_k}
\pv{x_i}
\xrightarrow{c_{k+1}}\ldots
\xrightarrow{c_{n}}
u_n$
containing $\pv{x_i}$ into the two parts which are separated by
$\pv{x_i}$: 
$u_1\xrightarrow{c_1}
\ldots
\xrightarrow{c_k}
\pv{x_i}$ 
and
$
\pv{x_i}
\xrightarrow{c_{k+1}}\ldots
\xrightarrow{c_{n}}
u_n$. We do the same for \nv{x_i}. 
Let $I'$ be the conjunction of
the resulting summary constraints, and let $I$ be corresponding set 
of \utvpiint constraints. 
\begin{theorem}
$I$ is an interpolant for $(A,B)$.
\end{theorem}
\begin{proof}
(i)
As with Case 1, again, $A\models_{\utvpiint}I$. 
(ii)
Since we split the maximal $C_A$ paths as described above, 
the constraints in $I'$
and $C_B$ form a cycle matching the hypotheses of
Theorem~\ref{teo:utvpi_int}, from which 
$I\wedge B$ is \utvpiint-inconsistent.
(iii)
$\pv{x_i},\nv{x_i}$ occur in both $A'$ and $B'$ by
hypothesis, and every other variable $\pv{x_j},\nv{x_j}$ occurring in the
conjunction of the summary constraints is an end-point variable, 
so that 
$I'\preceq C_A$ and $I'\preceq C_B$, and thus 
$I\preceq A$ and $I\preceq B$.
\end{proof}

\begin{example}\label{ex:case2}
Consider again the set of constraints $S$ of Example~\ref{ex:case1}, 
partitioned into $A$ and $B$ as follows:
\begin{displaymath}
  A \left\{
    \begin{array}{l}
      (0 \leq x_3 - x_1 + 2) \\
      (0 \leq -x_6 - x_4) \\
      (0 \leq x_2 + x_6 -4) \\
      (0 \leq x_1 - x_2 + 4)
    \end{array}
  \right.
  \hspace{3cm}
  B \left\{
    \begin{array}{l}
      (0 \leq -x_2 - x_3 -5) \\
      (0 \leq x_5 + x_2 + 3) \\
      (0 \leq x_4 - x_5)
    \end{array}
  \right.
\end{displaymath}
and the zero-weight cycle $C$ of $G(A'\land B')$ shown in Figure~\ref{fig:case2}.
As in the previous example, there is a path $\nv{x_2} \leadsto \pv{x_2}$ of weight $-1$ 
and a path $\pv{x_2} \leadsto \nv{x_2}$ of weight $1$.
In this case there is only one maximal $C_A$ path, namely $\pv{x_4} \leadsto \pv{x_3}$.
Since the cycle obtained by replacing it with its summary constraint 
$(0 \leq \pv{x_3} - \pv{x_4}  + 2)$ does not contain $\pv{x_2}$,
we split $\pv{x_4} \leadsto \pv{x_3}$
into two paths, $\pv{x_4} \leadsto \pv{x_2}$ and $\pv{x_2} \leadsto \pv{x_3}$,
whose summary constraints are 
$(0 \leq \pv{x_2} - \pv{x_4} -4)$ and $(0 \leq \pv{x_3} - \pv{x_2} + 6)$ respectively.
By replacing the two paths above with the two summary constraints, we get
a zero-weight cycle which still contains the two odd paths 
$\nv{x_2} \leadsto \pv{x_2}$ and $\pv{x_2} \leadsto \nv{x_2}$. 
Therefore, $I \defas (0 \leq x_2 - x_4 -4) \land (0 \leq x_3 - x_2 + 6)$ is an interpolant for $(A, B)$.

Notice that the \utvpiint-formula $J \defas (0 \leq x_3 - x_4 + 2)$ 
corresponding to the summary constraint of the maximal $C_A$ path $\pv{x_4} \leadsto \pv{x_3}$
is not an interpolant, 
since $J \land B$ is not \utvpiint-inconsistent.
In fact, if we replace the maximal $C_A$ path $\pv{x_4} \leadsto \pv{x_3}$ 
with the summary constraint $\pv{x_4} \xrightarrow{2} \pv{x_3}$, 
the cycle we obtain has still weight zero, but it contains no odd path 
between two variables $\pv{x_i}$ and $\nv{x_i}$.
\end{example}

\begin{figure}[t]
\vspace{-1mm}
  \begin{minipage}[t]{0.5\linewidth}
    \centering
    \scalebox{0.45}{\input{utvpiint_case3.pstex_t}}
    \caption{\utvpiint interpolation, Case 3.
      \label{fig:case3}}
  \end{minipage}
  \begin{minipage}[t]{0.5\linewidth}
    \centering
    \scalebox{0.45}{\input{utvpiint_case4.pstex_t}}
    \caption{\utvpiint interpolation, Case 4.
      \label{fig:case4}}
  \end{minipage}
\end{figure}

\utvpicase{Case 3:} 
\emph{$x_i$ occurs in $A$ but not in $B$, 
and one of the paths 
$\pv{x_i} \leadsto \nv{x_i}$ or $\nv{x_i} \leadsto \pv{x_i}$ in $C$
contains only constraints of $C_A$.}
In this case, $\pv{x_i}$ and $\nv{x_i}$ occur in $A'$ but not in $B'$.
Suppose that $\nv{x_i} \leadsto \pv{x_i}$ consists only of constraints of $C_A$
({the case $\pv{x_i} \leadsto \nv{x_i}$ is analogous}).

Let $2k+1$ be the weight of the path $\nv{x_i} \leadsto \pv{x_i}$ 
(which is odd by hypothesis),
and let $\overline{C}$ be the cycle obtained by replacing such path with the edge
$\nv{x_i} \xrightarrow{2k} \pv{x_i}$ in $C$. In the following, we 
call such a replacement \emph{tightening summarization}.
Since $C$ has weight zero, $\overline{C}$ has negative weight.
Let $C^P$ be the set of \dl-constraints in the path $\nv{x_i} \leadsto \pv{x_i}$.
Let $I'$ be the $\dl$-interpolant computed from $\overline{C}$ for 
$(C_A\setminus C^P\cup\{(0 \le \pv{x_i} - \nv{x_i} + 2k)\}, C_B)$,
and let $I$ be the corresponding $\utvpiint$ formula.
\begin{theorem}
$I$ is an interpolant for $(A, B)$.
\end{theorem}

\begin{proof}
(i)
Let $P$ be the set of \utvpiint constraints in the path $\nv{x_i} \leadsto \pv{x_i}$. 
Since the weight $2k+1$ of such path is odd, 
we have that $P \models_\utvpiint (0 \leq x_i + k)$
(cf. page \pageref{utvpi_tightening}).
Since $P \subseteq A$, therefore, $A \models_\utvpiint (0 \leq x_i + k)$.
By observing that $(0 \leq \pv{x_i} - \nv{x_i} + 2k)$ 
is the $\dl$-constraint corresponding to $(0 \leq x_i + k)$
we conclude that 
$C_A\setminus C^P \cup (0 \leq \pv{x_i} - \nv{x_i} + 2k) \models_\dl I'$
implies that $A \setminus P \cup (0 \leq x_i + k) \models_\utvpiint I$, 
and so that $A \models_\utvpiint I$.

(ii)
Since all the constraints in $C_B$ occur in $\overline{C}$, 
we have that $B \land I$ is \utvpiint-inconsistent.

(iii)
Since by hypothesis all the constraints in the path $\nv{x_i} \leadsto \pv{x_i}$ occur in $C_A$, 
from $I' \preceq (C_A \setminus C^P \cup \{(0 \leq \pv{x_i}-\nv{x_i}+2k)\})$
we have that $I \preceq A$.
Finally, since all the constraints in $C_B$ occur in $\overline{C}$, 
we have that $I\preceq B$.
\end{proof}

\begin{example}\label{ex:case3}
Consider again the set $S$ of constraints of Example~\ref{ex:case1}, 
this time partitioned into $A$ and $B$ as follows:
\begin{displaymath}
  A \left\{
    \begin{array}{l}
      (0 \leq x_1 - x_2 + 4) \\
      (0 \leq x_3 - x_1 + 2) \\
      (0 \leq -x_2 - x_3 -5) \\
      (0 \leq x_2 + x_6 -4)
    \end{array}
  \right.
  \hspace{3cm}
  B \left\{
    \begin{array}{l}
      (0 \leq x_5 + x_2 + 3) \\
      (0 \leq -x_6 - x_4) \\
      (0 \leq x_4 - x_5)
    \end{array}
  \right.
\end{displaymath}

Figure~\ref{fig:case3} shows a zero-weight cycle $C$ of $G(A'\land B')$. 
The only maximal $C_A$ path is $\nv{x_6} \leadsto \nv{x_2}$. 
Since the path $\pv{x_2} \leadsto \nv{x_2}$ has weight $1$, 
we can add the tightening edge \mbox{$\pv{x_2} \xrightarrow{1-1} \nv{x_2}$} 
to $G(A'\land B')$ (shown in dots and dashes in Figure~\ref{fig:case3}), 
corresponding to the constraint $(0 \leq \nv{x_2} - \pv{x_2})$.
Since all constraints in the path $\pv{x_2} \leadsto \nv{x_2}$ belong to $A'$,
$A' \models (0 \leq \nv{x_2} - \pv{x_2})$. 
Moreover, the cycle obtained by replacing the path $\pv{x_2} \leadsto \nv{x_2}$
with the tightening edge $\pv{x_2} \xrightarrow{0} \nv{x_2}$
has a negative weight ($-1$). 
Therefore, we can generate a $\dl$-interpolant 
$I'\defas (0 \leq \nv{x_2} - \nv{x_6} -4)$ from such cycle,
which corresponds to the $\utvpiint$-interpolant $I \defas (0 \leq -x_2 + x_6 -4)$.

Notice that, similarly to Example~\ref{ex:case2}, also in this case
we cannot obtain an interpolant 
from the summary constraint $(0 \leq \nv{x_2} - \nv{x_6} -3)$ 
of the maximal $C_A$ path $\nv{x_6} \leadsto \nv{x_2}$,
as $(0 \leq -x_2 + x_6 -3) \land B$ is not \utvpiint-inconsistent.
\end{example}

\utvpicase{Case 4:} 
{\em $x_i$ occurs in $A$ but not in $B$,
and neither the path $\pv{x_i} \leadsto \nv{x_i}$ nor the path $\nv{x_i} \leadsto \pv{x_i}$ in $C$ 
consists only of constraints of $C_A$.}
As in the previous case, $\pv{x_i}$ and $\nv{x_i}$ occur in $A'$ but not in $B'$,
and hence they occur in  $C_A$ but not in $C_B$.
In this case, however, 
we can apply a tightening summarization neither to $\pv{x_i} \leadsto \nv{x_i}$ nor to $\nv{x_i} \leadsto \pv{x_i}$,
since none of the two paths consists only of constraints of $C_A$. 
We can, however, perform a \emph{conditional tightening summarization} as follows.
Let $C^P_A$ and $C^P_B$ be the sets of constraints of $C_A$ and $C_B$ respectively 
occurring in the path $\nv{x_i} \leadsto \pv{x_i}$,
and let $\overline{C}^P_A$ and $\overline{C}^P_B$ be the sets of summary constraints 
of maximal paths in $C^P_A$ and $C^P_B$.
From $\overline{C}^P_A \cup \overline{C}^P_B$, we can derive $\nv{x_i}\xrightarrow{2k}\pv{x_i}$ (cf. Case 3), 
where $2k+1$ is the weight of the path $\nv{x_i}\leadsto \pv{x_i}$. 
Therefore, $\overline{C}^P_A\cup \overline{C}^P_B \models (0 \leq \pv{x_i} - \nv{x_i} + 2k)$,
and thus $\overline{C}^P_A \models \overline{C}^P_B \rightarrow (0 \leq \pv{x_i} - \nv{x_i} + 2k)$.
We say that $(0 \leq \pv{x_i} - \nv{x_i} + 2k)$ is the summary constraint 
for $\nv{x_i} \leadsto \pv{x_i}$ \emph{conditioned to} $\overline{C}^P_B$.

Using conditional tightening summarization, we generate an interpolant as follows.
By replacing the path $\nv{x_i}\leadsto \pv{x_i}$ 
with $\nv{x_i}\xrightarrow{2k}\pv{x_i}$, we obtain a negative-weight cycle $\overline{C}$,
as in Case 3. 
Let $I'$ be the \dl-interpolant computed from $\overline{C}$ for
$(C_A\setminus C^P_A\cup \{(0\leq \pv{x_i}-\nv{x_i}+2k)\}, C_B\setminus C^P_B)$,
and let $I$ be the corresponding \utvpiint formula.
Finally, let $\overline{P}_B$ be the conjunction of \utvpiint constraints corresponding to $\overline{C}^P_B$. 
\begin{theorem}
$(\overline{P}_B \rightarrow I)$ is an interpolant for $(A, B)$.
\end{theorem}

\begin{proof}
(i)
We know that 
$C_A\setminus C^P_A\cup \{(0\leq \pv{x_i}-\nv{x_i}+2k)\} \models I'$,
because $I'$ is a \dl-interpolant.
Moreover, 
$\overline{C}^P_A\cup \overline{C}^P_B \models (0 \leq \pv{x_i} - \nv{x_i} + 2k)$,
and so $C^P_A\cup \overline{C}^P_B \models (0 \leq \pv{x_i} - \nv{x_i} + 2k)$.
Therefore, $C_A\cup \overline{C}^P_B \models I'$, and thus 
$A \cup \overline{P}_B \models_\utvpiint I$, 
from which $A \models_\utvpiint (\overline{P}_B \rightarrow I)$.

(ii)
Since $I'$ is a \dl-interpolant for 
$(C_A\setminus C^P_A\cup \{(0\leq \pv{x_i}-\nv{x_i}+2k)\}, C_B\setminus C^P_B)$,
$I' \land (C_B\setminus C^P_B)$ is \dl-inconsistent, 
and thus $I \land B$ is \utvpiint-inconsistent.
Since by construction $B \models_\utvpiint \overline{P}_B$, 
$(\overline{P}_B\rightarrow I) \land B$ is \utvpiint-inconsistent.

(iii)
From $I'\preceq C_B\setminus C^P_B$ we have that $I \preceq B$,
and from $I' \preceq C_A\setminus C^P_A\cup \{(0\leq \pv{x_i}-\nv{x_i}+2k)\}$
that $I \preceq A$. 
Moreover, all the variables occurring in the constraints in $\overline{C}^P_B$ 
are end-point variables, 
so that $\overline{C}^P_B\preceq C_A$ and $\overline{C}^P_B\preceq C_B$, and thus 
$\overline{P}_B\preceq A$ and $\overline{P}_B\preceq B$.
Therefore, $(\overline{P}_B \rightarrow I) \preceq A$ and $(\overline{P}_B \rightarrow I) \preceq B$.
\end{proof}

\begin{example}\label{ex:case4}
We partition the set $S$ of constraints of Example~\ref{ex:case1} into $A$ and $B$ as follows:
\begin{displaymath}
  A \left\{
    \begin{array}{l}
      (0 \leq x_1 - x_2 + 4) \\
      (0 \leq -x_2 - x_3 -5) \\
      (0 \leq x_5 + x_2 + 3) \\
      (0 \leq x_2 + x_6 -4)
    \end{array}
  \right.
  \hspace{3cm}
  B \left\{
    \begin{array}{l}
      (0 \leq x_3 - x_1 + 2) \\
      (0 \leq -x_6 - x_4) \\
      (0 \leq x_4 - x_5)
    \end{array}
  \right.
\end{displaymath}
Consider the zero-weight cycle $C$ of $G(A' \land B')$ shown in Figure~\ref{fig:case4}.
In this case, neither the path $\pv{x_2} \leadsto \nv{x_2}$ nor
the path $\nv{x_2} \leadsto \pv{x_2}$ consists only of constraints of $A'$,
and thus we cannot use any of the two tightening edges 
$\pv{x_2} \xrightarrow{1-1} \nv{x_2}$ and $\nv{x_2} \xrightarrow{-1-1} \pv{x_2}$
\emph{directly}
for computing an interpolant.
However, we can compute the summary $\nv{x_2} \xrightarrow{-2} \pv{x_2}$ 
for $\nv{x_2} \leadsto \pv{x_2}$
\emph{conditioned to} $\pv{x_5} \xrightarrow{0} \nv{x_6}$, 
which is the summary constraint of the $B$-path $\pv{x_5} \leadsto \nv{x_6}$, 
and whose corresponding \utvpiint constraint is $(0 \leq -x_6 - x_5)$.
By replacing the path $\nv{x_2} \leadsto \pv{x_2}$ with such summary, 
we obtain a negative-weight cycle $\overline{C}$,
from which we generate the \dl-interpolant $(0 \leq \pv{x_1} - \pv{x_3} - 3)$,
corresponding to the \utvpiint formula $(0 \leq x_1 - x_3 - 3)$.
Therefore, the generated \utvpiint-interpolant is 
\mbox{$(0 \leq -x_6 - x_5) \rightarrow (0 \leq x_1 - x_3 - 3)$}.

As in Example~\ref{ex:case3}, notice that we cannot generate an interpolant 
from the conjunction of summary constraints of maximal $C_A$ paths,
since the formula we obtain
(i.e. $(0 \leq x_1 + x_6) \land (0 \leq x_5 - x_3 -2)$)
is not inconsistent with $B$.
\end{example}


\section{Computing interpolants for combined theories via DTC}
\label{sec:dtc}

In this Section, we consider the problem of generating interpolants
for a pair of \tonetwo-formulas $(A, B)$, and propose a method based
on the Delayed Theory Combination (DTC) approach \cite{infocomp_dtc05}.
First, in \sref{sec:dtc_background} we provide some background on
Nelson-Oppen (NO) and DTC combination methods, and recall from
\cite{musuvathi_interpolation} the basics of interpolation for
combined theories using NO; then, we present our novel technique for
computing interpolants using DTC (\sref{sec:dtc_interpolation});
in \sref{sec:dtc_discussion} we discuss the advantages of the novel
method; finally, in \sref{sec:dtc_multipleinterpolants}, we show how
our novel technique can be used to generate multiple interpolants from
the same proof.

\subsection{Background}
\label{sec:dtc_background}

\subsubsection{Resolution proofs with NO vs. resolution proofs with DTC}
\label{sec:dtc_background_proofs}

One of the typical approaches to the SMT problem in combined theories, 
SMT(\tonetwo), is that of combining the solvers for $\T_1$ and for
$\T_2$ with the Nelson-Oppen (NO) integration schema \cite{NO79}.
The NO framework works for combinations of stably-infinite, signature-disjoint theories $\T_i$ with equality.
Moreover, it requires the input
formula to be \emph{pure} (i.e., s.t. all the atoms contain only
symbols in one theory): if not, a
\emph{purification} step is performed, by recursively labeling terms $t$ with
fresh variables $v_t$, and by conjoining the definition atom $(v_t = t)$ to
the formula. 
This process is linear in the size of the
input formula.~%
\footnote{{As shown in \cite{barrett_frocos}, the purification step
    is not strictly necessary. However, in the rest we shall assume that it is
  performed (as it is traditionally done in papers on combination of theories),
  since it makes the exposition easier.} }
For instance, the formula
$(f(x+3y)= g(2x-y))$ can be purified into
$(f(v_{x+3y})= g(v_{2x-y}))\wedge(v_{x+3y}=x+3y) \wedge (v_{2x-y}=2x-y))$.

In the NO setting, the two decision procedures for $\T_1$ and $\T_2$
cooperate by deducing and exchanging \emph{interface
equalities}\footnote{They deduce and exchange \emph{disjunctions} of
interface equalities if the theory is not convex.}, that is,
equalities between variables appearing in atoms of different theories
(\emph{interface variables}).

With an NO-based SMT solver, resolution proofs for formulas in a
combination \tonetwo of theories have the same structure as those for
formulas in a single theory \T. The only difference is that theory
lemmas in this case are the result of the NO-combination of $\T_1$ and
$\T_2$ (i.e., they are \tonetwo-lemmas)
(Figure~\ref{fig:no_dtc_proofs} left). From the point of view of
interpolation, the difference with respect to the case of a single
theory \T is that the \tonetwo-interpolants for the negations of the
\tonetwo-lemmas can be computed with the combination method of
\cite{musuvathi_interpolation} whenever it applies 
(see \sref{sec:dtc_background_intno}).

\medskip
Recently, an alternative approach for combining theories in SMT has
been proposed, called Delayed Theory Combination (DTC)
\cite{infocomp_dtc05}. With DTC, the solvers for $\T_1$ and $\T_2$ do
not communicate directly. The integration is performed by the SAT
solver, by augmenting the Boolean search space with up to all the
possible interface equalities, so that each truth assignment on both
original atoms and interface equalities is checked for consistency
independently on both theories.  DTC has several advantages wrt. NO,
{in terms of versatility, efficiency, and restrictions imposed to
\T-solvers} \cite{infocomp_dtc05,AMAI08_dtc}, so that many current SMT tools
implement variants and evolutions of DTC.

\begin{figure}[t]
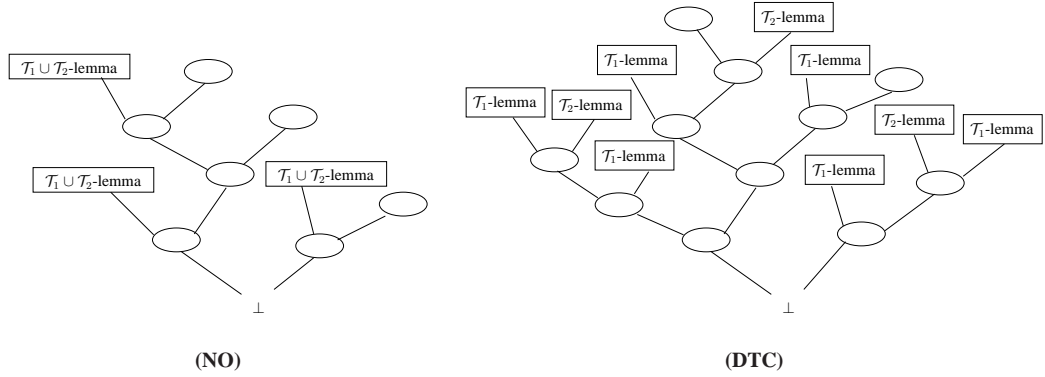

  \centering
  \begin{tabular}{c@{\hspace{5mm}}c}
    \scalebox{0.5}{\input{no_proof.pstex_t}} &
    \scalebox{0.5}{\input{dtc_proof.pstex_t}} \\[3mm]
    \textbf{(NO)} & \textbf{(DTC)}
  \end{tabular}
  \caption{Different structures of resolution proofs of unsatisfiability for
    \tonetwo-formulas, using NO (left) and DTC (right).
    \label{fig:no_dtc_proofs}}
\end{figure}

With DTC, resolution proofs are quite different from those obtained
with NO. There is no
\tonetwo-lemma anymore, because the two $\T_i$-solvers don't communicate
directly. Instead, the proofs contain both $\T_1$-lemmas and $\T_2$-lemmas
(Figure~\ref{fig:no_dtc_proofs} right), and -- importantly -- \emph{they
  contain also interface equalities}. 
(Notice that $\T_i$-lemmas derive either from $\T_i$-conflicts and
  from $\T_i$-propagation steps.)
In this case, the combination of
theories is encoded directly in the proofs (thanks to the presence of
interface equalities), and not ``hidden'' in the \tonetwo-lemmas as with
NO. This observation is at the heart of our DTC-based interpolant combination
method. 

\begin{example}\label{ex:no_dtc_proof}
  Consider the following formula $\phi$: 
  \begin{displaymath}
    \begin{split}
      \phi \defas ~&(a_1=f(a_2)) \wedge (b_1=f(b_2)) \wedge \\
      &(y-a_2=1) \wedge (y-b_2=1) \wedge (a_1+y=0) \wedge (b_1+y=1).
    \end{split}
  \end{displaymath}  
  $\phi$ is expressed over the combined theory $\euf\cup\laR$: the first two
  atoms belong to $\euf$, while the last four belong to $\laR$.  

  Using the NO combination method, $\phi$ can be proved unsatisfiable as follows:
  \begin{enumerate}
  \item From the conjunction $(y-a_2=1) \wedge (y-b_2=1)$, 
    the $\laR$-solver deduces the interface equality $(a_2=b_2)$, 
    which is sent to the $\euf$-solver;

  \item From $(a_2=b_2)$ and the conjunction 
    $(a_1=f(a_2)) \wedge (b_1=f(b_2))$ 
    the $\euf$-solver deduces the interface equality $(a_1=b_1)$, 
    which is sent to the $\laR$-solver;

  \item Together with the conjunction $(a_1+y=0) \wedge (b_1+y=1)$, $(a_1=b_1)$
    causes an inconsistency in the $\laR$-solver;

  \item The $\euf\cup\laR$ conflict-set generated is 
    $\{(y-a_2=1), (y-b_2=1), (a_1=f(a_2)), (b_1=f(b_2)), (a_1+y=0), (b_1+y=1)\}$,
    corresponding to the $\euf\cup\laR$-lemma 
    $C \defas \neg(y-a_2=1) \lor \neg(y-b_2=1) \lor \neg(a_1=f(a_2)) \lor \neg(b_1=f(b_2)) \lor \neg(a_1+y=0) \lor \neg(b_1+y=1)$.
  \end{enumerate}
  
  The corresponding NO proof of unsatisfiability for $\phi$ is thus:
  \begin{small2}
    \begin{displaymath}
      \newcommand{\pdots}{\cdots\phantom{X}}
      \infer{\bot}{
        \infer{\pdots}{
          \infer{\pdots}{
            \infer{\pdots}{
              \infer{\pdots}{
                \infer{\pdots}{
                  C & (b_1+y=1)
                } & (a_1+y=0)
              } & (y-b_2=1)
            } & (y-a_2=1)
          } & (b_1=f(b_2))
        } & (a_1=f(a_2))
      }
    \end{displaymath}
  \end{small2}

With DTC, the Boolean search space is augmented with the set of all possible
interface equalities 
$Eq \defas \{(a_1=a_2), (a_1=b_1), (a_1=b_2), (a_2=b_1), (a_2=b_2), (b_1=b_2)\}$,
so that the DPLL engine can branch on them.
If we suppose that the negative branch is explored first
(and we assume for simplicity that the $\T$-solvers do not perform deductions), 
using the DTC combination method $\phi$ can be proved unsatisfiable as follows:
\begin{enumerate}
\item Assigning $(a_2=b_2)$ to false causes an inconsistency in the $\laR$-solver, 
  which generates the $\laR$-lemma 
  $C_1 \defas \neg (y-a_2=1) \lor \neg (y-b_2=1) \lor (a_2=b_2)$. 
  $C_1$ is used by the DPLL engine to backjump and unit-propagate $(a_2=b_2)$;

\item After such propagation, assigning $(a_1=b_1)$ to false causes an
  inconsistency in the $\euf$-solver, 
  which generates the $\euf$-lemma
  $C_2 \defas \neg(a_1=f(a_2)) \lor \neg(b_1=f(b_2)) \lor \neg(a_2=b_2) \lor (a_1=b_1)$.
  $C_2$ is used by the DPLL engine to backjump and unit-propagate $(a_1=b_1)$;

\item This propagation causes an inconsistency in the $\laR$-solver,
  which generates the $\laR$-lemma
  $C_3 \defas \neg(y-a_2=1) \lor \neg(y-b_2=1) \lor \neg(a_1=b_1)$;

\item 
  After learning $C_3$, the DPLL engine detects the unsatisfiability of $\phi$.
\end{enumerate}

The corresponding DTC proof of unsatisfiability for $\phi$ is thus:
  \begin{small2}
    \begin{displaymath}
      \newcommand{\ca}{C_1}
      \newcommand{\cb}{(y-a_2=1)}
      \newcommand{\cd}{(y-b_2=1)}
      \newcommand{\ce}{C_2}
      \newcommand{\cg}{(b_1=f(b_2))}
      \newcommand{\ch}{C_3}
      \newcommand{\ck}{(b_1+y=1)}
      \newcommand{\cm}{(a_1+y=0)}
      \newcommand{\cn}{(a_1=f(a_2))}
      \newcommand{\pdots}{\cdots\phantom{X}}
      \infer{\bot}{
        \infer{\pdots}{
          \infer{\pdots}{
            \infer{\pdots}{
              \infer{\pdots}{
                \infer{\pdots}{
                  \infer{\pdots}{
                    \infer{\pdots}{
                      \ca & \cb
                    } & \cd
                  } & \ce
                } & \cg
              } & \ch
            }  & \ck
          } & \cm
        } & \cn
      }
    \end{displaymath}
  \end{small2}
\end{example}

{
An important remark is in order. 
It is relatively easy
to implement DTC in such a way that, if both $\T_1$ and $\T_2$ are
convex, then all $\T$-lemmas generated contain at most one positive
interface equality.  This is due to the fact that for convex theories
\T{} it is possible to implement efficient \Tsolvers which generates
conflict sets containing at most one negated equality between
variables \cite{jar_sat05}.
\footnote{We recall that, if \T is convex, then 
$\abmixmu\wedge \bigwedge_i\neg l_i\Tmodels\bot$ iff
$\abmixmu\wedge \neg l_i\Tmodels\bot$ for some $i$, where the $l_i$'s are positive literals.
}  
(E.g., this is true for all the $\T_i$-solvers on convex theories 
implemented in \mathsat.)
Thus, since we restrict to convex theories, in the rest of this paper we 
  can assume w.l.o.g. that every \Tlemma occurring as leaf in a
  resolution proof \PP of 
unsatisfiability deriving from DTC contains at most one 
positive interface equality.
}

\subsubsection{{Interpolation with Nelson-Oppen}}
\label{sec:dtc_background_intno}

The work in \cite{musuvathi_interpolation} gives a method for
generating an interpolant for a pair $(A, B)$ of \tonetwo-formulas
s.t.  $A\wedge B \models_{\tonetwo}\bot$ by means of the NO schema.
As in \cite{musuvathi_interpolation}, we assume that $A$ and $B$ have
been purified using disjoint sets of auxiliary variables.  We recall
from \cite{musuvathi_interpolation} a couple of definitions.
 
\begin{defn}[$AB$-mixed equality]
An equality between variables  $(a=b)$ 
is an $AB$-mixed equality iff $a \not\preceq B$ and $b
    \not\preceq A$ (or vice versa). 
\end{defn}

\begin{defn}[Equality-interpolating theory]\label{def:equality_interpolating}
A theory
    $\T$ is said to be equality-interpolating iff, for all $A$ and $B$ in $\T$
    s.t. $A\wedge B \models_\T (a=b)$ and for all $AB$-mixed equalities
    $(a=b)$,  
    there exists a term $t$ such that $A \wedge B
    \models_\T (a=t) \wedge (t=b)$ and $t \preceq A$ and $t \preceq B$.
\end{defn}

\noindent
The work in~\cite{musuvathi_interpolation} describes procedures for
computing the term $t$ from an $AB$-mixed interface equality $(a=b)$
for some convex theories of interest, including \euf{}, \larat{}, and
the theory of lists.

Notationally, with the letters $x$, $x_i$, $y$, $y_i$, $z$ we 
denote generic variables, whilst with the letters 
$a$, $a_i$, and
$b$, $b_i$ we denote variables s.t. 
$a_i \not\preceq B$ and $b_i
    \not\preceq A$;
hence, 
with the letters $e_i$ we denote generic 
$AB$-mixed interface equalities in the
form $(a_i=b_i)$;
 with the letters $\noabmixmu$, $\noabmixmu_i$ we denote
conjunctions of literals where no  $AB$-mixed interface equality
occurs, and with the letters $\abmixmu$, $\abmixmu_i$ we denote conjunctions of literals
where  $AB$-mixed interface equalities may occur. 
If $\abmixmu_i$  (resp $\noabmixmu_i$) is $\bigwedge_i l_i$, 
we write $\neg\abmixmu_i$ (resp. $\neg\noabmixmu_i$) for the clause $\bigvee_i
\neg l_i$. 

Let $A\land B$  be a \tonetwo-inconsistent conjunction
of \tonetwo-literals,
such that $A \defas A_1 \land A_2$ and $B \defas B_1 \land B_2$
where each $A_i$ and $B_i$ is $\T_i$-pure.
The NO-based method of \cite{musuvathi_interpolation} 
computes an interpolant for $(A, B)$ by combining $\T_i$-specific
interpolants for subsets of $A$, $B$ and the set of entailed interface equalities $\{e_j\}_j$
that are exchanged between the $\T_i$-solvers for deciding the unsatisfiability
of $A\land B$.
In particular, 
let $Eq \defas \{e_j\}_j$ be the set of entailed interface equalities.
Due to the fact that both $\T_1$ and $\T_2$ are equality-interpolating,
it is possible to assume w.l.o.g. that $Eq$ does not contain
$AB$-mixed equalities, 
{because instead of deducing an $AB$-mixed interface equality $(a=b)$,
a \Tsolver can always deduce the two corresponding equalities $(a=t)\wedge(t=b)$.}
{(Notice that the other \Tsolver treats the term $t$ as if it were a
   variable \cite{musuvathi_interpolation}.) 
}
{Let $A' \defas A \cup (Eq \downarrow A)$ and $B' \defas B \cup (Eq \downarrow B)$.}
Then, $\T_i$-specific partial interpolants are combined 
according to the following inductive definition:
\begin{equation}
  \label{eq:no_interpolant}
  I_{A,B}(e) \defas \left\{
    \begin{array}{ll}
      \bot & \text{if $e \in A$} \\
      \top & \text{if $e \in B$} \\
      (I^i_{A',B'}(e) \lor \bigvee_{e_a\in A'}I_{A,B}(e_a)) \land \bigwedge_{e_b\in B'} I_{A,B}(e_b) & \text{otherwise},
    \end{array}
  \right.
\end{equation}
where $e$ is either an entailed interface equality or $\bot$, and 
$I^i_{A',B'}(e)$ is a $\T_i$-interpolant 
for $(A' \cup \neg e, B')$ if $e \preceq A$,
and for $(A', B' \cup \neg e)$ otherwise (if $e \preceq B$).
{The computed interpolant for $(A, B)$ is then $I_{A,B}(\bot)$}.
{We refer the reader to \cite{musuvathi_interpolation} for
  more details.}

\subsection{From DTC solving to DTC Interpolation}
\label{sec:dtc_interpolation}

We now discuss how to extend the DTC method to interpolation. As with
\cite{musuvathi_interpolation}, we can handle the case that $\T_1$ and
$\T_2$ are convex and equality-interpolating.
The approach to generating interpolants for combined theories starts
from the proof generated by DTC.
Let $Eq$ be the set of all interface equalities occurring in a DTC
refutation proof for a \tonetwo-unsatisfiable formula $\phi\defas
A\wedge B$. 

In the case $Eq$ does not contain $AB$-mixed equalities, that
is, $Eq$ can be partitioned into two sets $(Eq \setminus B) \defas \{ (x=y) |
(x=y) \preceq A \text{~and~} (x=y) \not\preceq B \}$ and $(Eq \downarrow B) \defas
\{(x=y) | (x=y) \preceq B \}$,
\emph{no
interpolant-combination method is needed}: the combination is already
encoded in the proof of unsatisfiability, and a direct application of
Algorithm 1 to such proof yields an interpolant for the combined
theory \tonetwo.
{Notice that this fact holds} despite the fact that the
interface equalities in $Eq$ 
occur neither in $A$ nor in $B$, but might be introduced in the
resolution proof \PP by \T-lemmas. {In fact, as
observed in \cite{mcmillan_interpolating_prover},} as long as for an
atom $p$ either $p \preceq A$ or $p \preceq B$ holds, it is possible
to consider it part of $A$ (resp. of $B$) simply by assuming the
tautology clause $p \vee \neg p$ to be part of $A$ (resp. of
$B$). Therefore, we can treat the interface equalities in $(Eq
\setminus B)$ as if they appeared in $A$, and those in $(Eq \downarrow
B)$ as if they appeared in $B$.

{
When $Eq$ contains $AB$-mixed equalities, instead,  a proof-rewriting step is
performed in order to obtain a proof which is free from $AB$-mixed equalities,
that is amenable for interpolation as described above.
The idea is similar to that used in \cite{musuvathi_interpolation} in
the case of NO: using the fact that $\T_1$ and $T_2$ are
equality-interpolating, we reduce this case to the previous one by
``splitting'' every $AB$-mixed interface equality $(a_i=b_i)$ into the
conjunction of two parts $(a_i=t_i)\wedge(t_i=b_i)$, such that
$(a_i=t_i) \preceq 
A$ and $(t_i=b_i) \preceq B$. The main difference is that we do this
\emph{a posteriori}, after the construction of the {resolution} proof
of unsatisfiability \PP.
In order to do this, we traverse \PP and split each $AB$-mixed
equality, performing also the necessary manipulations to ensure that
the result is still a resolution proof of unsatisfiability.
}

{
We describe this process in two steps. 
In \sref{sec:dtc_ielocal} we introduce a particular kind of
resolution proofs of unsatisfiability, called \ielocal, and show how to 
 eliminate $AB$-mixed interface
equalities from  \ielocal proofs;
in \sref{sec:dtc_ielocal_generation}
 we show how to implement a variant of DTC so that to generate
\ielocal proofs. 
}

\subsubsection{{Eliminating $AB$-mixed equalities by exploiting ie-locality}}
\label{sec:dtc_ielocal}
\label{sec:dtc_mixed_general}
\label{sec:dtc_mixed_analysis}

\begin{defn}[\ielocal proof]
\label{def:ielocal}
  A resolution proof of unsatisfiability \PP is local with respect to interface
  equalities (\ielocal) iff the interface equalities occur only in subproofs
  \ieproof{i} of \PP, such that within each \ieproof{i}:
    \begin{renumerate}
    \item all  leaves  are also
    \Tlemma leaves of \PP; 
    \item all the pivots are interface
    equalities;
    \item the root contains no interface equality; 
    \item every right premise of an inner node is a leaf \Tlemma
    containing exactly one positive interface equality.
\footnote{\label{foot:leftright}
We have adopted the graphical convention that 
at each resolution step in a \ieproof{i} subproof,
if $(a_i=b_i)$ is the pivot, then
the premises containing $\neg(a_i=b_i)$ and $(a_i=b_i)$ 
are the left and right premises respectively.%
}

    \end{renumerate}
\end{defn}

\noindent
As a consequence of this definition, we also have that, within each
\ieproof{i} in \PP:
\begin{renumerate}
\addtocounter{cntrenumerate}{4}
    \item all nodes  are \tonetwo-valid; {({\em Proof sketch}: they result from
    Boolean resolution steps from \Tone-valid and \Ttwo-valid clauses,  
 hence they are \tonetwo-valid.)}
    \item the only leaf \Tlemma which is a left premise contains no positive
    interface equality.~%
({{\em Proof sketch:} we notice that, 
in a resolution step $\frac{C_1\
  C_2}{C_3}$, if $C_3$ contains no positive interface equality, at
least one between $C_1$ and $C_2$ contains no positive interface
equality; since by (iv) the right premise contains one positive
interface equality, only the left premise contains no positive interface
equality. Thus the leftmost leaf \T-lemma of \ieproof{i}
contains no positive interface equality.})

\item 
{if an interface equality $e_j$ occurs negatively in some \Tlemma $C_j$,
then $e_j$ occurs positively in a leaf \Tlemma $C_k$ which is the right premise of 
  a resolution step whose left premise derives from $C_j$ and other \Tlemmas.
}
  (\emph{Proof sketch:} suppose that $\neg e_j$ occurs in $C_j$ 
  but $e_j$ does not occur in any such $C_k$. 
  Then $e_j$ can not be a pivot, 
  hence $\neg e_j$ occurs in the root of $\PPie_i$, thus violating (iii).)
\end{renumerate}

Intuitively, in \ielocal proofs of unsatisfiability all the reasoning on
interface equalities is circumscribed within \ieproof{i} subproofs,
which are {linear sub-proofs} involving only \Tlemmas as leaves, 
starting from the one  containing no positive interface equality, 
each time eliminating one negative interface equality by resolving it
against the only positive one occurring in another leaf \Tlemma.

\begin{example}\label{ex:ielocal_proof}
  Consider the $\euf\cup\larat$ formula $\phi$ of Example~\ref{ex:no_dtc_proof},
  and the \Tlemmas $C_1, C_2$ and $C_3$ introduced by DTC 
  to prove its unsatisfiability.
  The proof \PP of Example~\ref{ex:no_dtc_proof} is not \ielocal,
  because resolution steps involving interface equalities are interleaved with
  resolution steps involving other atoms. 
  The following proof $\PP'$, instead, is \ielocal: 
  all the interface equalities are used as pivots 
  in the $\Pi^\ie$ subproof:
  \begin{small2}
    \begin{displaymath}
        \newcommand{\ca}{C_3}
        \newcommand{\cb}{C_2}
        \newcommand{\cc}{\ldots\phantom{X}}
        \newcommand{\cd}{C_1}
        \newcommand{\cq}{\ldots\phantom{X}}
        \newcommand{\ce}{(a_2+z=1)}
        \newcommand{\cf}{\ldots\phantom{X}}
        \newcommand{\cg}{(a_1+z=0)}
        \newcommand{\ch}{(z-x_2=1)}
        \newcommand{\ci}{\ldots\phantom{X}}
        \newcommand{\cj}{\ldots\phantom{X}}
        \newcommand{\ck}{(a_1=f(x_1))}
        \newcommand{\cl}{\ldots\phantom{X}}
        \newcommand{\cm}{(a_2=f(x_2))}
        \newcommand{\cn}{(z-x_1=1)}
        \newcommand{\co}{\ldots\phantom{X}}
        \newcommand{\cp}{\bot}
        \begin{array}{l}
      \infer{\cp}{
        \infer{\co}{
          \infer{\cl}{
            \infer{\cj}{
              \infer{\ci}{
                \infer{\cf}{
                  \fbox{
                  \infer[\text{[pivot $(a_2=b_2)$]}]{\cq}{
                    \infer[\text{[pivot $(a_1=b_1)$]}]{\cc}{
                      \ca & \cb
                    } & \cd
                  }}^{\Pi^\ie} & \ce
                } & \hspace{-1cm} \cg
              } & \hspace{-1cm}\ch
            }  & \hspace{-1cm} \ck
          } & \hspace{-1cm} \cm
        } & \hspace{-1cm} \cn
      } \\
          C_1 \defas (\underline{a_2 = b_2}) \vee \neg(y - a_2 = 1) \vee \neg(y - b_2 = 1) \\
          C_2 \defas (\underline{a_1 = b_1}) \vee \neg(b_1 = f(b_2)) \vee
      \neg(a_1 = f(a_2)) \vee \neg(\underline{a_2 = b_2}) \\
          C_3 \defas \neg(a_1 + y = 0) \vee \neg(b_1 + y = 1) \vee
          \neg(\underline{a_1 = b_1}).
      \end{array}
    \end{displaymath}
  \end{small2}
\end{example}

{%
If \PP is an \ielocal proof containing $AB$-mixed interface equalities,
 then it is possible to eliminate all of them
 from \PP
by applying Algorithm 2 to every \ieproof{i} subproof of \PP.~%
In a nutshell, each  \ieproof{i} subproof is explored bottom-up,
starting from the right premise of the root,
 each time expanding the rightmost side \T-lemma in the form 
$C_i\defas (a_i=b_i)\vee\neg\noabmixmu_i$ 
{s.t. $(a_i=b_i)$ is $AB$-mixed}
into the (implicit) conjunction of two novel   \T-lemmas 
$C'_i\defas (a_i=t_i)\vee\neg\noabmixmu_i$ and
$C''_i\defas (t_i=b_i)\vee\neg\noabmixmu_i$ 
(step (4)), where $t_i$ is the $AB$-pure term 
computed from $C_i$ as described in \sref{sec:dtc_background_intno}.
Then the resolution step against $C_i$ is substituted with 
the concatenation of two resolution steps against $C'_i$ and $C''_i$ (step (5)) 
and then the substitution $\neg(a_i=b_i) \longmapsto
\neg(a_i=t_i)\vee\neg(t_i=b_i)$ is propagated bottom-up along the left
subproof $\Pi$. 
Notice that $C'_i$ and $C''_i$ are still \Ti-valid because \Ti is
Equality-interpolating and $\noabmixmu_i$ does not contain other $AB$-mixed 
interfaced equalities. 
}

\begin{figure}[t]
  \centering
  \begin{minipage}[h]{\linewidth}
{\fontsize{10}{11}\selectfont
    \vspace{3mm} \hrule \vspace{2mm} \textbf{Algorithm 2: Rewriting of \PPie
      subproofs} \vspace{1mm} \hrule \vspace{2mm}
    \begin{enumerate}
    \item Let $\sigma$ be a mapping from negative $AB$-mixed interface
      equalities to a disjunction of two negative interface
      equalities, such that 
      \mbox{$\sigma[\neg(a_i=b_i)] \mapsto \neg(a_i=t_i) \lor \neg(t_i=b_i)$}
      and $t_i$ is an $AB$-pure term as described in \sref{sec:dtc_background_intno}. 
      Initially, $\sigma$ is empty.

    \item Let $C_i \defas (a_i=b_i) \vee \neg \abmixmu_i$ be the right premise
      \Tlemma of the root of the \PPie subproof.     

    \item Replace each $\neg(a_j=b_j)$ in $C_i$ with
      $\sigma[\neg(a_j=b_j)]$, to obtain 
      \mbox{$C^*_i \defas (a_i=b_i) \lor \neg\noabmixmu_i$}.
      {%
        If $(a_i=b_i)$ is not $AB$-mixed, then let $\Pi$ be the subproof
        rooted in the left premise, and go to step (7).
      }

    \item Split $C^*_i$ into $C'_i \defas (a_i=t_i) \lor
      \neg\noabmixmu_i$ and $C''_i \defas (t_i=b_i) \lor \neg\noabmixmu_i$.

    \item Rewrite the subproof 
      \begin{displaymath}
\vcenter{
        \infer{\neg\abmixmu_k \lor \neg\abmixmu_i}{
          \infer{\neg (a_i=b_i) \lor \neg \abmixmu_k}{\vdots} & C_i
        }}
\text{~~into~~}
\vcenter{
        \infer{\neg\noabmixmu_k \lor \neg\noabmixmu_i}{
          \infer{\neg(t_i=b_i)\lor \neg\noabmixmu_k \lor \neg\noabmixmu_i}{
            \fbox{\infer{\neg(a_i=t_i) \lor \neg(t_i=b_i) \lor \neg\abmixmu_k}{\vdots}}^\Pi
            & C'_i
          } & C''_i
        }
}
      \end{displaymath}
      where $\neg\noabmixmu_k$ is obtained by $\neg\abmixmu_k$ 
      by substituting each negative $AB$-mixed interface equality $\neg(a_j=b_j)$
      with $\sigma[\neg(a_j=b_j)]$.

    \item Update $\sigma$ by setting $\sigma[\neg(a_i=b_i)]$ to $\neg(a_i=t_i)
      \lor \neg(t_i=b_i)$.

    \item If $\Pi$ is of the form $\dfrac{~\vdots \quad C_j}{\cdots}$,
      set $C_i$ to $C_j$ and go to step (3).

    \item Otherwise, 
      $\Pi$ is the leaf $\neg(a_i=t_i) \lor \neg(t_i=b_i) \lor \neg\abmixmu_k$. 
      In this case, replace each $\neg(a_j=b_j)$ in $\neg\abmixmu_k$ with 
      $\sigma[\neg(a_j=b_j)]$, and then exit.
  \end{enumerate}
  \vspace{1mm} \hrule \vspace{3mm}
\normalsize} %
\end{minipage}
\end{figure}

\begin{example}\label{ex:split_ie}
  Consider the formula $\phi$ of Example~\ref{ex:no_dtc_proof} and its
  \ielocal proof of unsatisfiability of Example~\ref{ex:ielocal_proof}. 
  Suppose that $\phi$ is partitioned  as follows:
  \begin{displaymath}
    \begin{split}
      \phi &\defas A \wedge B \\
      A &\defas (a_1 = f(a_2)) \wedge (y - a_2 = 1) \wedge (a_1 + y = 0) \\
      B &\defas (b_1 = f(b_2)) \wedge (y - b_2 = 1) \wedge (b_1 + y = 1)
    \end{split}
  \end{displaymath}
  In this case, both interface equalities $(a_1 = b_1)$ and $(a_2 = b_2)$ are
  $AB$-mixed. Consider the \PPie subproof of Example~\ref{ex:ielocal_proof}:
  \begin{small2}
    \begin{displaymath}
      \begin{array}{l}
        \begin{array}{lr}
          \begin{array}[b]{l}
          C_1 \defas (\underline{a_2 = b_2}) \vee \neg(y - a_2 = 1) \vee \neg(y - b_2 = 1) \\
          C_2 \defas (\underline{a_1 = b_1}) \vee \neg(b_1 = f(b_2)) \vee
      \neg(a_1 = f(a_2)) \vee \neg(\underline{a_2 = b_2}) \\
          C_3 \defas \neg(a_1 + y = 0) \vee \neg(b_1 + y = 1) \vee
          \neg(\underline{a_1 = b_1})
          \end{array} & \hspace{2cm}
          \fbox{
      \infer{\Theta_2}{
        \infer{\Theta_1}{
          C_3 & C_2
        } & C_1
      }}^{\Pi^\ie}
       \end{array}
       \\
      \Theta_1 \defas \neg(a_1 + y = 0) \vee \neg(b_1 + y = 1) \vee \neg(b_1 = f(b_2)) \vee \neg(a_1 = f(a_2)) \vee \neg(a_2 = b_2) \\
      \Theta_2 \defas  \neg(a_1 + y = 0) \vee \neg(b_1 + y = 1)\vee \neg(b_1 = f(b_2)) \vee \neg(a_1 = f(a_2)) \vee \neg(y - a_2 = 1) \vee \neg(y - b_2 = 1)
      \end{array}
    \end{displaymath}
  \end{small2}
  The first \T-lemma processed by Algorithm 2 is $C_1$. Using the technique of
  \cite{musuvathi_interpolation}, $(a_2=b_2)$ is split into $(a_2=y-1) \wedge
  (y-1=b_2)$ (step (4)), thus obtaining $C_1'$, $C_1''$ and the new proof 
  (in step (5)):
  \begin{small2}
    \begin{displaymath}
        \begin{array}{lr}
          \begin{array}[b]{rl}
            C_1' \defas & (a_2 = y-1) \vee \neg(y - a_2 = 1) \vee \neg(y - b_2 = 1) \\
            C_1'' \defas & (y-1 = b_2) \vee \neg(y - a_2 = 1) \vee \neg(y - b_2 = 1) \\
            \Theta_2' \defas & \neg(y-1=b_2) \vee \neg(a_1 + y = 0) \vee \neg(b_1 + y = 1)\vee \neg(b_1 = f(b_2)) \vee \\
            & \neg(a_1 = f(a_2)) \vee\neg(y - a_2 = 1) \vee \neg(y - b_2 = 1)
          \end{array} & \hspace{1cm}
          \fbox{
      \infer{\Theta_2}{
        \infer{\Theta_2'}{
          \infer{\Theta_1}{
            C_3 & C_2
          } & C_1'
        } & C_1''
      }}
       \end{array}
    \end{displaymath}
  \end{small2}
  Then, $\sigma[\neg(a_2=b_2)]$ is set to $\neg(a_2=y-1)\lor\neg(y-1=b_2)$
  (step (6)),
  and a new iteration of the loop (3)-(7) is performed, this time 
  processing $C_2$.
  First, $\neg(a_2=b_2)$ is replaced by $\neg(a_2=y-1)\vee\neg(y-1=b_2)$
  (step (3)). 
  Then, $(a_1=b_1)$ can be split into $(a_1=f(y-1)) \wedge (f(y-1)=b_1)$
  (step (4)). After the rewriting of step (5), the proof is:
  \begin{small2}
    \begin{displaymath}
        \begin{array}{lr}
          \begin{array}[c]{rl}
            C_2' \defas & (a_1 = f(y-1)) \vee \neg(b_1 = f(b_2)) \vee \neg(a_1 =
            f(a_2)) \vee \neg(a_2 = y-1) \vee \\ & \neg(y-1=b_2)\\
            C_2'' \defas & (f(y-1) = b_1) \vee \neg(b_1 = f(b_2)) \vee \neg(a_1 = f(a_2)) \vee \neg(a_2 = y-1) \vee \\ & \neg(y-1=b_2) \\
            \Theta_1' \defas & \neg(a_1 + y = 0) \vee \neg(b_1 + y = 1) \vee \neg(b_1 = f(b_2)) \vee \neg(a_1 = f(a_2)) \vee \\ & \neg(a_2 = y-1) \vee \neg(y-1=b_2) \\
            \Theta_1'' \defas & \neg(a_1=f(y-1)) \vee \neg(a_1 + y = 0) \vee \neg(b_1 + y = 1) \vee \neg(b_1 = f(b_2)) \vee \\ & \neg(a_1 = f(a_2)) \vee \neg(a_2 = b_2) \\
          \end{array} & \hspace{5mm}
          \fbox{
      \infer{\Theta_2}{
        \infer{\Theta_2'}{
          \infer{\Theta_1'}{
            \infer{\Theta_1''}{
              C_3 & C_2'
            } & C_2''
          } & C_1'
        } & C_1''
      }}
       \end{array}
    \end{displaymath}
  \end{small2}
  Finally, $C_3$ is processed in step (8), $\neg(a_1=b_1)$ gets replaced with
  $\neg(a_1=f(y-1))\lor\neg(f(y-1)=b_1)$, 
  and the following final proof $\Pi'^\ie$ is generated:
  \begin{displaymath}
      \infer{\Theta_2}{
        \infer{\Theta_2'}{
          \infer{\Theta_1'}{
            \infer{\Theta_1''}{
              C_3'
              & C_2'
            } & C_2''
          } & C_1'
        } & C_1''
      }    
  \end{displaymath}
  such that $C_3' \defas C_3[\neg(a_1=b_1)\mapsto \neg(a_1=f(y-1))\lor\neg(f(y-1)=b_1)]$.
\end{example}

The following theorem states that Algorithm 2 is correct. 

\begin{theorem}
  Let \PP be a \PPie subproof, 
  and let $\PP'$ be the result of applying Algorithm 2 to \PP.
  Then:
  \begin{enumerate}[(a)]
  \item $\PP'$ does not contain any $AB$-mixed interface equality; and
  \item $\PP'$ is a valid subproof with the same root as \PP.
  \end{enumerate}
\end{theorem} 
\begin{proof} \
  \begin{enumerate}[(a)]
  \item 
    Consider the \Tlemma $C_i$ of Step (3). 
    By item (vii) of Definition~\ref{def:ielocal}, 
    all negative interface equalities occurring in $C_i$
    occur positively in leaf \Tlemmas that are closer to the root of \PP.
    For the same reason, 
    the first $\T$-lemma $C_i$ analyzed in step (2) contains no
    {negative} $AB$-mixed interface equalities.
    Therefore, it follows by induction that all negative $AB$-mixed interface 
    equalities in $C_i$ must have been split in Step (4) 
    of a previous iteration of the loop (3)-(7) of Algorithm 2, 
    and thus they occur in $\sigma$.
    The same argument can be used to show also that at steps (5) and (8)
    every negative $AB$-mixed interface equality in $\neg\abmixmu_k$ occurs in $\sigma$.\\

  \item 
    We show that:
    \begin{enumerate}[(i)]
    \item Every substep $\dfrac{\Theta'\quad \Theta''}{\Theta'''}$ of $\PP'$
      is a valid resolution step;
    \item every leaf of $\PP'$ is a \Tlemma; and
    \item the root of $\PP'$ is the same as that of $\PP$.
    \end{enumerate}
    \smallskip
 
    \begin{enumerate}[(i)]
    \item The only problematic case is the resolution step
      \begin{displaymath}
        \infer{\neg(t_i=b_i)\lor \neg\noabmixmu_k \lor \neg\noabmixmu_i}{
          \neg(a_i=t_i) \lor \neg(t_i=b_i) \lor \neg\abmixmu_k & C'_i
        }
      \end{displaymath}
      introduced in step (5) of Algorithm 2.  In this case, we have to show
      that at the end of the algorithm, all the negative $AB$-mixed interface
      equalities in $\neg\abmixmu_k$ have been replaced such that the result
      is identical to $\neg\noabmixmu_k$.  We already know that all negative
      $AB$-mixed equalities in $\neg\abmixmu_k$ occur in $\sigma$, thus we
      only have to show that $\sigma[\neg e_j]$ cannot change between the
      time when $\neg e_j$ was rewritten to obtain $\neg \noabmixmu_k$ and the
      time in which it is rewritten in $\neg\abmixmu_k$.  The negative
      equality $\neg e_j$ is replaced in $\neg\abmixmu_k$ at the next
      iteration of the algorithm (in step (5) for inner nodes, and in step (8)
      for the final leaf).  In the meantime, the only update to $\sigma$ is
      performed in step (6), but it involves the negative equality
      $\neg(a_i=b_i)$, which does not occur in $\neg\abmixmu_k$.

    \item 
      Let $C_i$ be a \Tlemma in \PP.
      First, we observe that if $C_i \equiv \neg(a_i=b_i)\lor \neg\abmixmu_i$,
      then for any $t_i$ also the clause $C_i^* \defas \neg(a_i=t_i) \lor \neg(t_i=b_i) \lor \neg\abmixmu_i$ is a \Tlemma, 
      since $(a_i=t_i) \land (t_i=b_i) \models_\T (a_i=b_i)$ by transitivity.
      Therefore, it follows by induction on the number of substitutions that 
      the clauses obtained in steps (3) and (8) of Algorithm 2 are still \Tlemmas. 
      Finally, since we are considering equality-interpolating theories, 
      after step (4) of Algorithm 2 both $C_i'$ and $C_i''$ are \Tlemmas.

    \item Since the root of \PP does not contain any interface equality
      (item (iii) of Definition~\ref{def:ielocal}), in step (5)
      $\neg\noabmixmu_i \equiv \neg\abmixmu_i$ and $\neg\noabmixmu_k \equiv
      \neg\abmixmu_k$, and therefore the root does not change.  
   \end{enumerate}
\hspace{.9\textwidth} \hfill  \qed
  \end{enumerate}

\end{proof}

Clearly, Algorithm 2 operates in linear time on the number of \T-lemmas, 
and thus of $AB$-mixed interface equalities. 
Moreover, every time an interface equality is split, 
only two new nodes are added to the proof 
(a right leaf and an inner node), 
and therefore the size of $\PP'$ is linear in that of \PP.

{
The advantage of having \ielocal proofs is that they ease significantly the
process of eliminating $AB$-mixed interface equalities.  First, since
all the reasoning involving interface equalities is confined in \PPie
subproofs, only such subproofs -- which typically constitute only a
small fraction of the whole proof -- need to be traversed and
manipulated.  
Second, the simple structure of \PPie subproofs allows
for an efficient application of the rewriting process
of steps (5) and (3), preventing any explosion in size of the proof.
In fact, e.g., if in step (5) the right premise of the last
  step were instead the root of some subproof $\Pi_i$ with $C_i$ as a leaf,
then two copies of  $\Pi'_i$ and $\Pi''_i$ would be produced,
in which each instance of $(a_i=b_i)$ bust be replaced with 
$(a_i=t_i)$ and $(t_i=b_i)$ respectively. 
}  

\subsubsection{Generating \ielocal proofs in DTC}
\label{sec:dtc_ielocal_generation}
\newcommand{\uprope}[2]{\ensuremath{e_{#1}^{#2}\xspace}}
\newcommand{\antcl}[2]{\ensuremath{C_{#1}^{#2}\xspace}}
\newcommand{\antmu}[2]{\ensuremath{\abmixmu_{#1}^{#2}\xspace}}
\newcommand{\dpllmain}{DPLL-1\xspace}
\newcommand{\dpllaux}{DPLL-2\xspace}
\begin{figure}[t]
\scalebox{0.7}{\input{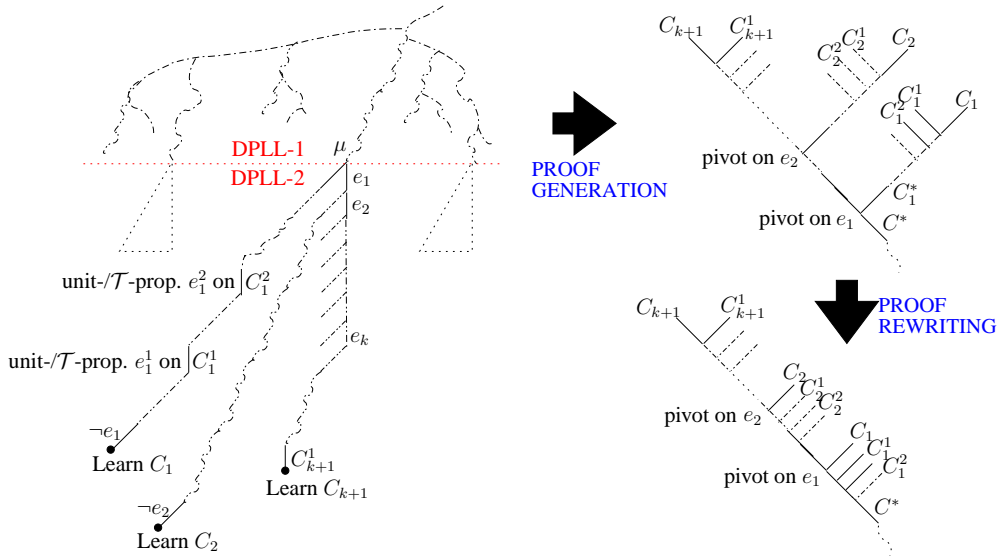}}
\label{fig:DTC_proof}
\caption{\label{fig:rewriting_ielocal}
{Simple strategy for generating \ielocal proofs. }
Left: DTC search;   
top-right: corresponding (sub)proof;
bottom-right:  $\PPie$ (sub)proof after rewriting.
}
\end{figure}
{
In this section we show how to implement a variant of DTC so that to generate
\ielocal proofs of unsatisfiability. 
}
For the sake of simplicity, we describe first a simplified algorithm
which makes use of two distinct DPLL engines.
We then describe how to avoid the need of a second DPLL engine
with the use of a particular search strategy for DTC.

The simplified algorithm uses two distinct DPLL engines, 
a \emph{main} one and an \emph{auxiliary} one, 
which we shall call \dpllmain and \dpllaux respectively.
{Consider Figure~\ref{fig:rewriting_ielocal}, left.}
\dpllmain receives in input the clauses of the input problem $\phi$ 
(which we assume pure and \Tonetwo-inconsistent), 
but no interface equality, which are instead given to \dpllaux. 
\dpllmain enumerates {total} Boolean models $\abmixmu$ of $\phi$, 
and invokes the two $\T_i$-solvers separately on the subsets $\abmixmu_{\T_1}$
and $\abmixmu_{\T_2}$ of $\abmixmu$. 
If one \Tisolver reports an inconsistency, then \dpllmain{} backtracks. 
Otherwise, both $\abmixmu_{\T_i}$ are $\T_i$-consistent, 
and \dpllaux is invoked on the list of unit clauses composed of the
\tonetwo-literals in $\abmixmu$, to check its \tonetwo-consistency. 

{
\dpllaux branches only on interface equalities, 
assigning them always to false first. 
Some interface equalities $\uprope{1}{j}$, however, may be assigned to true by
unit-propagation on previously-learned clauses in the form 
$\antcl{1}{j}\defas \neg\antmu{1}{j}\vee \uprope{1}{j}$,
or by \T-propagation on deduction clauses $\antcl{1}{j}$ in the same form; 
we call $\antcl{1}{j}$  the {\em antecedent clause} of $\uprope{1}{j}$.~%
\footnote{{%
Notationally, $\uprope{i}{j}$ denotes the $j$-th most-recently
unit-propagated interface equality in the branch in which $C_i$ is learned, 
and $\antcl{i}{j}\defas \neg\antmu{i}{j}\vee \uprope{i}{j}$
 denotes the antecedent clause of $\uprope{i}{j}$.
}}
(As in \cite{AMAI08_dtc}, we assume that when a
  \T-propagation step $\antmu{i}{j}\models_{\T} \uprope{i}{j}$ occurs,
  $\antmu{i}{j}$ being a subset of the current branch, the 
  deduction clause $\antcl{i}{j}\defas \neg\antmu{i}{j}\vee \uprope{i}{j}$ is learned, 
  either temporarily or permanently; 
  if so, we can see this step as a  unit-propagation on \antcl{i}{j}.) 
When all the interface equalities have been assigned a truth value, 
the propositional model $\abmixmu' \equiv \abmixmu_{\T_1} \cup \abmixmu_{\T_2} \cup \abmixmu_\ie$ 
is checked for \Tonetwo-consistency by invoking each of the $\T_i$-solvers 
on $\abmixmu_{\T_i}\cup\abmixmu_\ie$.
\footnote{{%
In fact, it is not necessary to wait for all interface equalities to have a value
before invoking the $\T_i$-solvers. 
Rather, the standard \emph{early pruning} optimization (see \sref{sec:algorithms_smt}) can be applied.
}
}
Since $\phi$ is inconsistent, one of the two $\T_i$-solvers detects an
inconsistency (if both do, we consider only the first). 
Therefore a $\T_i$-lemma $C_1$ is generated. 
As stated at the end of \sref{sec:dtc_background_proofs}, we
  can assume w.l.o.g. that  
$C_1$ contains at most one positive interface equality $e_1$. 
(Notice also that all negative  interface equalities $\neg \uprope{1}{j}$ in $C_1$, 
if any, have been assigned by unit-propagation or \T-propagation 
on some antecedent clause $\antcl{1}{j}$.)
\dpllaux then learns $C_1$ and uses it as conflicting clause to backjump: 
starting from $C_1$, it eliminates from the clause every $\neg \uprope{1}{j}$ 
by resolving the current clause against its  antecedent clause $\antcl{1}{j}$,
until no negated equality occurs in the final clause $C_1^*$.~%
\footnote{In order to determine the order
  in which to eliminate the interface equalities, 
  the \emph{implication graph} of the auxiliary DPLL engine can be used. 
  This is a standard process in the conflict analysis 
  in modern SAT and SMT solvers (see, e.g., \cite{avg_sat07,seba_lazy_smt}).}  
}

If $C_1$ includes one positive interface equality $e_1$,
then also the final clause $C_1^*$  includes it, 
so that \dpllaux uses $C_1^*$ as a conflict clause 
to jump up to $\abmixmu$ and to unit-propagate $e_1$. 
Then \dpllaux{} starts exploring a new branch. 
This process is repeated on several branches, learning a sequence of
\Tlemmas $C_1,...,C_k$ each $C_i$ containing only one
{positive} interface equality $e_i$, 
until a branch causes the generation of a
\Tlemma $C_{k+1}$ containing no positive interface equalities.
Then $C_{k+1}$ is resolved backward against the antecedent clauses 
of its negative interface equalities, 
generating a final conflict clause $C^*$ which contains no interface equalities.

{%
Overall, \dpllaux has checked the \Tonetwo-unsatisfiability of $\abmixmu$, 
building a resolution (sub)proof $\PP^*$ whose root is
$C^*$. 
(Figure ~\ref{fig:rewriting_ielocal}, top right.)
Then the \tonetwo-lemma $C^*$ is %
passed to \dpllmain, which uses it as a blocking clause for the
assignment $\abmixmu$, it backtracks and continues the search. 
When the empty clause is obtained, 
it generates a proof of unsatisfiability in the usual way 
(see e.g. \cite{avg_sat07}). 
}

{%
Since the main solver knows nothing about interface equalities, 
they can only appear inside the proofs of the blocking
clauses generated by the auxiliary solver (like $\PP^*$). 
Each $\PP^*$  is not yet a $\PPie$ subproof, 
since it complies only with items (i), (ii) and (iii) 
of Definition~\ref{def:ielocal} but not with item (iv).
The reason for the latter fact is that $\PP^*$  contains a set of
right branches $\Pi_{C_i}$, one of each \Tlemma $C_i$ in
$\{C_{k+1},...,C_1\}$, representing the 
resolution steps to resolve away the interface equalities introduced
by unit-propagation/\T-propagation in each branch. 
Each such sub-branch $\Pi_{C_i}$, however, can be reduced to length one
by moving downwards the resolution steps with 
the antecedent clauses $\antcl{i}{1},\antcl{i}{2},...$ which 
$C_i$  encounters in the branch. (Figure~\ref{fig:rewriting_ielocal}, bottom right.)
This is done by recursively applying the following rewriting step
 to $\Pi_{C_i}$, until it reduces to the single clause $C_i$: 
\newcommand{\simpleproofl}{\infer{\neg e_i\vee\neg\abmixmu'_i  }{\vdots}} 
\newcommand{\simpleproofc}{\overbrace{\neg\antmu{i}{j} \vee \uprope{i}{j}}^{\antcl{i}{j}}}
\newcommand{\simpleproofr}{\infer{\neg\abmixmu''_i \vee \neg \uprope{i}{j} \vee e_i}{\antcl{i}{j-1}&\infer{\vdots}{\antcl{i}{1} \quad C_i}}}
\begin{equation}
\label{eq:postpone11}
\infer
{ 
 \neg\abmixmu'_i \vee \neg\antmu{i}{j} \vee \neg\abmixmu''_i
}
{ 
 \simpleproofl
 &
 \fbox{
 \infer
 {
  \neg\antmu{i}{j} \vee \neg\abmixmu''_i \vee e_i
 }
 {
  \simpleproofc
  & 
  \simpleproofr
 }
 }^{\Pi_{C_i}}
} 
\hspace{-.3cm}
\Longrightarrow
\infer
{ 
 \neg\abmixmu'_i \vee \neg\antmu{i}{j} \vee \neg\abmixmu''_i
}
{ 
 \infer{\neg\abmixmu'_i\vee\neg\abmixmu''_i\vee\neg \uprope{i}{j}}
 {
 \simpleproofl
 & 
 \fbox{\simpleproofr}^{\Pi'_{C_i}}
 } 
 & 
 \simpleproofc 
} 
\end{equation}
As a result, each  
 $\PP^*$ is transformed into a $\PPie$ subproof,
 so that the final proof is \ielocal.
}

\smallskip
In an actual implementation, there is no need of having two distinct
\dpll solvers for constructing \ielocal proofs. 
In fact, we can obtain the same result by adopting a variant 
of the DTC Strategy 1 of \cite{AMAI08_dtc}. 
We never select an interface equality for case splitting if
there is some other unassigned atom, and we always assign false to interface
equalities first. 
{%
Moreover, we ``delay'' \T-propagation of interface equalities until all
the original atoms have been assigned a truth value.
}
Finally, when splitting on interface equalities, we restrict 
both the backjumping and the learning procedures of the DPLL engine as follows.
Let $d$ be the depth in the DPLL tree at which the first interface
equality is selected for case splitting. 
If during the exploration of the current DPLL branch
we have to backjump above $d$, then we generate by resolution 
a conflict clause that does not contain any interface equality, 
and ``deactivate'' all the $\T$-lemmas containing some interface equality
--- that is, we do not use such $\T$-lemmas for performing unit propagation ---
and we re-activate them only 
when we start splitting on interface equalities again.
Using such strategy, we obtain the same effect as in the simple algorithm
using two DPLL engines: 
the search space is partitioned in two distinct subspaces,
the one of original atoms and the one of interface equalities,
and the generated proof of unsatisfiability reflects such partition.

Finally, we remark that what described above is only \emph{one} 
possible strategy for generating \ielocal proofs, and not necessarily 
the most efficient one. 
{Moreover, that of generating  \ielocal proofs is only a {\em sufficient} 
condition to obtain interpolants from DTC avoiding duplications of 
sub-proofs, and more general strategies may be conceived. 
}
The investigation of alternative strategies is part of ongoing and future work.

\subsection{Discussion}
\label{sec:dtc_discussion}

Our new DTC-based combination method has several advantages over the
traditional one of \cite{musuvathi_interpolation} based on NO:

\begin{enumerate}
\item {It inherits all the advantages of DTC over the
    traditional NO in terms of versatility, efficiency and restrictions
    imposed to \T-solvers \cite{infocomp_dtc05,AMAI08_dtc}. Moreover, it
    allows for using a more modern SMT solver, since many state-of-the-art
    solvers adopt variants or extensions of DTC instead of NO.}

\item {Instead of requiring an ``ad-hoc'' method for
  performing the combination,  it exploits the Boolean interpolation
  algorithm. In
    fact, thanks to the fact that interface equalities occur in the proof of
    unsatisfiability \PP, once the $AB$-mixed terms in \PP are split there is
    no need of any interpolant-combination method at all. In contrast, with
    the NO-based method of \cite{musuvathi_interpolation} interpolants for
    \tonetwo-lemmas are generated by combining ``theory-specific partial
    interpolants'' for the two $\T_i$'s with an algorithm that essentially
    duplicates the work that in our case is performed by the Boolean
    algorithm.
{   
    This allows also for potentially exploiting optimization techniques for
    Boolean interpolation which are or will be made available from the
    literature. }
}
\item {By splitting $AB$-mixed terms only \emph{after} the construction of
    the proof \PP, it allows for computing several interpolants for several
    different partitions of the input problem into $(A, B)$ \emph{from the
      same proof \PP}. This is particularly important for applications in
    abstraction refinement \cite{DBLP:conf/popl/HenzingerJMM04}. (This feature
    is discussed  in \S\ref{sec:dtc_multipleinterpolants}.)}
\end{enumerate}

The work of \cite{musuvathi_interpolation} can in principle deal with
non-convex theories. Our approach is currently limited to the case of
convex theories; however, we see no reason that would prevent from it
being extensible at least theoretically to the case of nonconvex
theories. Extending the approach to non-convex theories is part of
ongoing work. We also remark that implementing the algorithm of
\cite{musuvathi_interpolation} for non-convex theories is a
non-trivial task, and in fact we are not aware of any such
implementation. 

Another algorithm for computing interpolants in combined theories is given 
in \cite{sofronie_interpolation_ijcar}. 
Rather than a combination of theories with disjoint signatures, 
that work considers the interpolation problem for extensions of a base 
(convex) theory with new function symbols, 
and it is therefore orthogonal to ours.
The solution adopted is however similar to what we propose, in the sense that 
also the algorithm of \cite{sofronie_interpolation_ijcar} works by splitting 
$AB$-mixed terms. 
The difference is that our algorithm is tightly integrated in an SMT context,
as it is guided by the resolution proof generated by the DPLL engine.

\subsection{Generating multiple interpolants}
\label{sec:dtc_multipleinterpolants}

{
In \sref{sec:interpolation_smt} we remarked that 
a sufficient condition for generating multiple interpolants 
is that all the interpolants $I_i$'s are computed from the same proof
of unsatisfiability. %
}
{%
When generating interpolants with our DTC-based algorithm,
however, we generate a different proof of unsatisfiability $\PP_i$ 
for each partition of the input formula $\phi$ into $A_i$ and $B_i$.
In particular, every $\PP_i$ is obtained from the same ``base'' proof $\PP$,
by splitting all the $A_iB_i$-mixed interface equalities with the algorithm 
described in \sref{sec:dtc_interpolation}. %
In this section, we show that \eqref{eq:multiple_interpolants} (at \sref{sec:interpolation_smt}) holds 
also when each $\PP_i$ is obtained from the same \ielocal proof $\PP$
by the rewriting of Algorithm 2 of \sref{sec:dtc_ielocal}.
In order to do so, we need the following lemma.
}

\begin{lemma}\label{thm:dtc_interpolant}
  Let $\Theta$ be a \tonetwo-lemma, and let \PP be a \PPie proof for it
  which does not contain any $AB$-mixed term. 
  {Then the formula $I_\Theta$ associated to $\Theta$ in Algorithm 1}
  is an interpolant for $(\neg\Theta\setminus B,\neg\Theta\downarrow B)$.
\end{lemma}
\begin{proof}
  By induction on the structure of \PP, we have to prove that:
  \begin{enumerate}
  \item $\neg\Theta\setminus B \models I_\Theta$;
  \item $I_\Theta \wedge (\neg\Theta\downarrow B)\models\bot$;
  \item $I_\Theta$ contains only common symbols.
  \end{enumerate}

  \noindent
  The base case is when \PP is just a single leaf. Then, the lemma trivially
  holds by definition of $I_\Theta$ in this case (see Algorithm 1).

  \noindent
  For the inductive step, let $\Theta_1 \defas (x=y) \vee \phi_1$ and
  $\Theta_2 \defas \neg(x=y) \vee \phi_2$ be the antecedents of $\Theta$ in
  \PP. (So $\Theta \defas \phi_1 \vee \phi_2$). Let $I_{\Theta_1}$ and
  $I_{\Theta_2}$ be the interpolants for $\Theta_1$ and $\Theta_2$ (by the
  inductive hypothesis).
  \begin{itemize}
  \item If $(x = y) \not\preceq B$, then $I_\Theta \defas I_{\Theta_1} \vee
    I_{\Theta_2}$. 
    \begin{enumerate}
    \item By the inductive hypothesis, $(\neg\phi_1\wedge\neg(x=y))\setminus B
      \equiv (\neg\phi_1\setminus B)\wedge\neg(x=y) \models I_{\Theta_1}$, and
      $(\neg\phi_2\setminus B)\wedge (x=y) \models I_{\Theta_2}$. Then by
      resolution $(\neg\phi_1 \wedge \neg\phi_2)\setminus B \equiv \neg\Theta
      \setminus B \models I_\Theta$.

    \item By the inductive hypothesis, $I_{\Theta_1} \models \phi_1 \downarrow
      B$ and $I_{\Theta_2} \models \phi_2 \downarrow B$, so $I_{\Theta_1}\vee
      I_{\Theta_2} \models (\phi_1 \vee \phi_2) \downarrow B$, that is
      $I_\Theta\wedge(\neg\Theta\downarrow B)\models \bot$.

    \item By the inductive hypothesis both $I_{\Theta_1}$ and $I_{\Theta_2}$
      contain only common symbols, and so also $I_\Theta$ does.
    \end{enumerate}

  \item If $(x=y) \preceq B$, then $I_\Theta \defas I_{\Theta_1} \wedge
    I_{\Theta_2}$.
    \begin{enumerate}
    \item By the inductive hypothesis, $\neg\phi_1\setminus B \models
      I_{\Theta_1}$ and
      $\neg\phi_2 \setminus B\models I_{\Theta_2}$, so $(\neg\phi_1 \wedge
      \neg\phi_2)\setminus B \equiv \neg\Theta\setminus B \models
      I_\Theta$.

    \item By the inductive hypothesis, we also have that $I_{\Theta_1} \models
      \phi_1\downarrow B \vee (x=y)$ and $I_{\Theta_2} \models \phi_2
      \downarrow B \vee \neg(x=y)$. Therefore, $I_{\Theta_1}\wedge
      I_{\Theta_2} \models (\phi_1 \vee \phi_2)\downarrow B$, that is
      $I_\Theta \wedge (\neg\Theta\downarrow B) \models \bot$.

    \item Finally, also in this case both $I_{\Theta_1}$ and $I_{\Theta_2}$
      contain only common symbols, and so also $I_\Theta$ does.
      \qed
    \end{enumerate}
  \end{itemize}
\end{proof}

We now formalize the sufficient condition of
\cite{DBLP:conf/popl/HenzingerJMM04} that \eqref{eq:multiple_interpolants}
holds if the $I_i$'s are computed from the same \PP. The proof of it
will be useful for showing that \eqref{eq:multiple_interpolants} holds also if
the $I_i$'s are computed from $\PP_i$'s obtained from \PP by splitting the
$A_iB_i$-mixed interface equalities.

\newcommand{\ThetaB}{\ensuremath{(\Theta\setminus \phi_3)}\xspace}
\begin{theorem}\label{thm:interpolants_same_proof}
  Let $\phi \defas \phi_1 \wedge \phi_2 \wedge \phi_3$, and let \PP be a proof
  of unsatisfiability for it. Let $A' \defas \phi_1 $, $B'
  \defas \phi_2 \wedge \phi_3$, $A'' \defas \phi_1 \wedge \phi_2$ and $B''
  \defas \phi_3$, and let $I'$ and $I''$ be two interpolants for $(A',B')$
  and $(A'',B'')$ respectively, both computed from \PP. Then
  \begin{displaymath}
    I' \wedge \phi_2 \models I''.
  \end{displaymath}
\end{theorem}
\begin{proof}
  Let $\PP_\Theta$ be a proof whose root is the clause $\Theta$. We will
  prove, by induction on the structure of $\PP_\Theta$, that
  \begin{displaymath}
    I'_\Theta \wedge \phi_2 \models I''_\Theta \vee \ThetaB,    
  \end{displaymath}
  where $I_\Theta$ is defined as in Algorithm 1. The validity of the theorem
  follows immediately, by observing that the root of \PP is $\bot$.
  
  \smallskip\noindent
  We have to consider three cases:
  \begin{enumerate}
  \item The first is when $\Theta$ is an input clause. Then, we have three
    subcases:
    \begin{enumerate}
    \item If $\Theta \in \phi_3$, then $I'_\Theta \defas \top$, $I''_\Theta
      \defas \top$ and $\ThetaB \equiv \bot$, so the theorem holds.

    \item If $\Theta \in \phi_1$, then $I'_\Theta \defas (\Theta \downarrow
      (\phi_2\cup \phi_3))$, $I''_\Theta \vee \ThetaB \defas (\Theta
      \downarrow \phi_3) \vee \ThetaB \equiv \Theta$, so the theorem holds
      also in this case.
    
    \item If $\Theta \in \phi_2$, then  $I'_\Theta \wedge \phi_2 \equiv \phi_2$
      and $I''_\Theta \vee \ThetaB \equiv \Theta$, so again the implication
      holds.
    \end{enumerate}

  \item The second is when $\Theta$ is a \T-lemma. In this case, we have that
    $I'_\Theta$ is an interpolant for $(\neg\Theta \setminus
    (\phi_2\cup\phi_3), \neg\Theta\downarrow (\phi_2\cup\phi_3))$ and
    $I''_\Theta$ is an interpolant for $(\neg\Theta \setminus \phi_3,
    \neg\Theta\downarrow \phi_3)$. Therefore, by the definition of
    interpolant, $(\neg\Theta \setminus (\phi_2\cup \phi_3)) \models
    I'_\Theta$ and $(\neg\Theta \setminus \phi_3) \models
    I''_\Theta$. Therefore, $I'_\Theta \vee (\Theta \setminus (\phi_2\cup
    \phi_3))$ and $I''_\Theta \vee \ThetaB$ are valid clauses, and so the
    implication trivially holds.

  \item In this case $\Theta$ is obtained by resolution from $\Theta_1 \defas
    \phi \vee p$ and $\Theta_2 \defas \psi \vee \neg p$. If $p \in \phi_1$ or
    $p \in \phi_3$, then by the inductive hypotheses that $I'_{\Theta_i}
    \wedge \phi_2 \models I''_{\Theta_i} \vee (\Theta_i\setminus \phi_3)$, we
    have that $I'_\Theta \wedge \phi_2 \models I''_\Theta \vee \ThetaB$.

    If $p \in \phi_2$, then $I'_\Theta \defas I'_{\Theta_1} \wedge
    I'_{\Theta_2}$ and $I''_\Theta \defas I''_{\Theta_1} \vee
    I''_{\Theta_2}$. Again, by the inductive hypotheses $I'_\Theta \wedge
    \phi_2 \models I''_\Theta \vee \ThetaB$ holds.
    \qed
  \end{enumerate}
\end{proof}

\begin{theorem}\label{thm:dtc_many_interpolants}
  Let $\phi \defas \phi_1 \wedge \phi_2 \wedge \phi_3$. Let $A' \defas
  \phi_1$, $A'' \defas \phi_1 \wedge \phi_2$, $B' \defas \phi_2 \wedge
  \phi_3$, and $B'' \defas \phi_3$. Let \PP be a proof of unsatisfiability for
  $\phi$, and let $\PP'$ and $\PP''$ be obtained from \PP by splitting all
  the $A'B'$-mixed and $A''B''$-mixed interface equalities respectively. Let
  $I'$ be an interpolant for $(A',B')$ computed from $\PP'$, and $I''$ be an
  interpolant for $(A'',B'')$ computed from $\PP''$. Then
  \begin{displaymath}
    I' \wedge \phi_2 \models I''.
  \end{displaymath}
\end{theorem}
\begin{proof}
  We observe that $\PP'$ and $\PP''$ are identical except for some \PPie
  subproofs that contained some mixed interface equalities. Then, we can
  proceed as in Theorem~\ref{thm:interpolants_same_proof}, we just need to
  consider one more case, namely when $\Theta$ is a \tonetwo-lemma at
  the root of a \PPie subproof. In this case, thanks to
  Lemma~\ref{thm:dtc_interpolant} we have the same situation as in the second
  case of the proof of Theorem~\ref{thm:interpolants_same_proof}, and so we
  can apply the same argument.
  \qed
\end{proof}

Thus, due to Theorem~\ref{thm:dtc_many_interpolants}, we can use our
DTC-based interpolation method in the context of abstraction refinement
without any modification: it is enough to remember the original proof \PP, and
compute the interpolant $I_i$ from the proof $\PP_i$ obtained by
splitting the $A_iB_i$-mixed terms in \PP, for each partition of the input
formula $\phi$ into $A_i$ and $B_i$ as in \eqref{eq:phi_ai_bi}.


\section{Experimental evaluation}
\label{sec:expeval}

The techniques presented in previous sections have been implemented
within \mathsat 4 \cite{mathsat4} 
\mathsat is an SMT solver supporting a wide range of theories and their combinations. 
In the last SMT solvers competition (SMT-COMP'08), 
it has proved to be competitive with the other state-of-the-art solvers.
In this Section, we experimentally evaluate our approach.

\subsection{Description of the benchmark sets}
\label{sec:benchmarks}

We have performed our experiments on two different sets of benchmarks.
The first is obtained by running the \blast software
model checker \cite{blast} on some Windows
device drivers; these are similar to those used in
\cite{rybalchenko_interp}. This is one of the most important
applications of interpolation in formal verification, namely
abstraction refinement in the context of CEGAR. The problem represents
an abstract counterexample trace, and consists of a conjunction of
atoms. In this setting, the interpolant generator is called very
frequently, each time with a relatively simple input problem.

The second set of benchmarks originates from the
SMT-LIB~\cite{RanTin-SMTLIB}, and is composed of a subset of the unsatisfiable
problems used in recent SMT solvers competitions
(\url{http://www.smtcomp.org}). The instances have been converted to
CNF and then split in two consistent parts of approximately the same size. The
set consists of problems of varying difficulty and with a nontrivial
Boolean structure.

The experiments have been performed on a 3GHz Intel Xeon machine with 4GB of
RAM running Linux. All the tools were run with a timeout of 600 seconds and a
memory limit of 900 MB.
All the benchmark instances, the \mathsat executable, and the set of
scripts used to perform the experiments are available at
{\small\url{http://disi.unitn.it/~griggio/papers/tocl_itp.tar.bz2}}.%

\subsection{Comparison with the state-of-the-art tools available}
\label{sec:mathsat_vs_others}

\begin{figure}[t]
  \centering
  \begin{tabular}{|c|c|c|c|c|c|}
    \hline
    \textbf{Family} & \textbf{\# of problems} & \textbf{\mathsat} &
    \textbf{\foci} & \textbf{\clp} & \textbf{\csisat} \\
    \hline
    kbfiltr.i & 64 & 0.16 & 0.36 & 1.47 & 0.17 \\
    diskperf.i & 119 & 0.33 & 0.78 & 3.08 & 0.39 \\
    floppy.i & 235 & 0.73 & 1.64 & 5.91 & 0.86 \\
    cdaudio.i & 130 & 0.35 & 1.07 & 2.98 & 0.47 \\
    \hline
  \end{tabular}
  \caption{Comparison of execution times of \mathsat, \foci, \clp and
    \csisat on
    problems generated by \blast.
    \label{fig:mathsat_foci_clp_blast}}
\end{figure}

\begin{figure}[t]
  \centering
      \begin{tabular}{cc@{\hspace{1cm}}c}
        & Execution Time & Size of the Interpolant \\
        \rotatebox{90}{\hspace{24mm}\foci} &
        \includegraphics[scale=0.7,bb=95 50 300 251,clip]
        {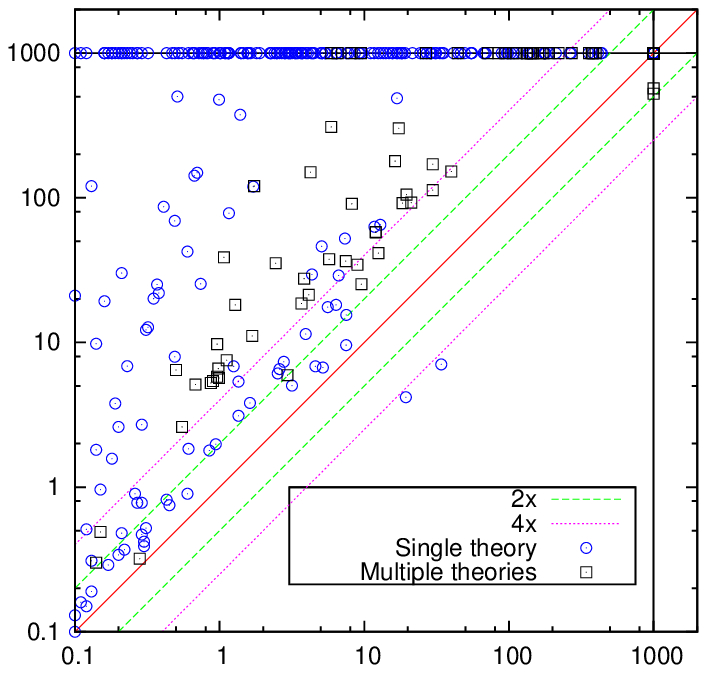} &
        \includegraphics[scale=0.7,bb=90 50 300 251,clip]
        {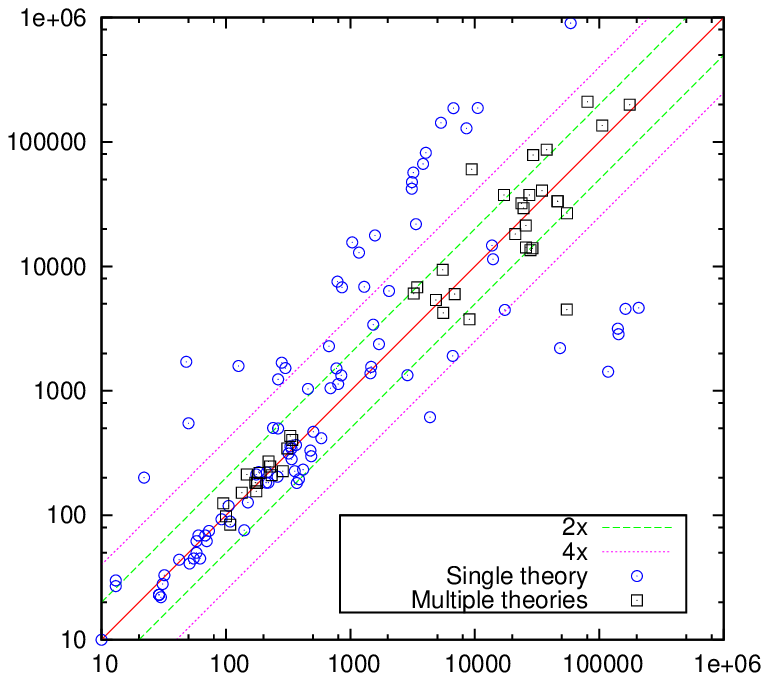} \\
        & \mathsat & \mathsat
      \end{tabular}  
  \caption{Comparison of \mathsat and \foci on SMT-LIB
    instances: execution time (left), and size of the interpolant
    (right). In the left plot, points on the horizontal and vertical lines are
    timeouts/failures. 
    \label{fig:mathsat_vs_foci}}    
\end{figure}

\begin{figure}[t]
  \centering
      \begin{tabular}{cc}
        & Execution Time \\
        \rotatebox{90}{\hspace{16mm}\clp} &
        \includegraphics[scale=0.8,bb=95 50 300 251,clip]
        {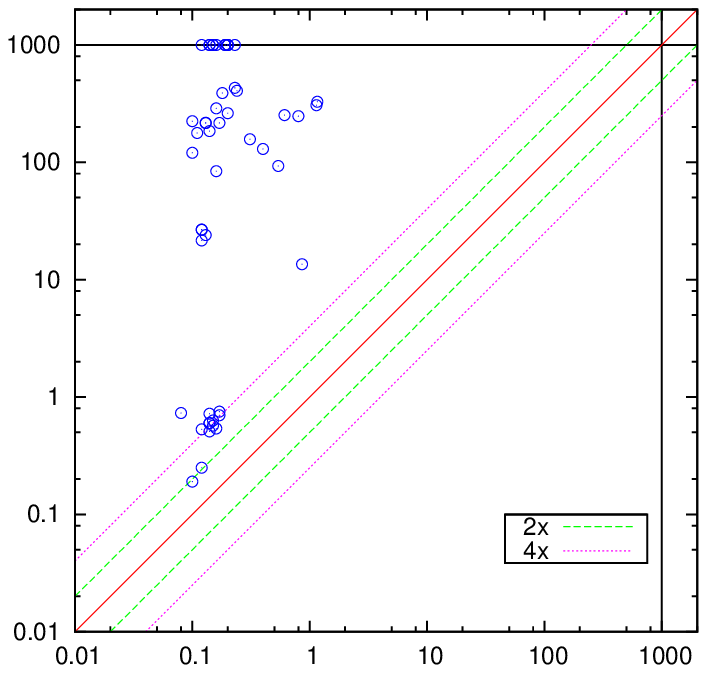} \\
        & \mathsat
      \end{tabular}
  \caption{Comparison of \mathsat and \clp on conjunctions of \laR atoms.
    \label{fig:mathsat_vs_clp}}      
\end{figure}

\begin{figure}[t]
  \begin{minipage}[t]{0.48\linewidth}
      \begin{tabular}{cc}
        & Execution Time \\
        \rotatebox{90}{\hspace{23mm}\csisat} &
        \includegraphics[scale=0.57,bb=110 50 360 302]
        {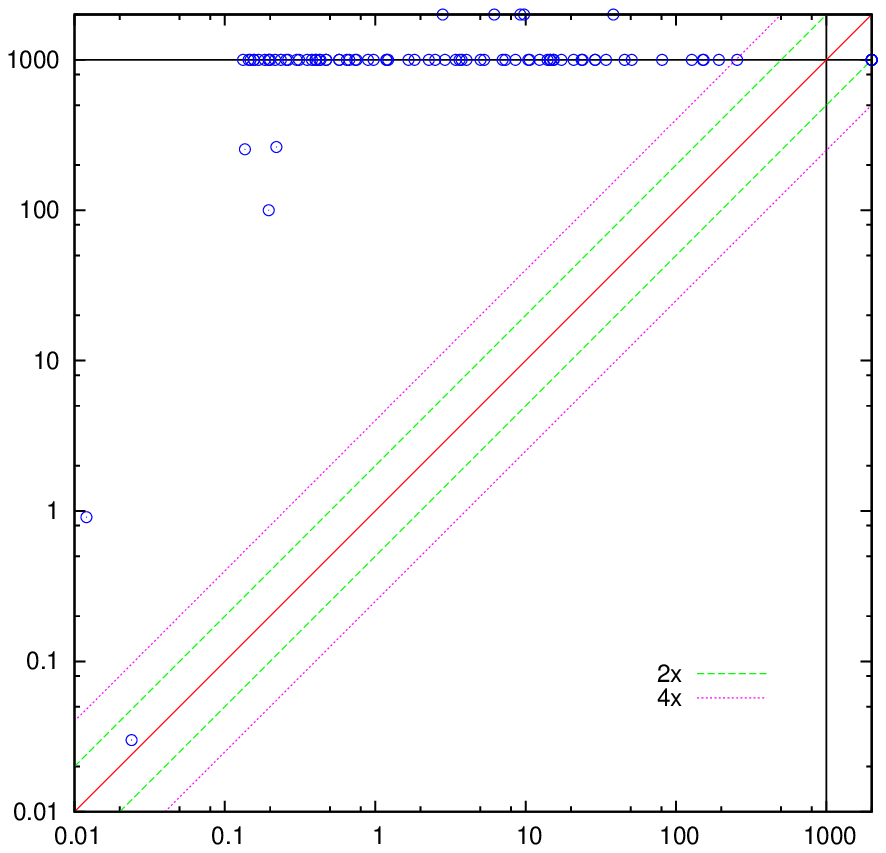} \\
        & \mathsat
      \end{tabular}
      \caption{Comparison of \mathsat and \csisat on SMT-LIB instances.
        \label{fig:mathsat_vs_csisat_smtlib}}
  \end{minipage}
  \hfill
  \begin{minipage}[t]{0.48\linewidth}
      \begin{tabular}{cc}
        & Execution Time \\
        \rotatebox{90}{\hspace{23mm}\csisat} &
        \includegraphics[scale=0.5]
        {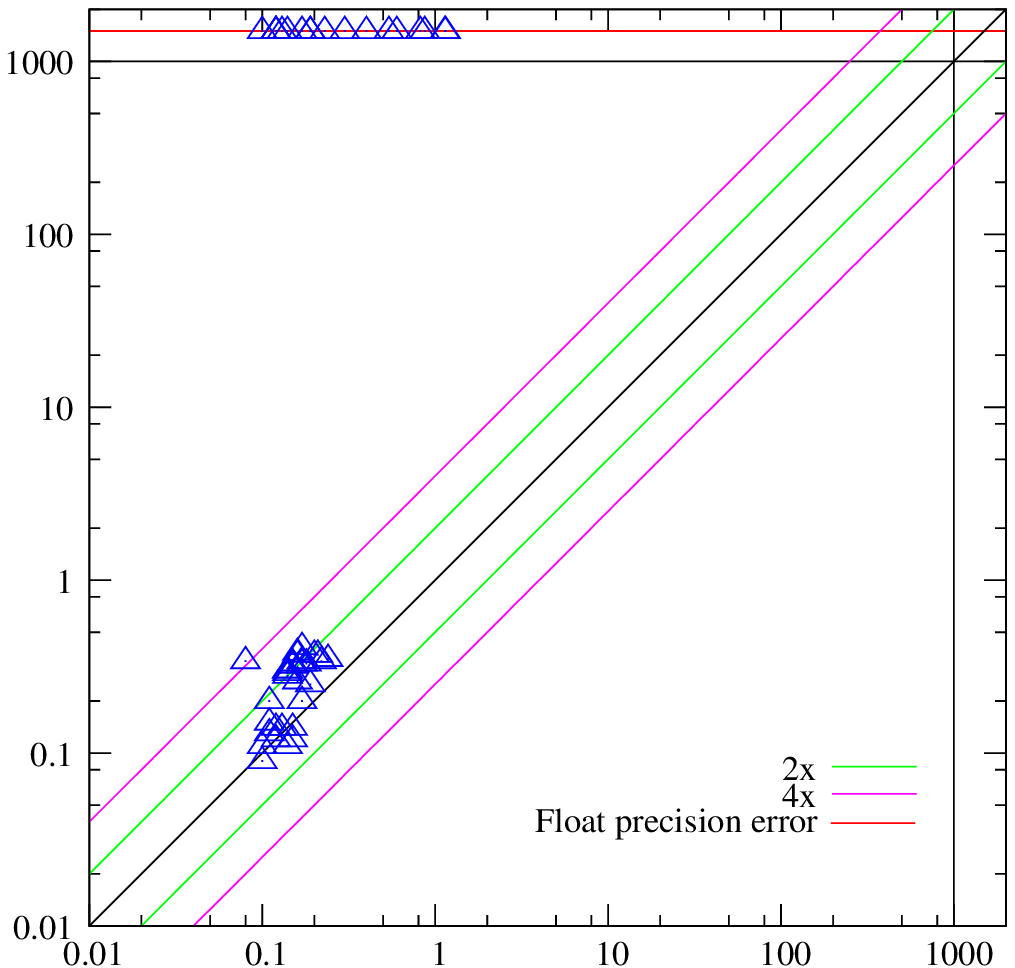} \\
        & \mathsat
      \end{tabular}
      \caption{Comparison of \mathsat and \csisat on conjunctions of \laR atoms.
        \label{fig:mathsat_vs_csisat}}
  \end{minipage}
\end{figure}

In this section, we compare with the other interpolant generators which are
available: \foci \cite{mcmillan_interpolating_prover,mcmillan_tacas06}, \clp
\cite{rybalchenko_interp} {and \csisat \cite{csisat}}.  Other natural
candidates for comparison would have been \zap~\cite{zap} and
\lifter~\cite{kroening_interp}; however, it was not possible to obtain them
from the authors.  
{We also remark that no comparison with \intjain \cite{jain_cav08} is
  possible, since the domains of applications of \mathsat and \intjain are
  disjoint: \intjain can handle \laint
  equations/disequations and modular equations but only conjunctions of
  literals, whereas \mathsat can handle formulas with arbitrary Boolean
  structure, but does not support \laint except for its fragments
  \dlint and \utvpiint.}

The comparison had to be adapted to the limitations of \foci, \clp
and \csisat. %
In fact, the current version of \foci 
which is publically available
does not handle the full \laR, but only
the \dlrat fragment\footnote{For example, it fails to detect {the
    \laR-}unsatisfiability of the following problem: $(0 \leq y - x + w)
  \wedge (0 \leq x - z - w) \wedge (0 \leq z - y - 1)$ .}. We also notice that
the interpolants it generates are not always \dlrat formulas. (See, e.g.,
Example~\ref{ex:no_dl_itp} of Section \ref{sec:dl}.)
\clp %
does handle the full \laR, but it accepts
only conjunctions of atoms, rather than formulas with arbitrary Boolean
structure.
\csisat, instead, can deal with $\euf\cup\larat$ formulas with
  arbitrary Boolean structure, but it does not support Boolean variables.  
These limitations made it impossible to compare all the four tools on
all the instances of our benchmark sets. Therefore, we perform the
following comparisons:

\begin{itemize}
\item[--] We compare all the four solvers on the problems generated by
  \blast;
\item[--] We compare \mathsat with \foci on SMT-LIB instances in the theories
  of \euf, \dlrat and their combination. In this case, we compare both the
  execution times and the sizes of the generated interpolants (in terms of
  number of nodes in the DAG representation of the formula). For computing
  interpolants in \euf, we apply the algorithm of
  \cite{mcmillan_interpolating_prover}, using an extension of the algorithm of
  \cite{NieuwenhuisOliverasIC2007} to generate \euf proof trees. The
  combination $\euf\cup\dlrat$ is handled with the technique described in
  \S\ref{sec:dtc};
\item[--] We compare \mathsat, \clp and \csisat on \laR problems
  consisting of conjunctions of atoms. These problems are single branches of
  the search trees explored by \mathsat for some \laR instances in the
  SMT-LIB. We have collected several problems that took more than $0.1$
  seconds to \mathsat to solve, and then randomly picked $50$ of them. In this
  case, we do not compare the sizes of the interpolants as they are always
  atomic formulas;
\item[--] We compare \mathsat and \csisat on the
    subset (Consisting of 78 instances of the about 400 collected)
    of the SMT-LIB instances without Boolean variables. 
\end{itemize}

\noindent
The results are collected in Figures~\ref{fig:mathsat_foci_clp_blast},
\ref{fig:mathsat_vs_foci}, \ref{fig:mathsat_vs_clp},
\ref{fig:mathsat_vs_csisat_smtlib}
 and \ref{fig:mathsat_vs_csisat}. We can observe the following facts:
\begin{itemize}
\item[--] Interpolation problems generated by \blast are trivial for all the
  tools. In fact, we even had some difficulties in measuring the execution
  times reliably. Despite this, \mathsat and \csisat seem to be a little
  faster than the others.
\item[--] For problems with a nontrivial Boolean structure, \mathsat
  outperforms \foci in terms of execution time. This is true even for problems
  in the combined theory $\euf\cup\dlrat$, despite the fact that the current
  implementation is still preliminary. 

  {
  As regards \csisat, it could solve
  (within the time and memory limits) only 5 of the 78 instances it could
  potentially handle, and in all cases \mathsat outperforms it.}

\item[--] In terms of size of the generated interpolants, the gap between
  \mathsat and \foci is smaller on average. However, the right plot of
  Figure~\ref{fig:mathsat_vs_foci} (which considers only instances for which
  both tools were able to generate an interpolant) shows that there are more
  cases in which \mathsat produces a smaller interpolant.
\item[--] On conjunctions of \laR atoms, \mathsat outperforms \clp, sometimes
  by more than two orders of magnitude. The performance of
    \mathsat and \csisat is comparable on such instances, 
    {with \mathsat being slightly faster. }
    However, there
    are several cases in which \csisat computes a wrong result, due to the use
    of floating-point arithmetic instead of infinite-precision arithmetic
    (which is used by \mathsat).
\end{itemize}
 
%
%


\section{Conclusions and Future Work}
\label{sec:concl}

In this paper, we have shown how to efficiently build interpolants
using state-of-the-art SMT solvers. Our methods encompass a wide range
of theories (including \euf, \dl,
\utvpi, and \la), and their
combination (based on the Delayed Theory Combination schema).
A thorough experimental evaluation shows that the proposed methods
retain the efficiency of the solvers, and are vastly superior to the
state of the art interpolants, both in terms of expressiveness, and in
terms of efficiency.

In the future, we plan to investigate the following issues. First, we
will improve the implementation of the interpolation method for
combined theories, that is currently rather na{\"i}ve, and limited to
the case of convex theories.
Second, we will investigate interpolation with other rules, in
particular Ackermann's expansion.
Finally, we will integrate our interpolator within a CEGAR loop based
on decision procedures, such as BLAST or the new version of NuSMV. In
fact, such an integration raises interesting problems related to
controlling the structure of the generated
interpolants~\cite{mcmillan_tacas06,mcmillan_cav07}, e.g. in order to
limit the number or the size of constants occurring in the proof.


\bibliographystyle{acmtrans}
\bibliography{bibliography}

\end{document}